\documentclass[acmtog,nonacm]{acmart}

\AtBeginDocument{%
  \providecommand\BibTeX{{%
    \normalfont B\kern-0.5em{\scshape i\kern-0.25em b}\kern-0.8em\TeX}}}

\usepackage{epsfig}
\usepackage{graphicx}
\usepackage{amsmath}
\usepackage{bm}
\usepackage[english]{babel}
\usepackage{commath} 
\usepackage{hhline} 
\usepackage{colortbl}
\usepackage[ruled]{algorithm2e} 
\renewcommand{\algorithmcfname}{ALGORITHM}
\SetAlFnt{\small}
\SetAlCapFnt{\small}
\SetAlCapNameFnt{\small}
\SetAlCapHSkip{0pt}
\IncMargin{-\parindent}
\usepackage{adjustbox}
\usepackage{multirow}
\UseRawInputEncoding
\usepackage{caption}
\usepackage[maxfloats=40]{morefloats}
\usepackage{comment}
\usepackage{enumitem}
\setlist{nosep}

\def\I{\mathbf{I}}
\def\II{\mathbf{II}}

\def\mat#1#2#3#4{\left(\!\!\begin{array}{c@{~~~}c} #1 & #2 \\ #3 & #4 \end{array}\!\!\right)}

\definecolor{LightGray}{gray}{0.9}

\definecolor{maroon}{rgb}{0.5, 0.0, 0.0}

\def\mathbi#1{\textbf{\em #1}}

\begin{document}

\title{Non-Photorealistic Rendering of Layered Materials: A Multispectral Approach}


\author{Corey Toler-Franklin}
\orcid{1234-5678-9012-3456}
\email{ctoler@cise.ufl.edu}
\author{Shashank Ranjan}
\affiliation{%
\institution{University of Florida}
\country{USA}}

\renewcommand{\shortauthors}{Toler-Franklin, C.  et al.}



\begin{abstract}

We present multispectral rendering techniques for visualizing layered materials found in biological specimens. We are the first to use acquired data from the near-infrared and ultraviolet spectra for non-photorealistic rendering (NPR).  Several plant and animal species are more comprehensively understood by multispectral analysis. However, traditional NPR techniques ignore unique information outside the visible spectrum. We introduce algorithms and principles for processing wavelength dependent surface normals and reflectance. Our registration and feature detection methods are used to formulate stylization effects not considered by current NPR methods including: \emph{Spectral Band Shading} which isolates and emphasizes shape features at specific wavelengths at multiple scales. Experts in our user study demonstrate the effectiveness of our system for applications in the biological sciences.

\end{abstract}

%
%


%
%



\begin{teaserfigure}
\centering
\includegraphics[height=1.3in]{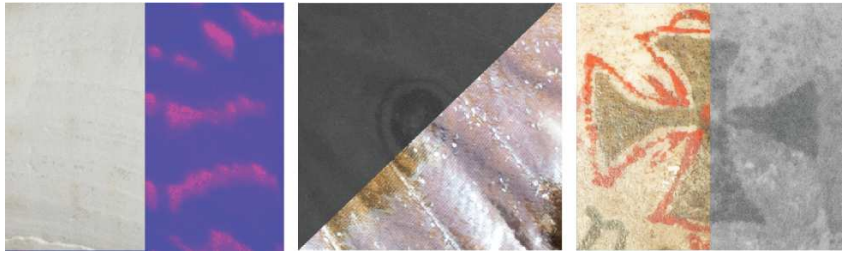}
  \caption{Examples of biological specimens that are analyzed with  multispectral analysis in life science research. (left) UV florescence reveals hidden color patterns in a shell fossil. (middle) Infrared imaging reveals circular patterns on the undersurface of a butterfly wing and (right) bone under red ceremonial paint. Specimen attribution (left to right) Conus delesertti fossil UF117269, Florida Museum of Natural History (FLMNH). Morpho sulkowsly butterfly, McGuire Center for Lepidoptera and Biodiversity, FLMNH. Painted cranium, catalog number VL/122, Courtesy of the Division of Anthropology, American museum of Natural History (AMNH).} 
\label{fig:teaser}

\end{teaserfigure}


\maketitle


\section{introduction}
\label{sec:intro}

We address the technical problem of extracting and analyzing shape in layered materials. Our work is inspired by research applications that use multispectral imaging to examine biological materials. It is often important to examine the surface of materials under layers of debris from erosion and other environmental conditions. The biological specimen itself is often composed of composite materials that require multispectral analysis for comprehensive study. For example, ultraviolet fluorescence reveals original color patterns on colorless fossils for species classification~\cite{Hendricks2015} (Figures~\ref{fig:teaser} \emph{left} and~\ref{fig:uvandsamples}). Infrared imaging permits study of subsurface materials hidden under pigments or dyes~\cite{baker2012} (Figure~\ref{fig:teaser} \emph{middle} and \emph{right}). These hidden features provide additional clues about the specimen's physiological characteristics. 

We introduce algorithms that combine multispectral imaging with NPR to communicate these otherwise invisible characteristics of our dataset. 
Our approach accentuates multispectral information at different material layers in RGBN images~\cite{rgbn07} for stylized rendering, biological illustration and scientific study. NPR is effective for scientific illustration~\cite{Gooch98} because it removes extraneous information (abstraction) while emphasizing relevant details (shape enhancement). Multispectral data has been used to improve detail and color fidelity~\cite{Darling2011} in realistic image synthesis, but has been overlooked in NPR. NPR algorithms still operate on visible-only photos and \hbox{3-D} geometry. NPR techniques for volumetric data~\cite{Joshi07,Corcoran2010} are also limited as they do not formulate stylization rules for signature enhancement across multiple wavelengths.  Our goal is to detect and maximize detailed information at different wavelengths. We incorporate contrast enhancement and spectral shape differences to increase fidelity of subtle near-infrared signatures~\cite{albanese2011}. We also record ultraviolet bispectral shape variations at different magnifications~\cite{Tabata1996} to enhance object scale features. Our contributions include:
\begin{itemize}
\item Principles and tools for processing multispectral data for NPR including: \hbox{2D} to \hbox{3D} alignment that minimizes shape distortion and near-infrared enhancement maps that detect and control shape detail in layered materials. (Section~\ref{sec:msprocessing}).  
\item Multispectral stylization techniques including: spectral band shading that reveals subsurface shape detail, bispectral shading that emphasizes object scale features at different magnifications and multiscale curvature shading and line drawing. (Section~\ref{sec:shading}). 
\item A user study that validates the utility of our approach for applications in biology and related life science research. (Section~\ref{sec:applications}). 
\end{itemize}

\section{Principles for Multispectral NPR}
\label{sec:principles} 

\noindent\textbf{Pilot Study} We  began our research by observing eighteen experts (practitioners and researchers) from four disciplines - biology ($5$), anthropology ($5$), forensics ($3$) and paleontology ($5$) -  in their work environments to understand challenges with real-world applications of multispectral analysis in life science research (see supplemental video). We identified three modes of analysis routinely used across these groups: HDR contrast imaging to enhance near-infrared detail while suppressing visible data; the use of fluorescent powders and ultraviolet radiation to reveal latent bio-materials; and micro imaging to examine biological structures. Our experts demonstrated advantages and limitations of these methods for critical research tasks. Inefficiencies included lack of fidelity when working with post processing software (indicating residual pigments with false color~\cite{Hendricks2015} is a well-known example), detail loss at longer wavelengths when denoising images, lack of user control over enhancement parameters and reliance on destructive DNA sequencing methods when micro analysis is ineffective. To address these challenges, we formulated principles and illustrative tools that combine flexible NPR parameter controls with measured multispectral properties, without the complexities (computation, time and expertise) of physically-based rendering solutions. Our principles follow these general processing rules:

\begin{enumerate}

\item \label{itm:filtering} Apply only minimal filtering to near-infrared  data (\hbox{2-D} spectra and normals) to prevent loss of subtle detail. 

\item \label{itm:contrastenhancement} Emphasize subsurface detail in regions where near-infrared contrast is greater than contrast in visible data. 

\item \label{itm:uvpatterndetection} Use bispectral reflectance to emphasize structural differences between materials.
 
\item \label{itm:microdetail} Incorporate object-scale shape from different magnifications for complex lighting effects. 

\item \label{itm:measureddata} Control stylization effects using 
measured parameters from real sources for authenticity.

\end{enumerate}

\noindent Additional rules are required for analyzing molecular and atomic structures. We focus on non-invasive visualization methods for understanding shape, material composition and anatomical structure.


\section{Overview}
\label{sec:overview}

\begin{figure}[ht]
\centering
\includegraphics[width=1.0\hsize]{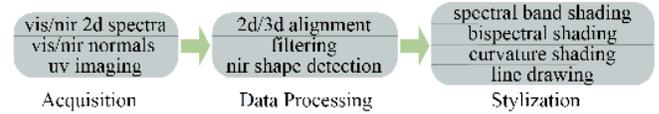}
\setlength\abovecaptionskip{-0.7\baselineskip}
\setlength{\belowcaptionskip}{-10pt} 
\caption{\label{fig:overview}%
Multispectral NPR Rendering Pipeline}
\end{figure}

Figure~\ref{fig:overview} presents a system overview. We compute the shape of biological materials in the visible, ultraviolet and infrared spectra. Our diverse dataset includes butterfly scales, bone, plants and  fossilized materials (cone shells) dating back millions of years. We first demonstrate how current shape from shading approaches fail for our application because they overlook variations in shape that occur at different material layers, and ignore spatially varying reflectance from surfaces with varying material composition. We then explain how we compute normals at different material layers as a function of the emission wavelength of our light source. Our experimental set-up quantitatively validates the accuracy of our normals. Next, we implement \hbox{2-D} to \hbox{3-D} alignment across spectral bands, filtering, per-pixel feature detection and near-infrared enhancement. These pre-processes are incorporated into novel NPR stylization methods. The results simulate reflectance properties that occur under different electromagnetic wavelengths which influence 
behavior, environmental adaptations~\cite{Kastberger2003} and species classification. Finally, we conduct a user study with experts in biology and related fields to demonstrate the utility of our approach for life science research.


\begin{figure}[t]
\centering
\includegraphics[width=1.0\hsize]{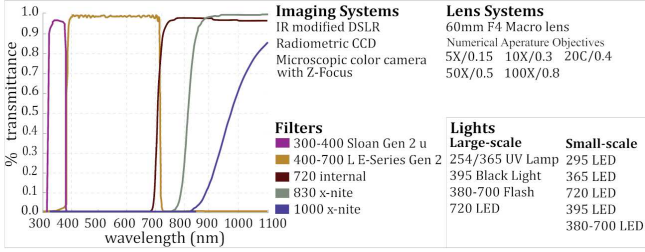}
\setlength\abovecaptionskip{-0.7\baselineskip}
\setlength{\belowcaptionskip}{-10pt} 
\caption{\label{fig:transcurves}%
Equipment: (Left) Imagers and filters. (Right) Light sources.}
\end{figure}

\begin{figure}[ht]
\centering
\includegraphics[width=0.8\hsize]{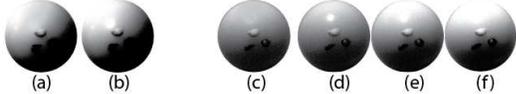}
\setlength\abovecaptionskip{3pt}
\setlength{\belowcaptionskip}{-14pt} 
\caption{\label{fig:highlight}%
Remove Highlight (a) result (b) original (c-f) result, $Ev_{0}$, $Ev_{1}$, $Ev_{2}$}
\end{figure}
\vspace{2mm}

\begin{figure}[ht]
\centering
\def\imh{0.4in}
\setlength{\tabcolsep}{0.2pt}
\newcolumntype{?}{!{\vrule width 0.25pt}}
\newcolumntype{C}[1]{>{\centering\let\newline\\\arraybackslash\hspace{0pt}}m{#1}}
\newcolumntype{R}[1]{>{\raggedright\let\newline\\\arraybackslash\hspace{0pt}}m{#1}}
\newcolumntype{L}[1]{>{\raggedleft\let\newline\\\arraybackslash\hspace{0pt}}m{#1}}
\begin{tabular}{C{0.2in}C{0.5in}R{0.45in}?C{0.53in}?C{0.53in}?L{0.45in}C{0.5in} C{0.2in}}

\multicolumn{1}{c}{\rotatebox{90}{}}
& \multicolumn{2}{c}{\multirow{2}{*}{\small GT}} 
& \multicolumn{1}{c}{\multirow{2}{*}{\small Traditional}} 
& \multicolumn{1}{c}{\multirow{2}{*}{\small Takatani13}} 
& \multicolumn{2}{c}{\small proposed} 
& \tabularnewline


{\rotatebox{90}{}}& & & & &\footnotesize $\mathbf{n}_{vis}$ &\footnotesize $\mathbf{n}_{nir}$ & \\

\multirow{3}{*}{\rotatebox{90}{\centering $spectralon$}}
&\includegraphics[height=\imh]{images/shapevalidation/spectralon/blank.eps}
&\includegraphics[height=\imh]{images/shapevalidation/spectralon/blank.eps}
&\includegraphics[height=\imh]{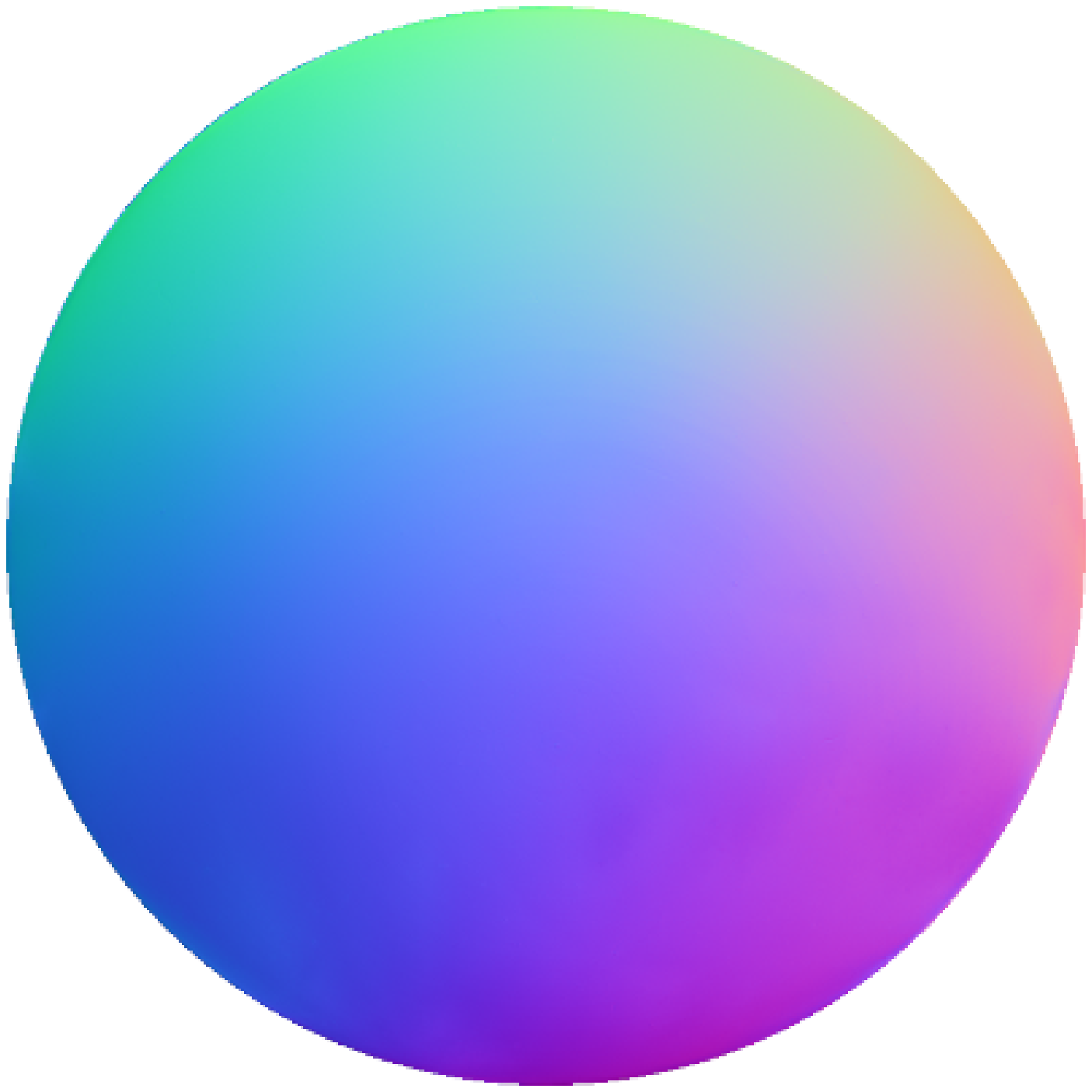}
&\includegraphics[height=\imh]{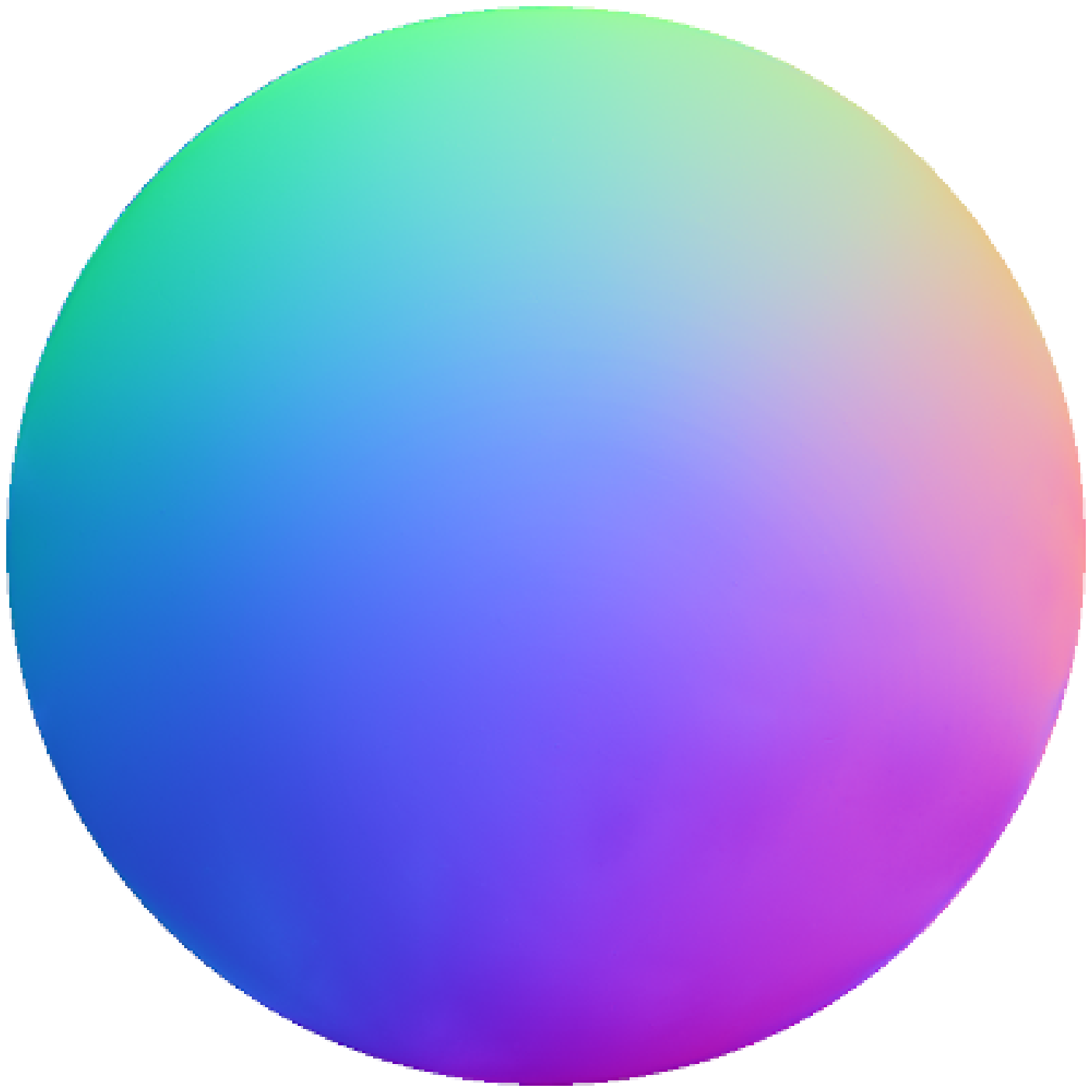}
&\includegraphics[height=\imh]{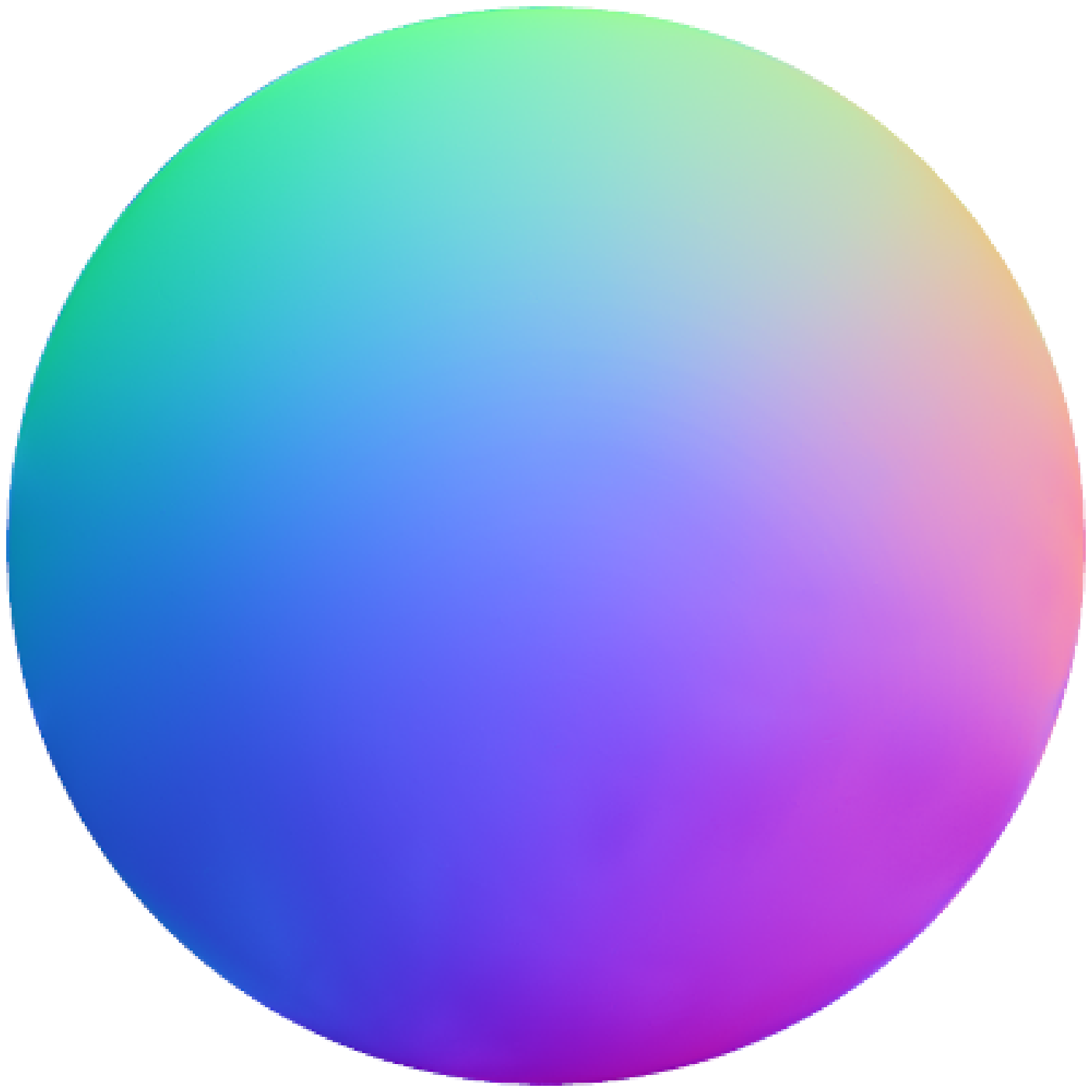}
&\includegraphics[height=\imh]{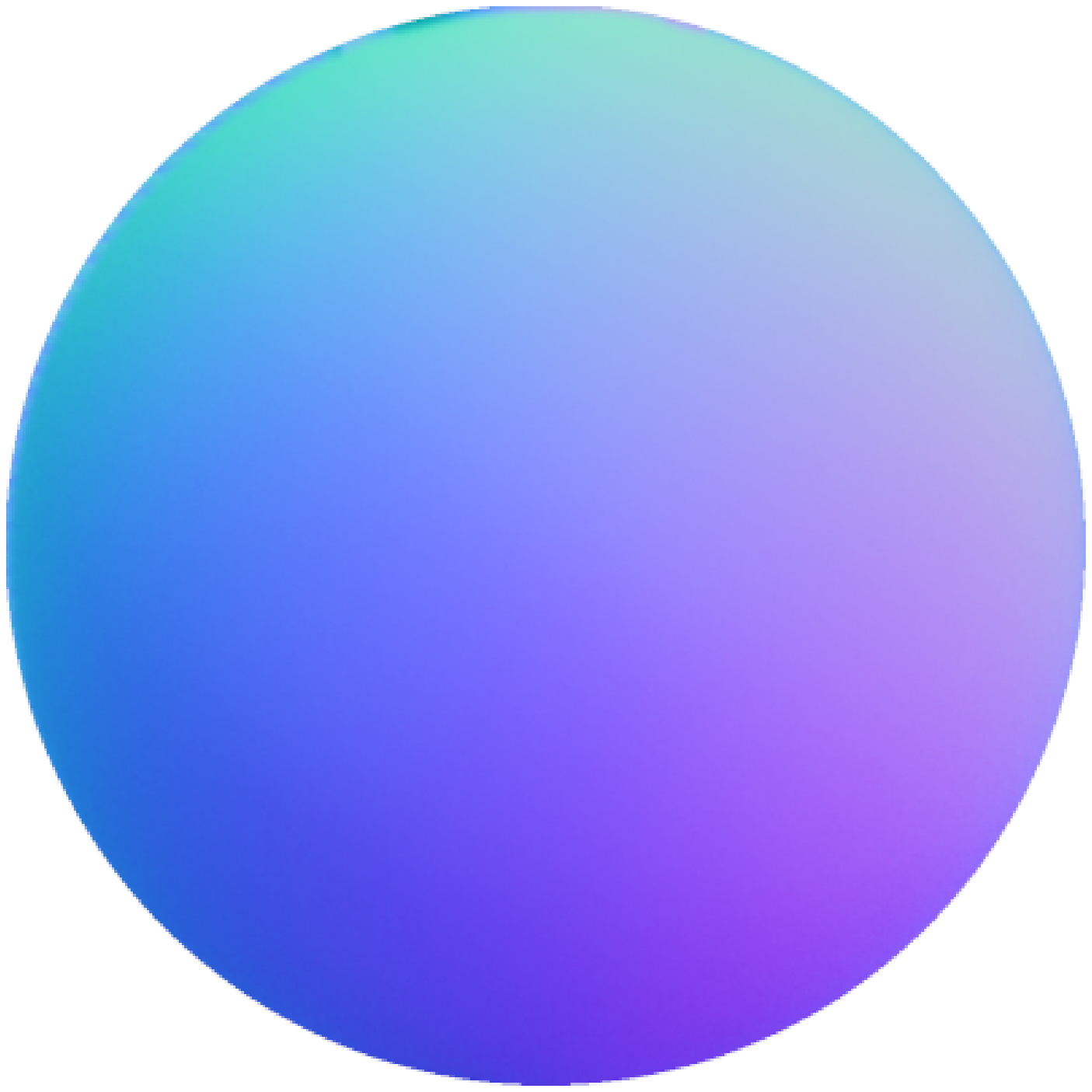}
&\multirow{3}{*}{\includegraphics[height=0.8in]{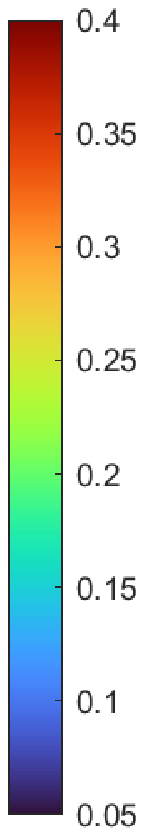}}
\\

&\includegraphics[height=\imh]{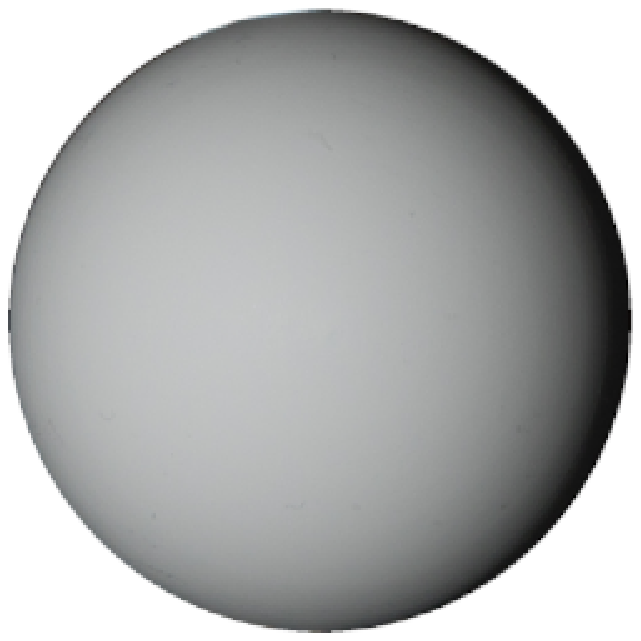}
&\includegraphics[height=\imh]{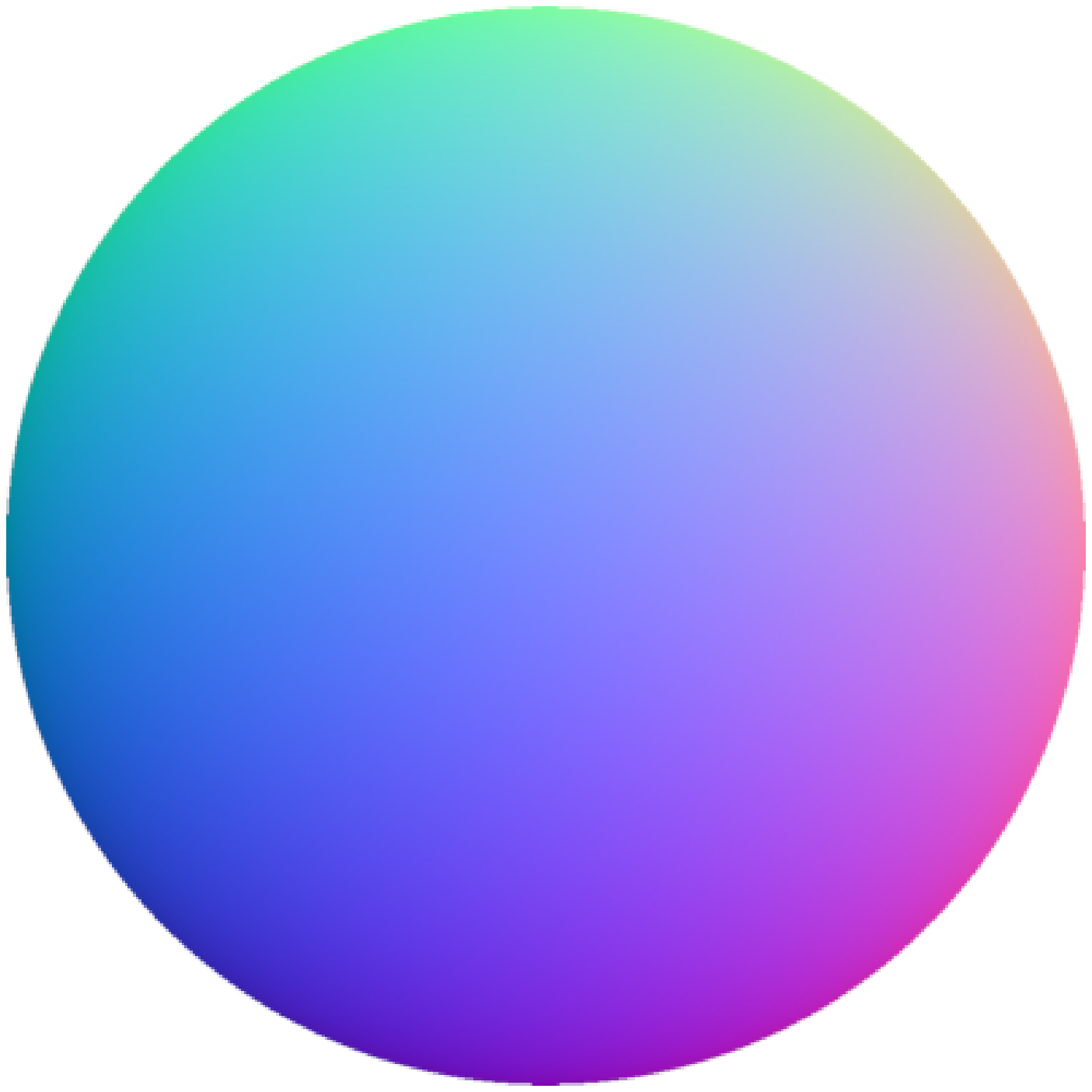}
&\includegraphics[height=\imh]{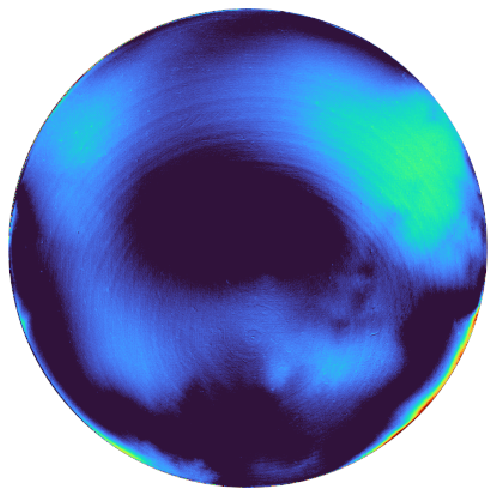}
&\includegraphics[height=\imh]{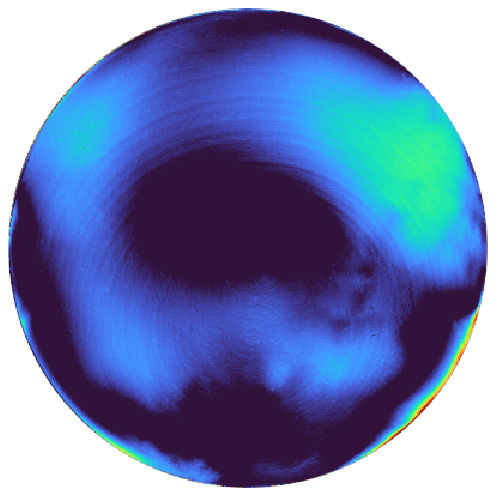}
&\includegraphics[height=\imh]{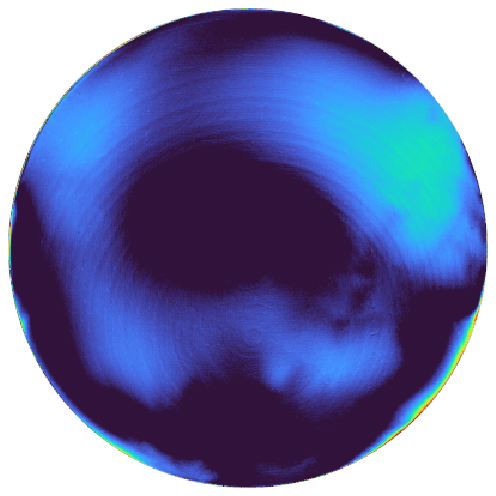}
&\includegraphics[height=\imh]{images/shapevalidation/spectralon/blank.eps}
&
\\

\vspace{8pt}

& & & $5.31$ & $5.29$ & $5.37$ & & \\


\multirow{5}{*}{\rotatebox[origin=c]{90}{$ptfe_{red}$}}
&\includegraphics[height=\imh]{images/shapevalidation/ptfe/blank.eps}
&\includegraphics[height=\imh]{images/shapevalidation/ptfe/blank.eps}
&\includegraphics[height=\imh]{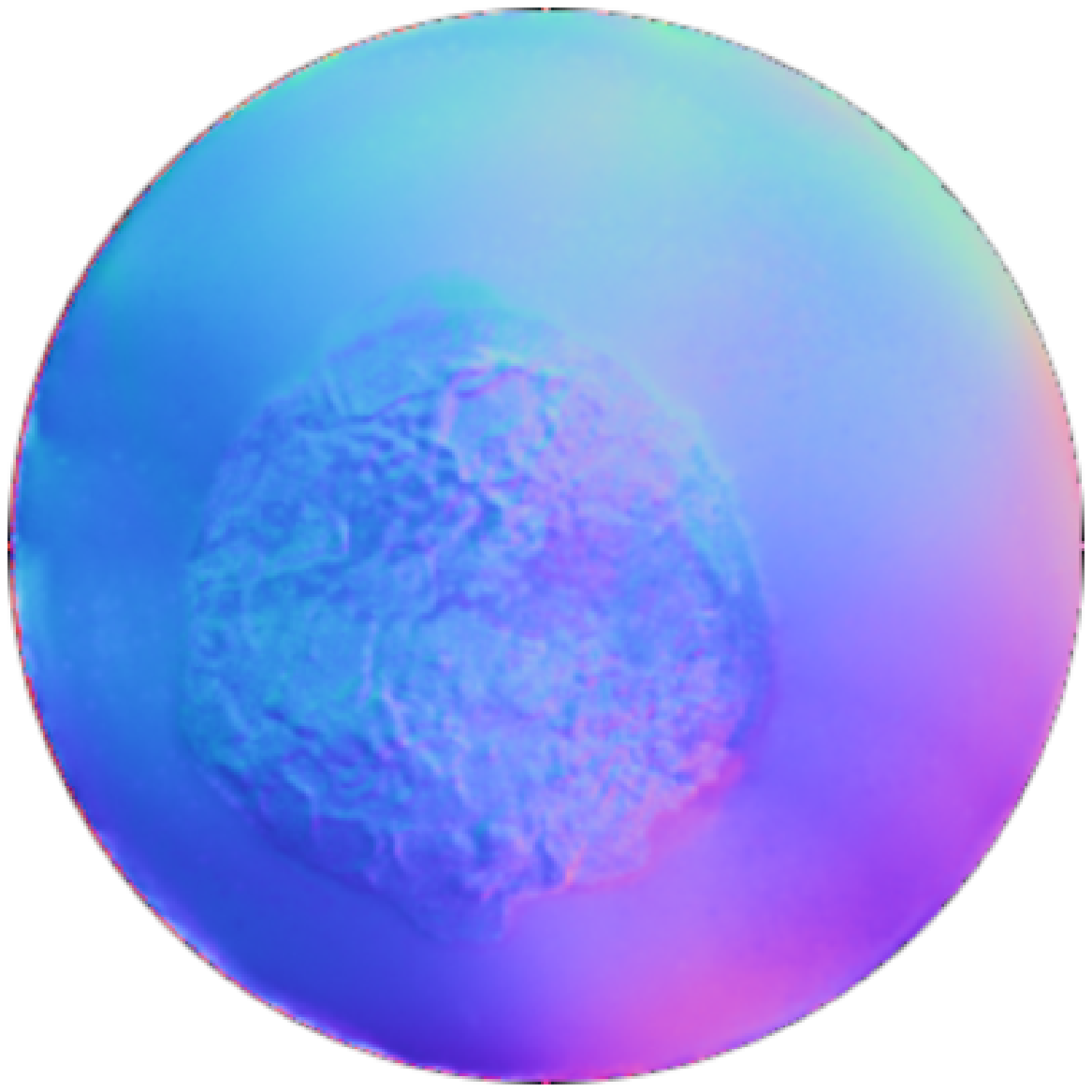}
&\includegraphics[height=\imh]{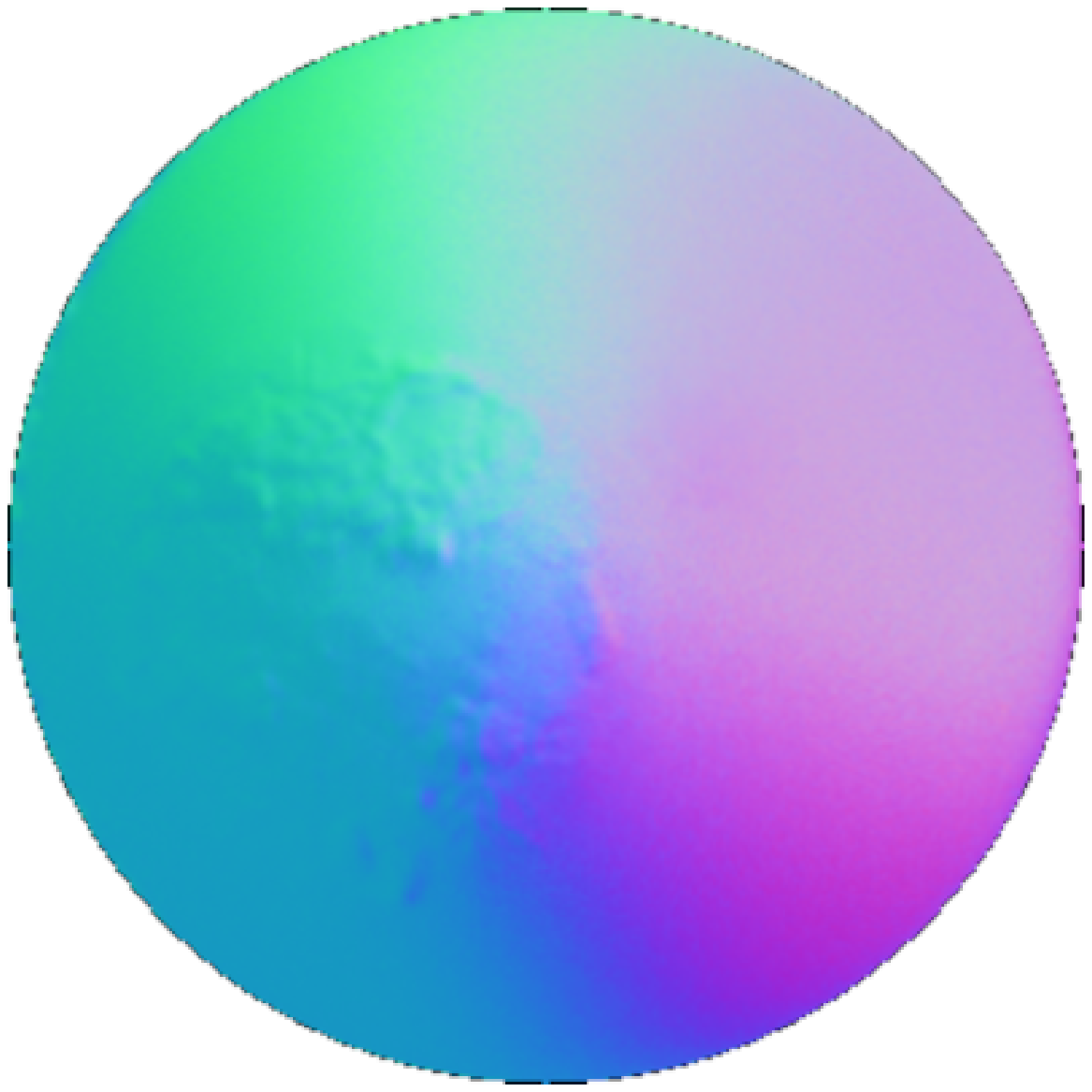}
&\includegraphics[height=\imh]{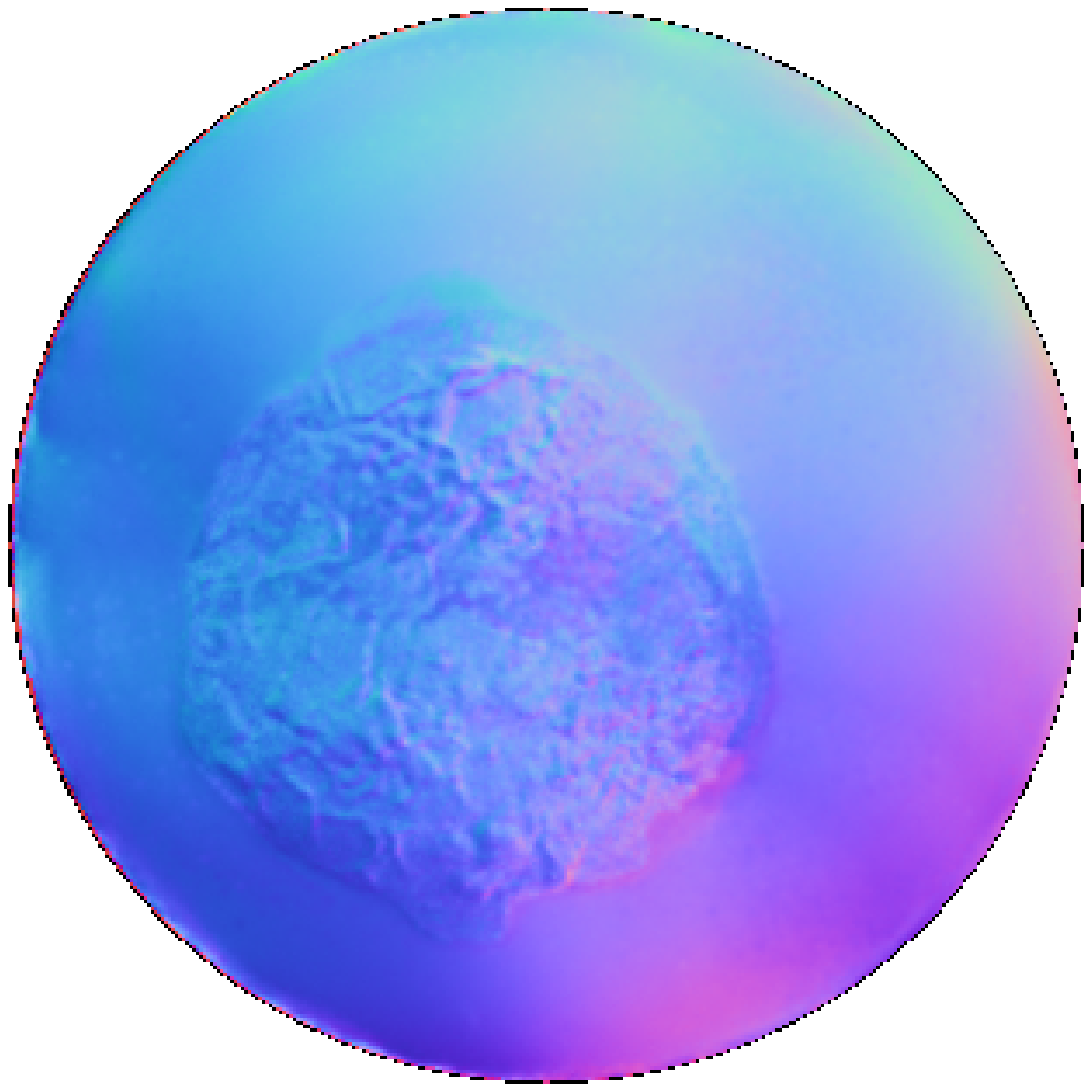}
&\includegraphics[height=\imh]{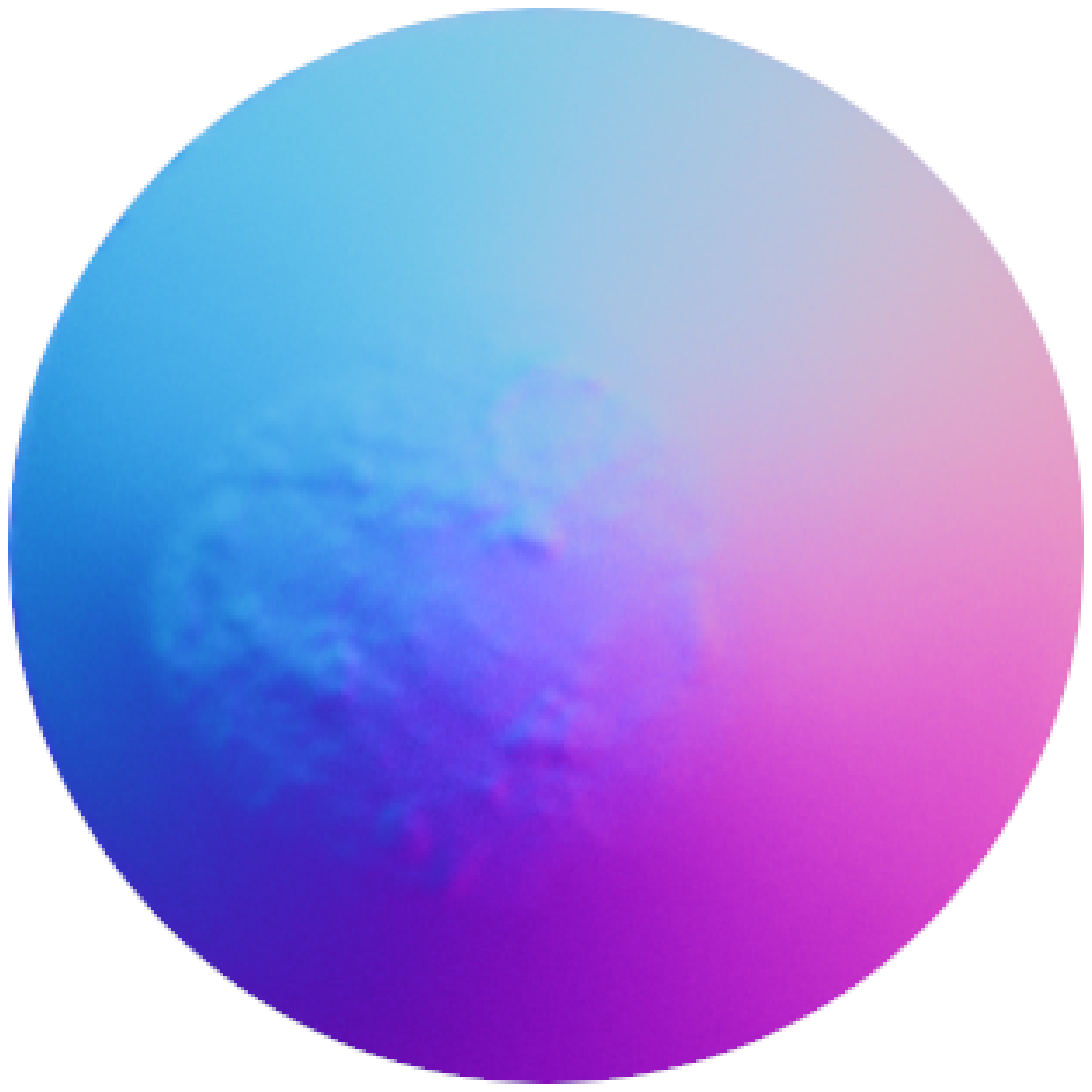}
&\multirow{5}{*}{\includegraphics[height=0.8in]{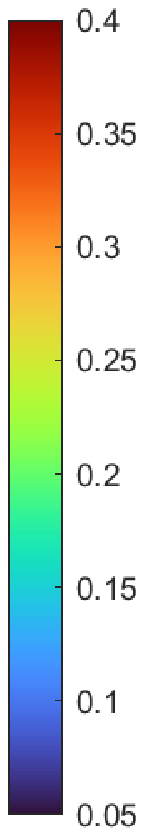}}
\\

&\includegraphics[height=\imh]{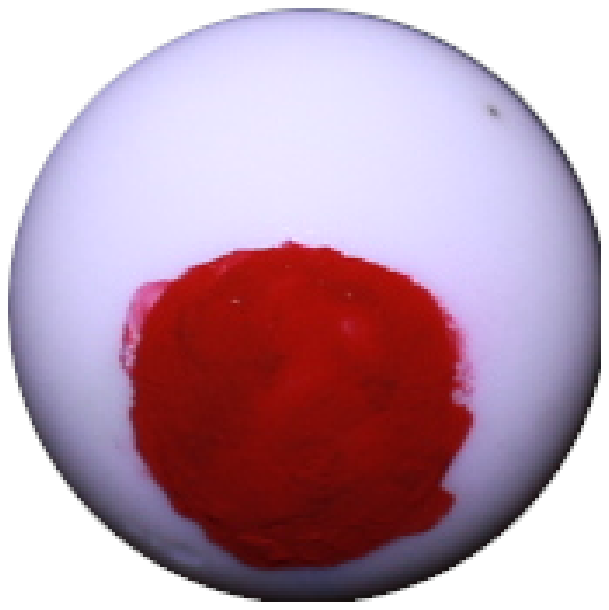}
&\includegraphics[height=\imh]{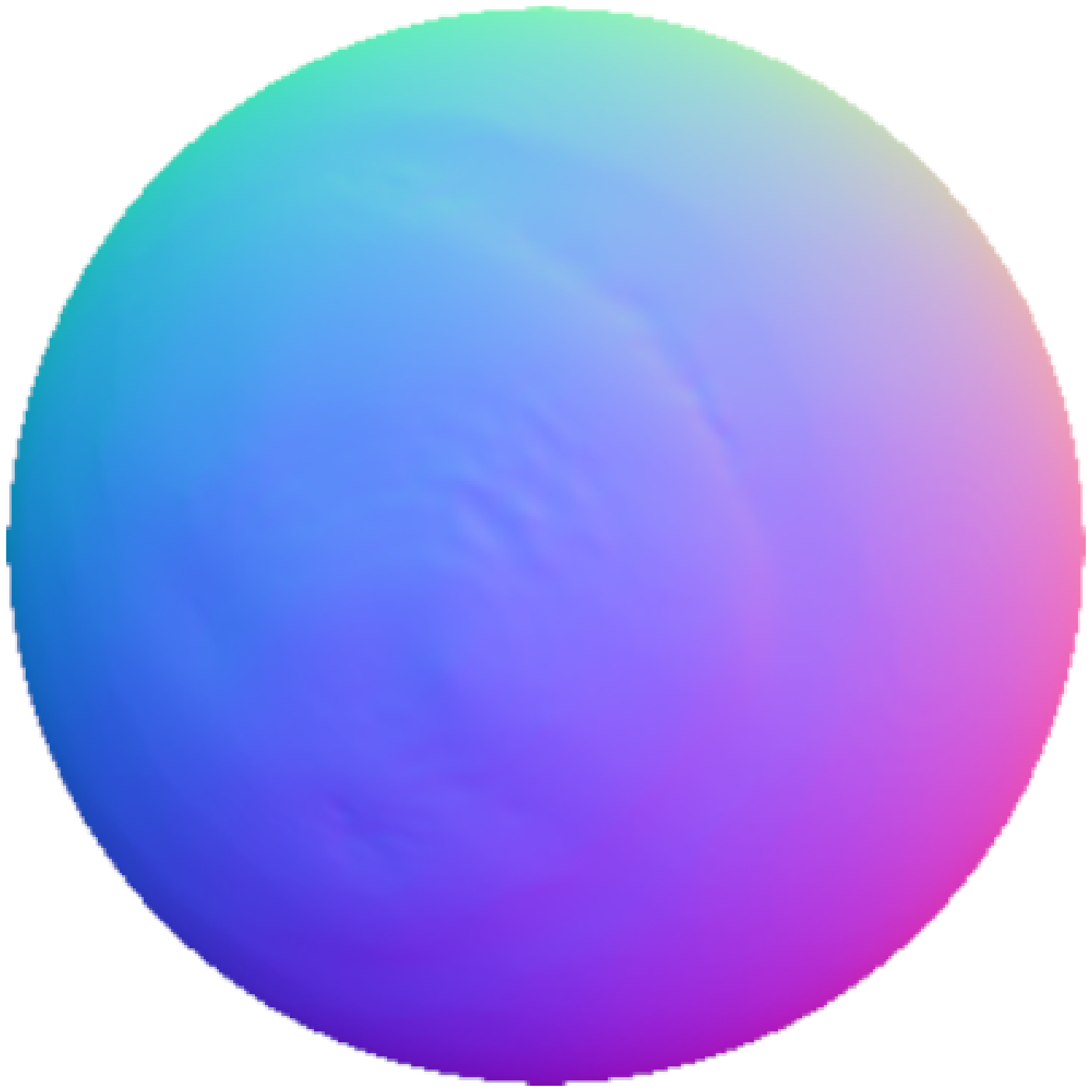}
&\includegraphics[height=\imh]{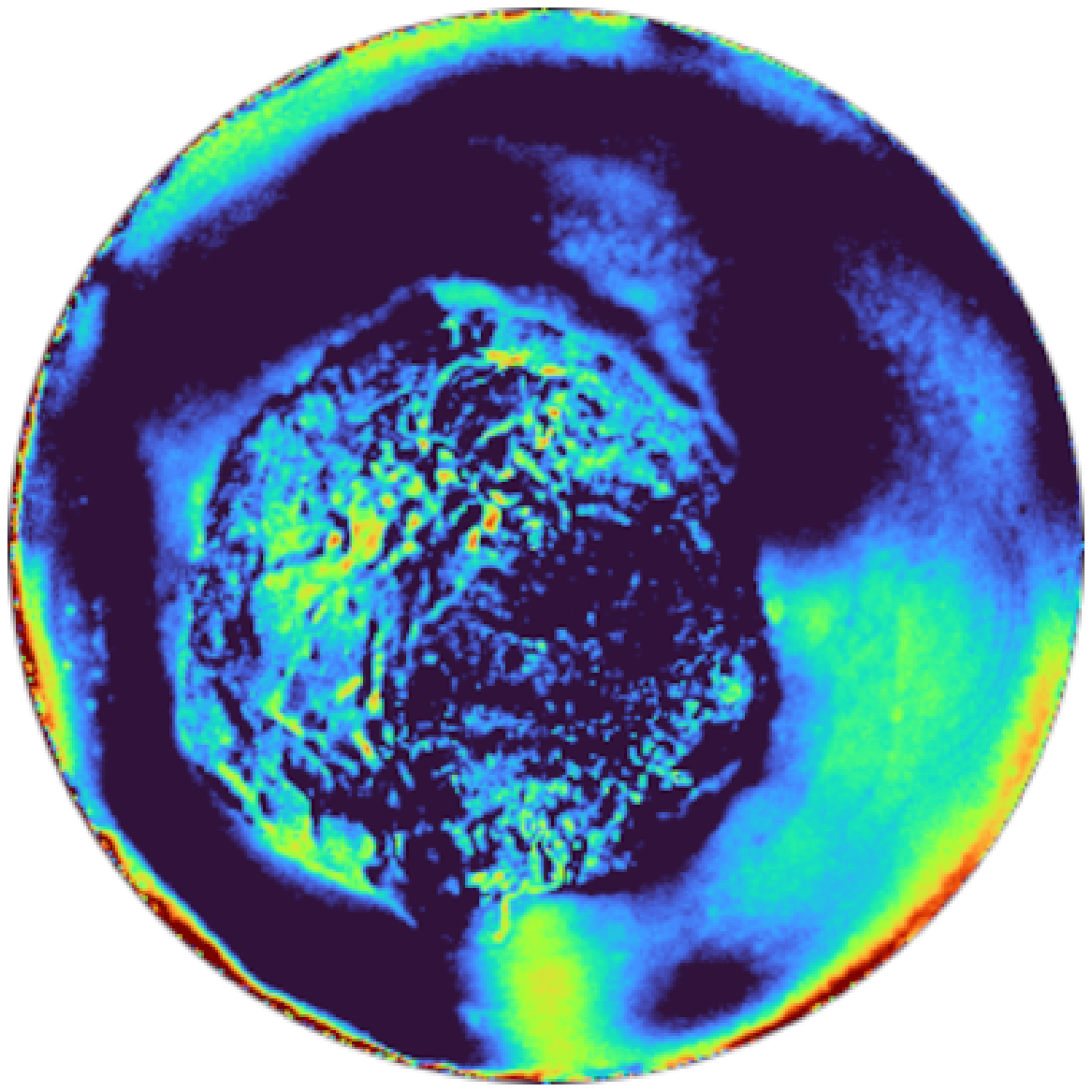}
&\includegraphics[height=\imh]{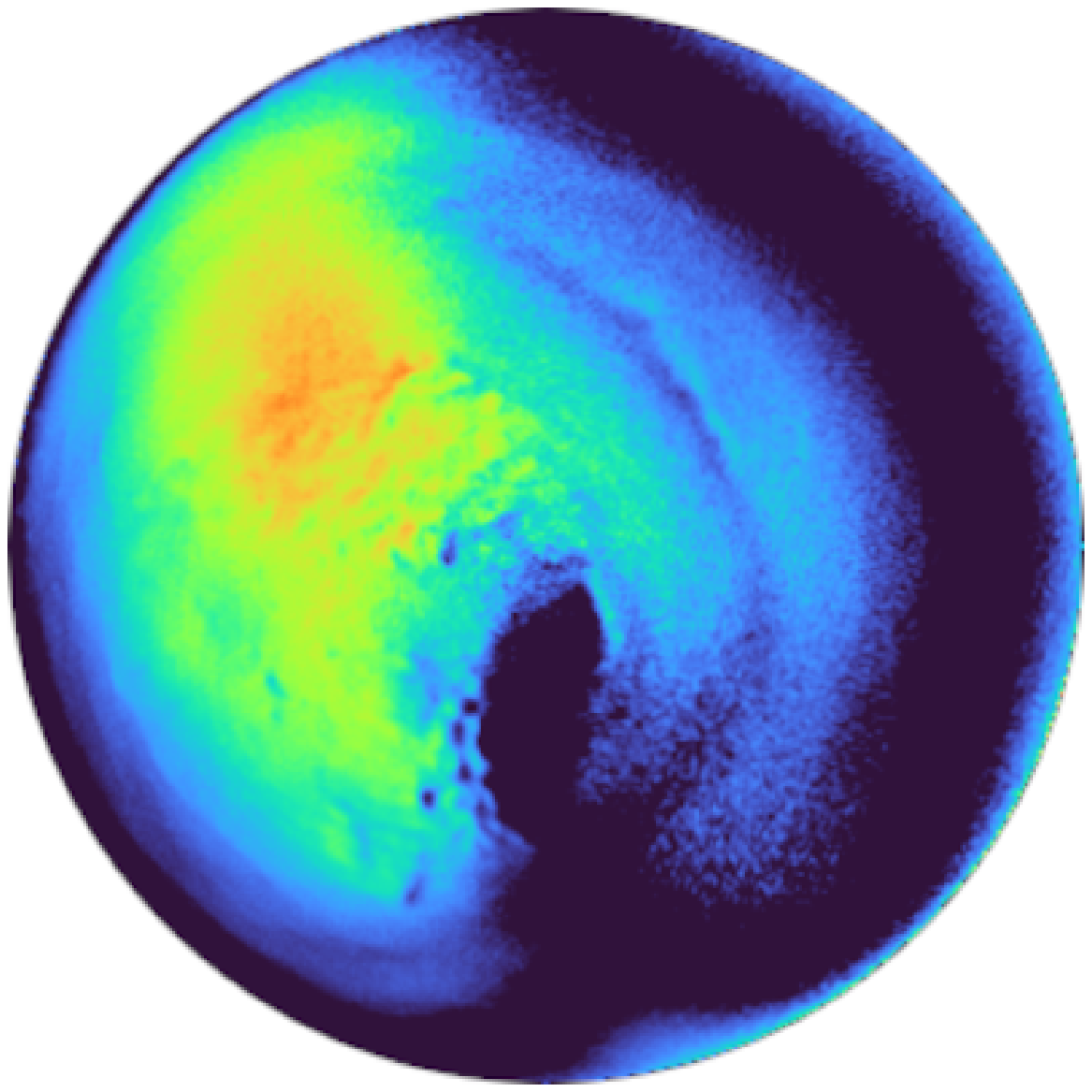}
&\includegraphics[height=\imh]{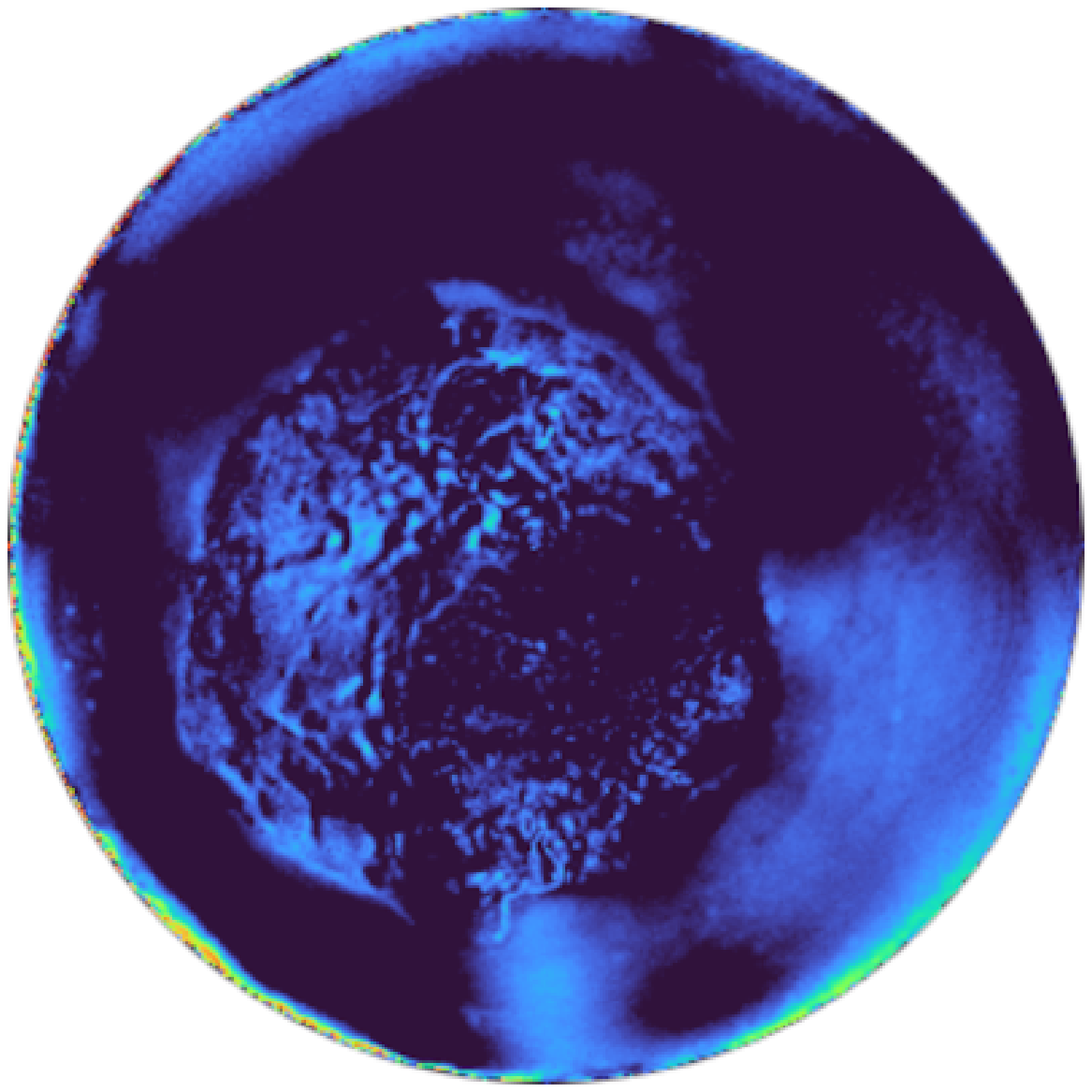}
&\includegraphics[height=\imh]{images/shapevalidation/ptfe/blank.eps}
&
\\

& & & $7.51$ & $15.67$ & $7.51$ & & \\

&\includegraphics[height=\imh]{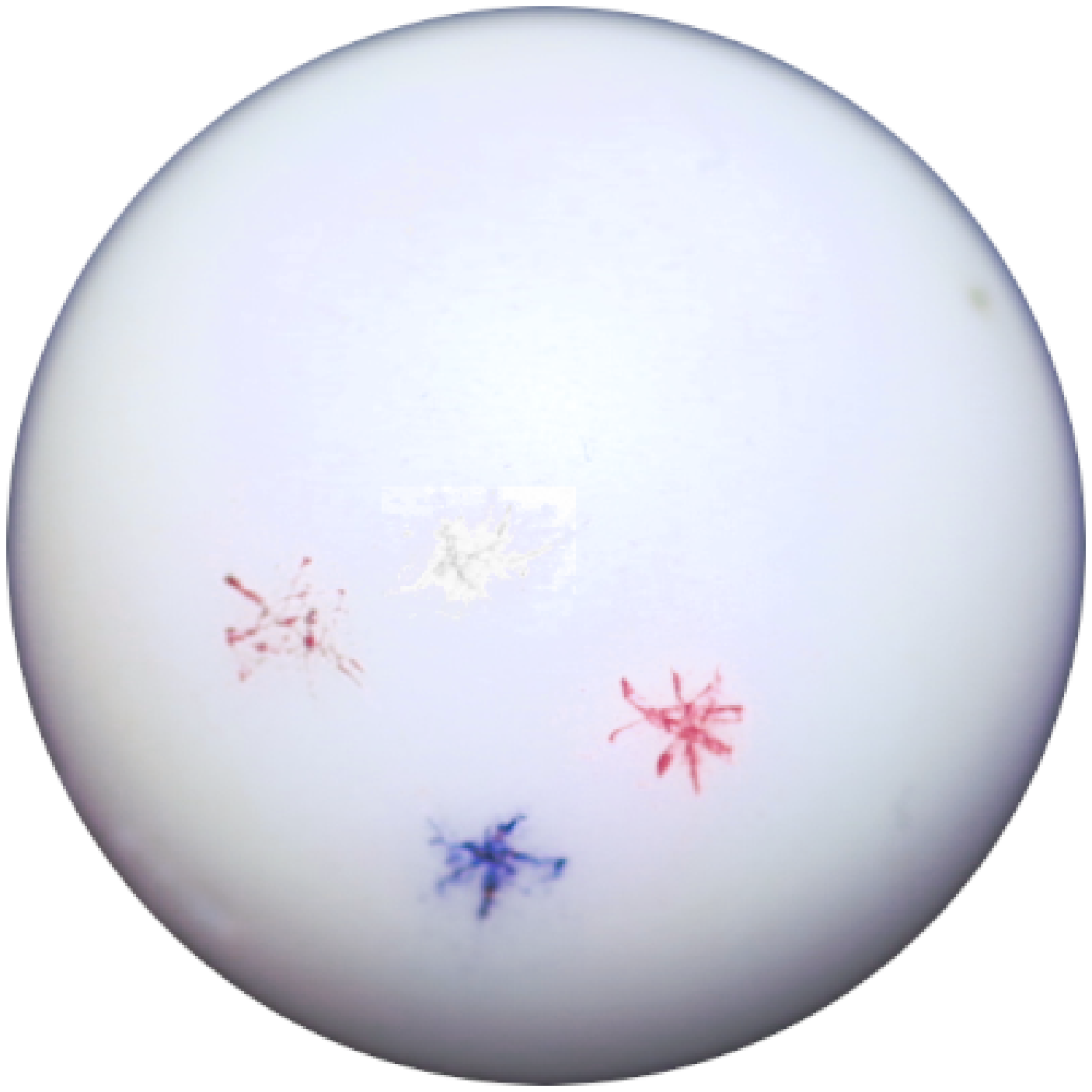}
&\includegraphics[height=\imh]{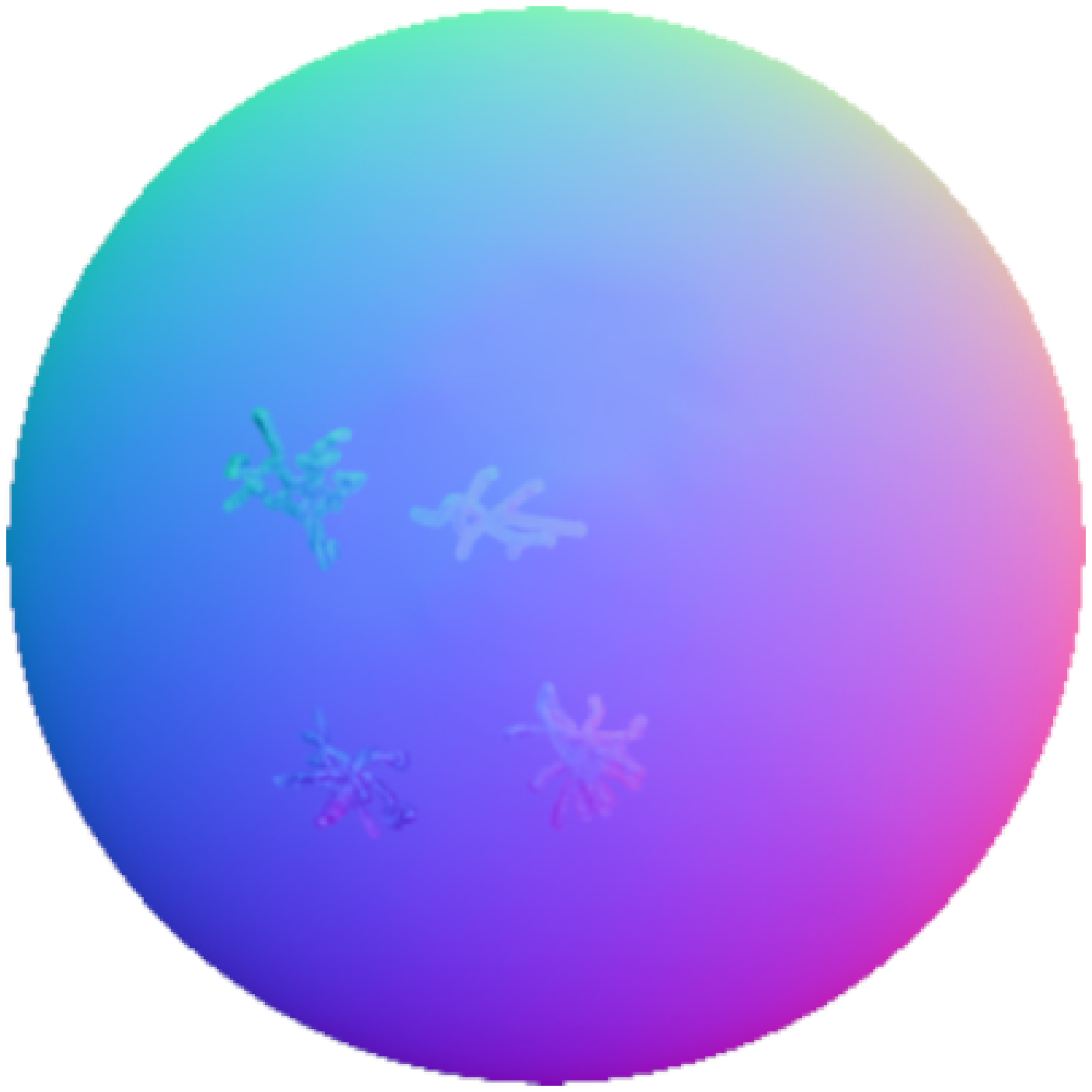}
&\includegraphics[height=\imh]{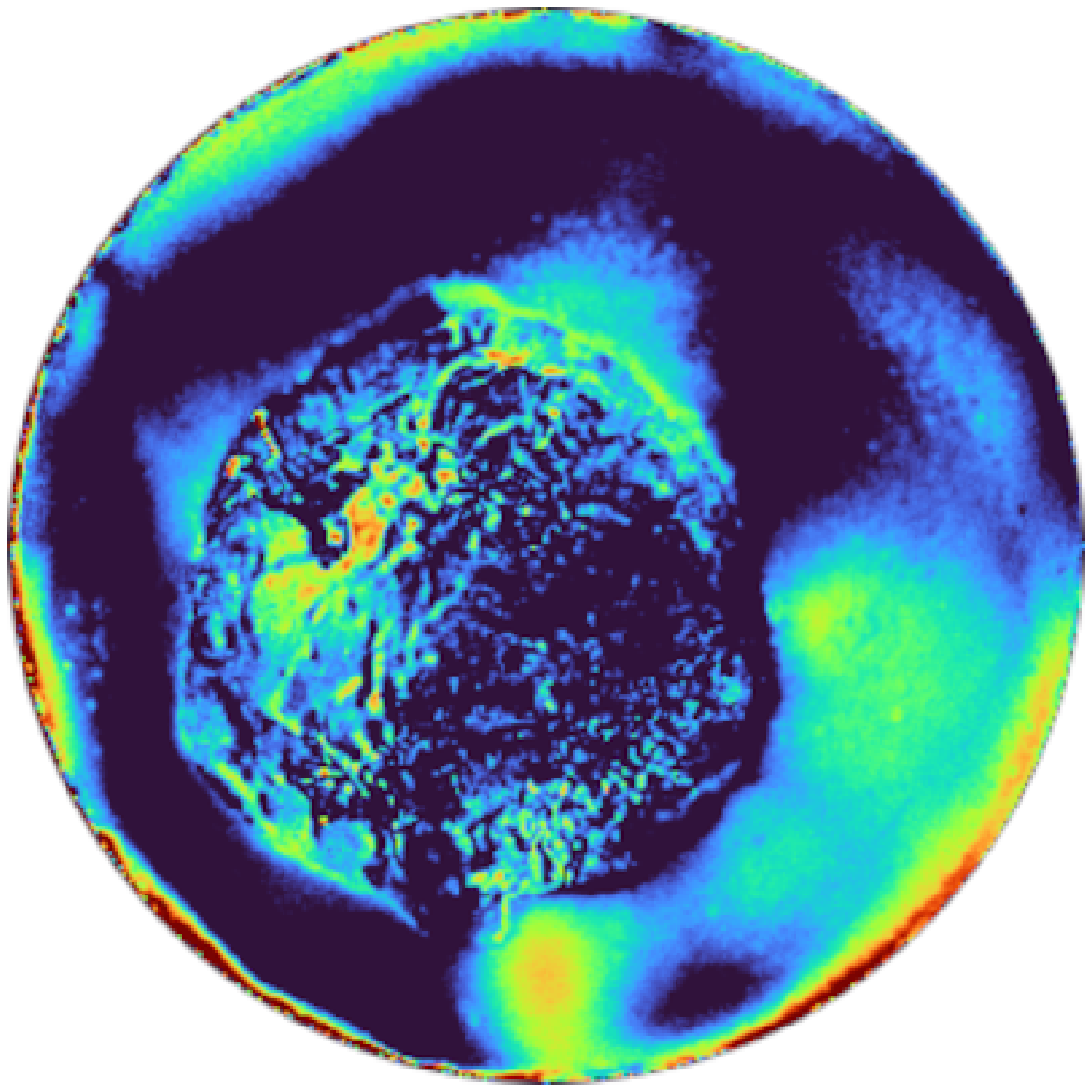}
&\includegraphics[height=\imh]{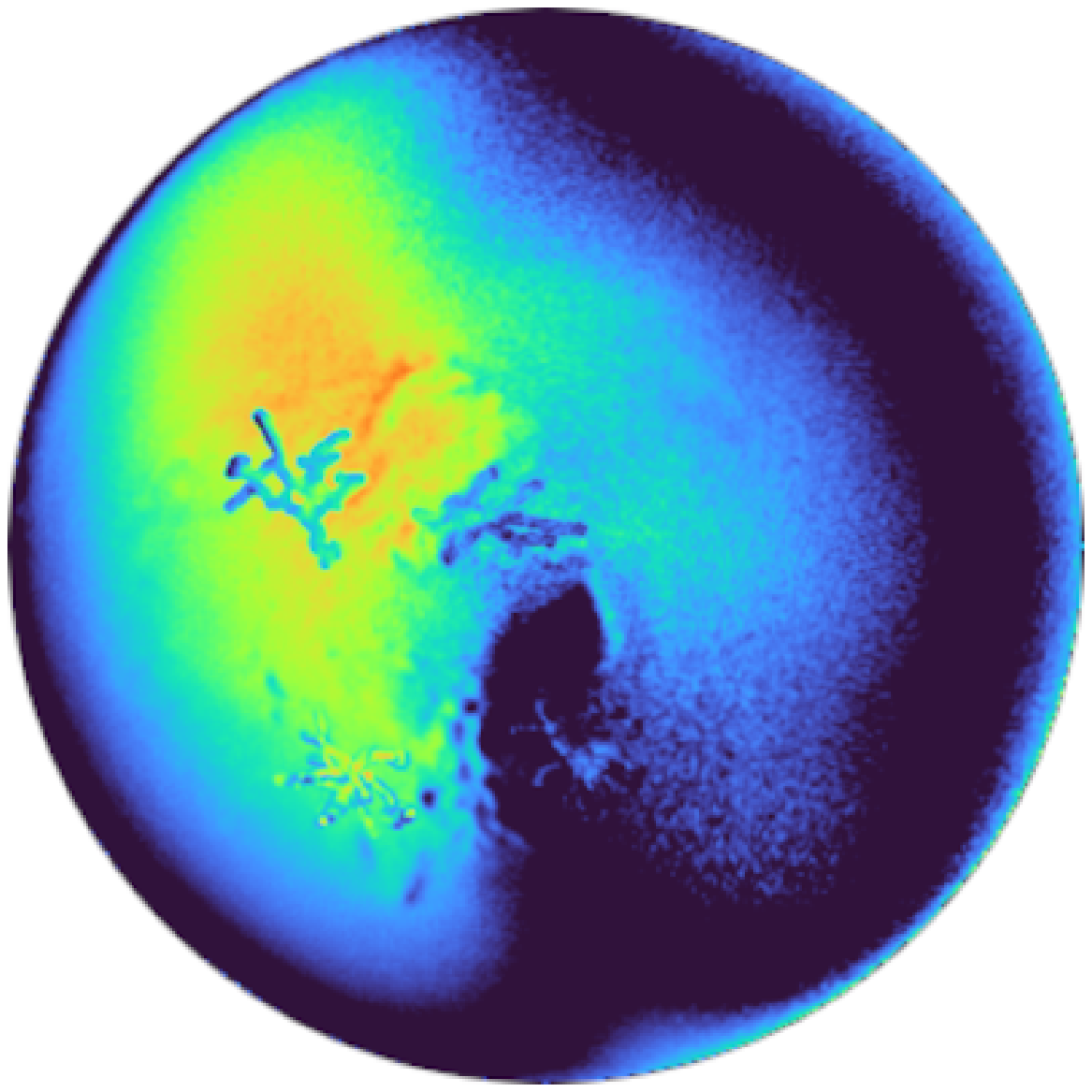}
&\includegraphics[height=\imh]{images/shapevalidation/ptfe/blank.eps}
&\includegraphics[height=\imh]{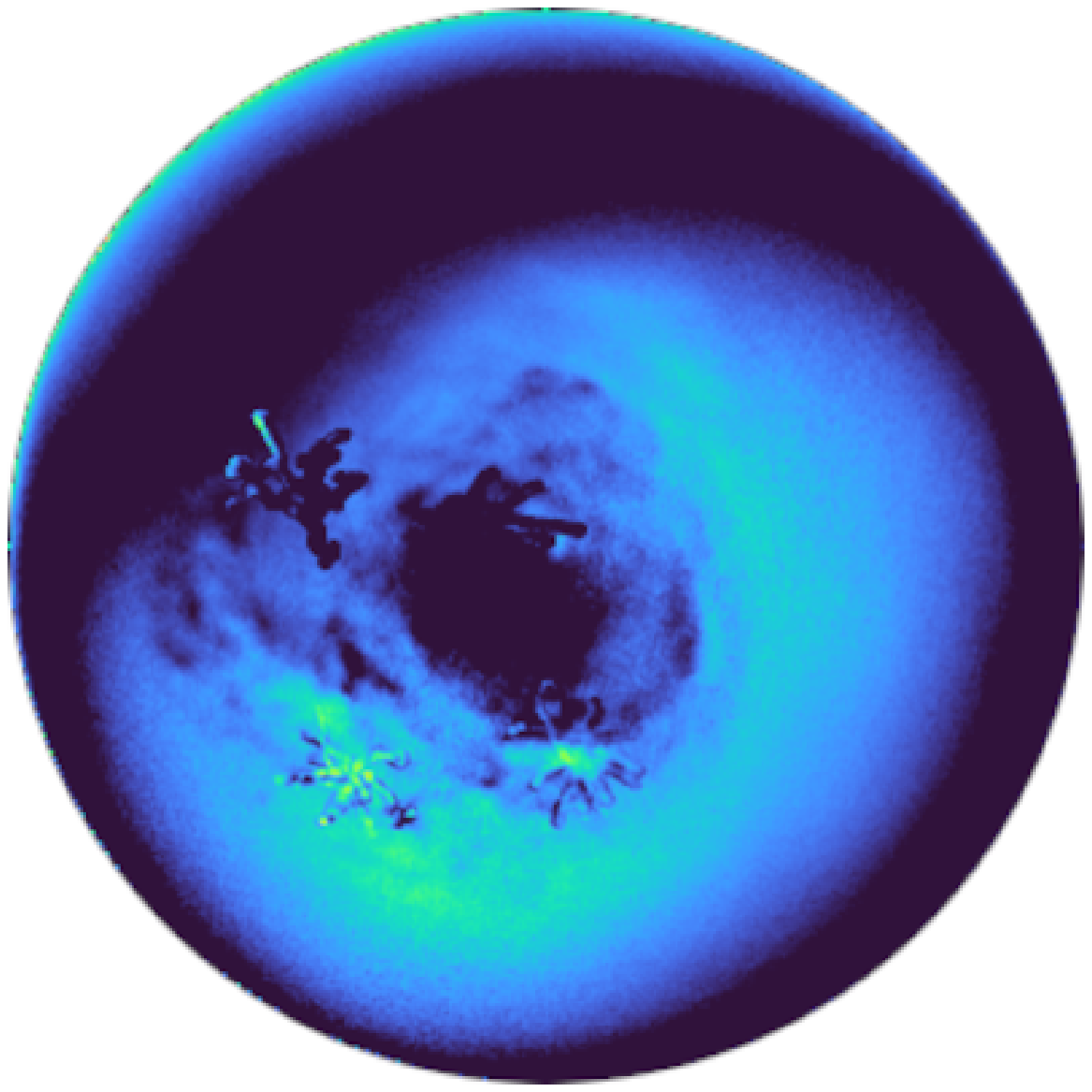}
&
\\

\vspace{8pt}

& & & $7.61$ & $15.58$ & & $7.61$ & \\

\multirow{5}{*}{\rotatebox{90}{\centering $acetal_{red}$}}
&\includegraphics[height=\imh]{images/shapevalidation/acetal-original/blank.eps}
&\includegraphics[height=\imh]{images/shapevalidation/acetal-original/blank.eps}
&\includegraphics[height=\imh]{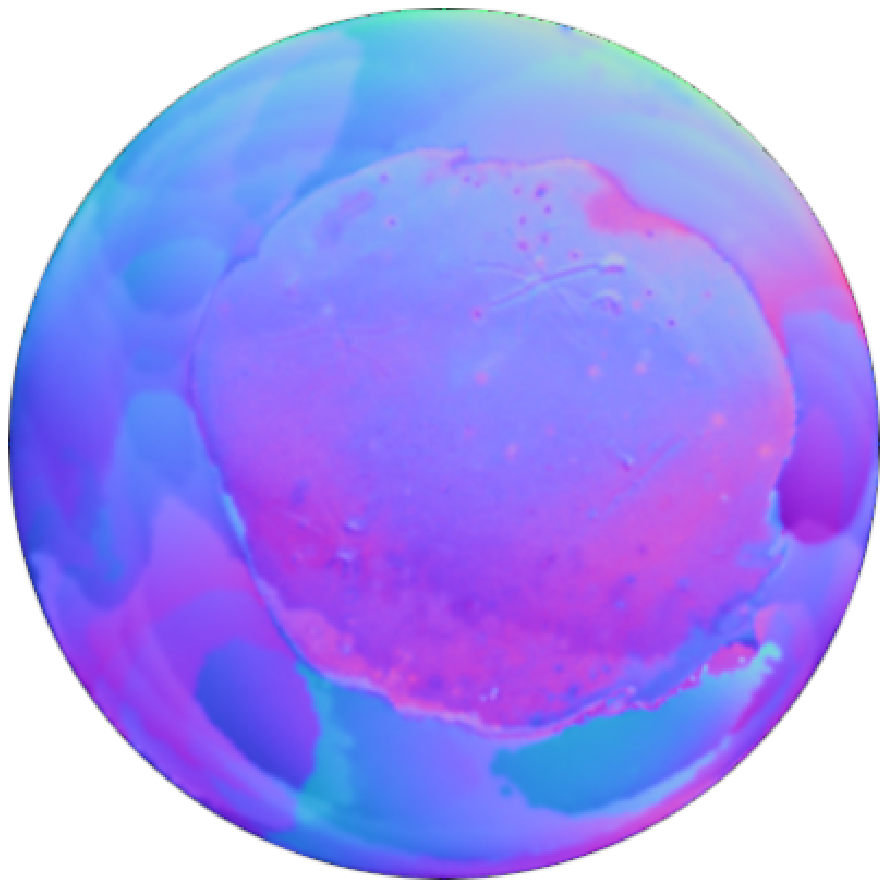}
&\includegraphics[height=\imh]{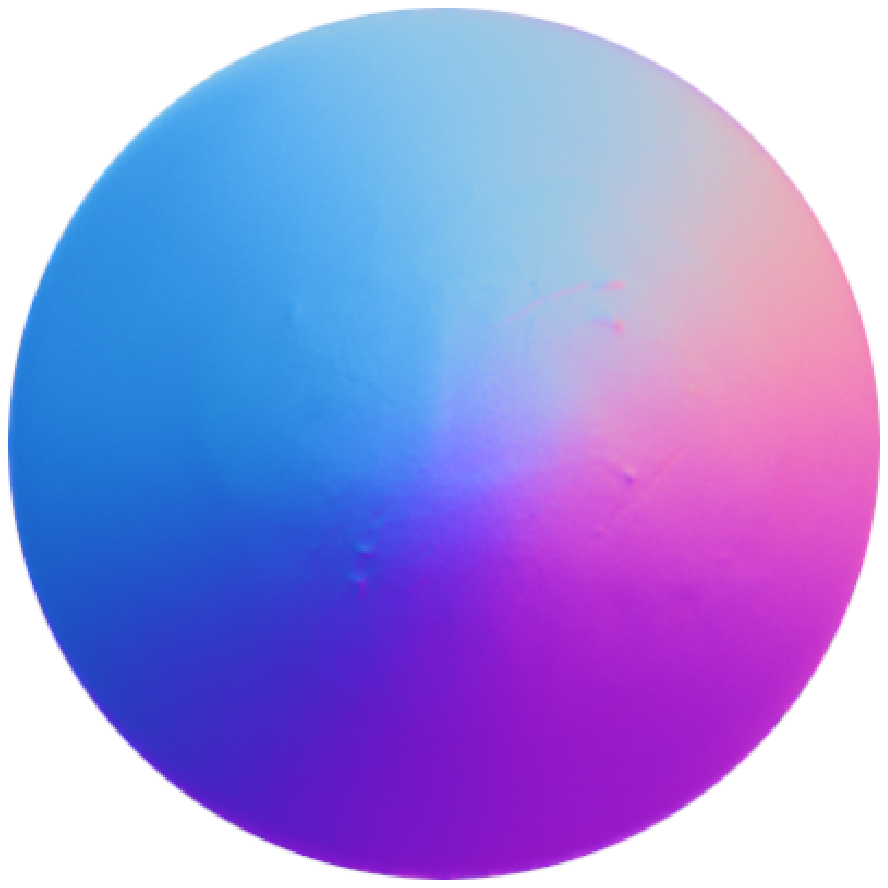}
&\includegraphics[height=\imh]{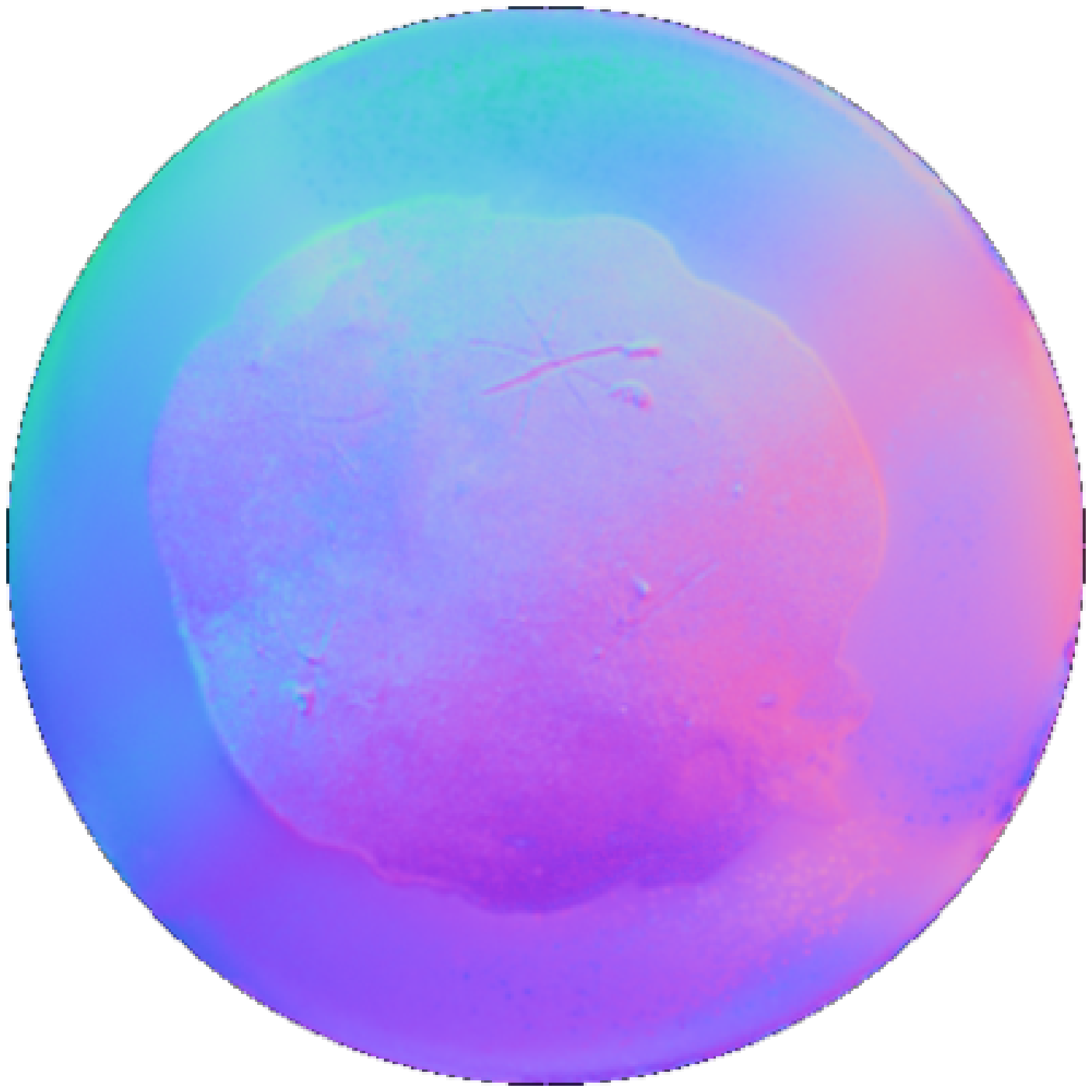}
&\includegraphics[height=\imh]{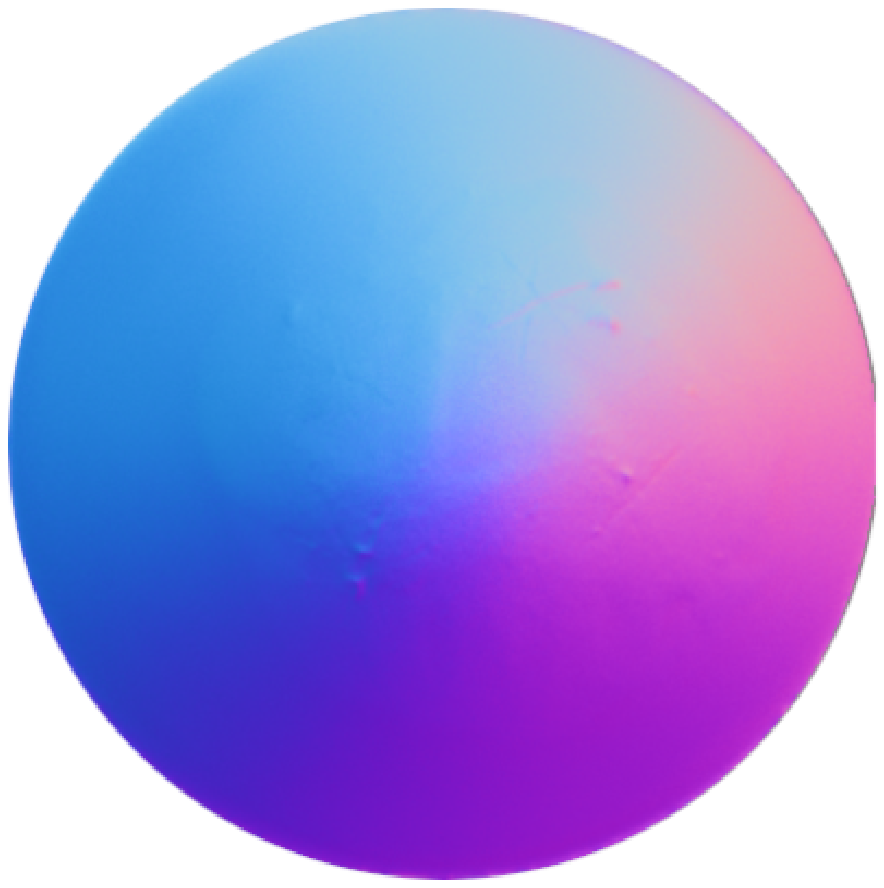}
&\multirow{5}{*}{\includegraphics[height=0.8in]{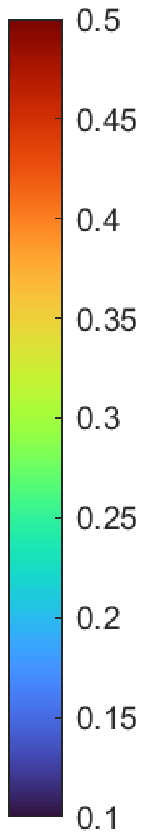}}
\\

&\includegraphics[height=\imh]{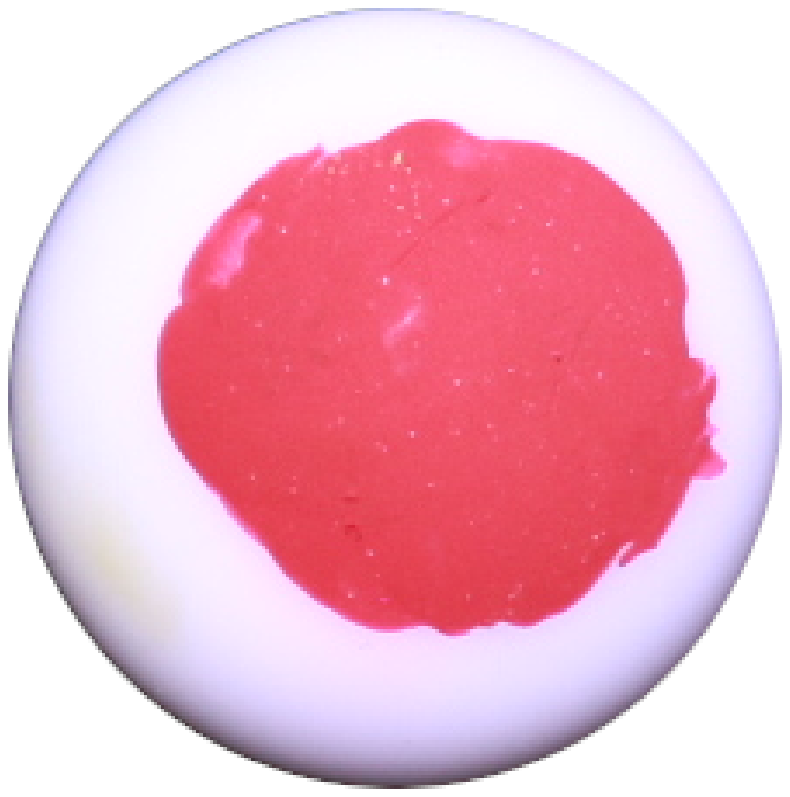}
&\includegraphics[height=\imh]{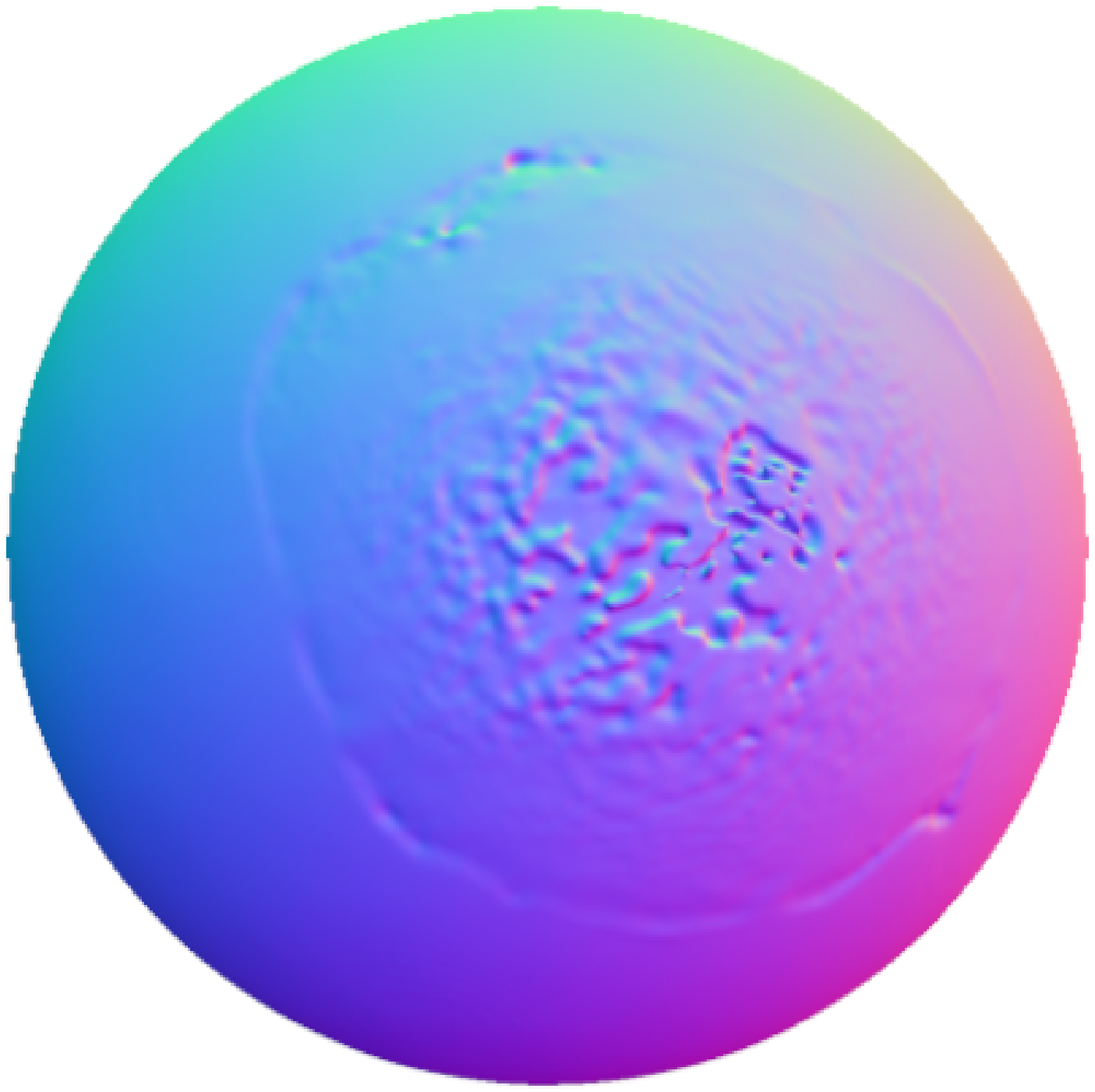}
&\includegraphics[height=\imh]{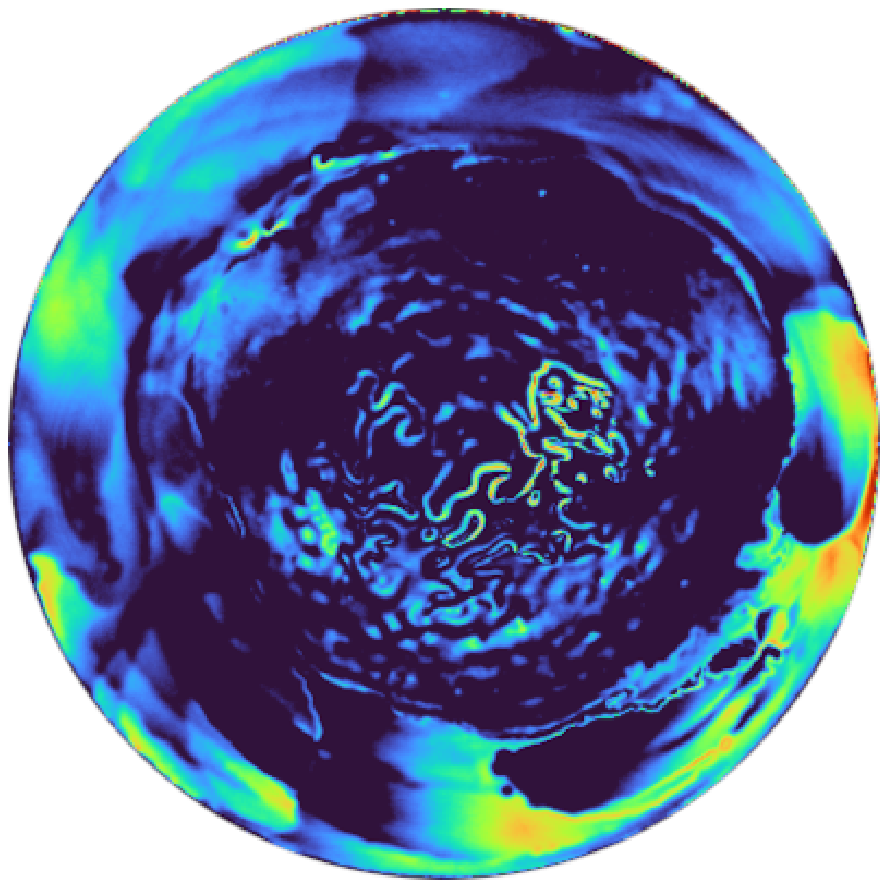}
&\includegraphics[height=\imh]{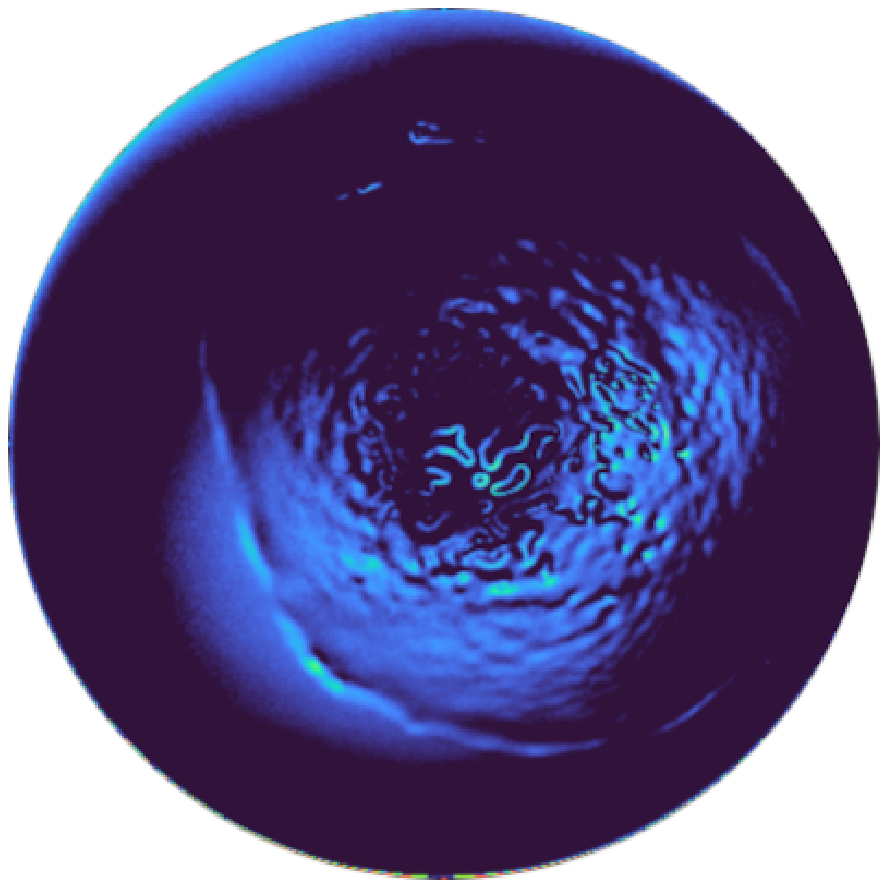}
&\includegraphics[height=\imh]{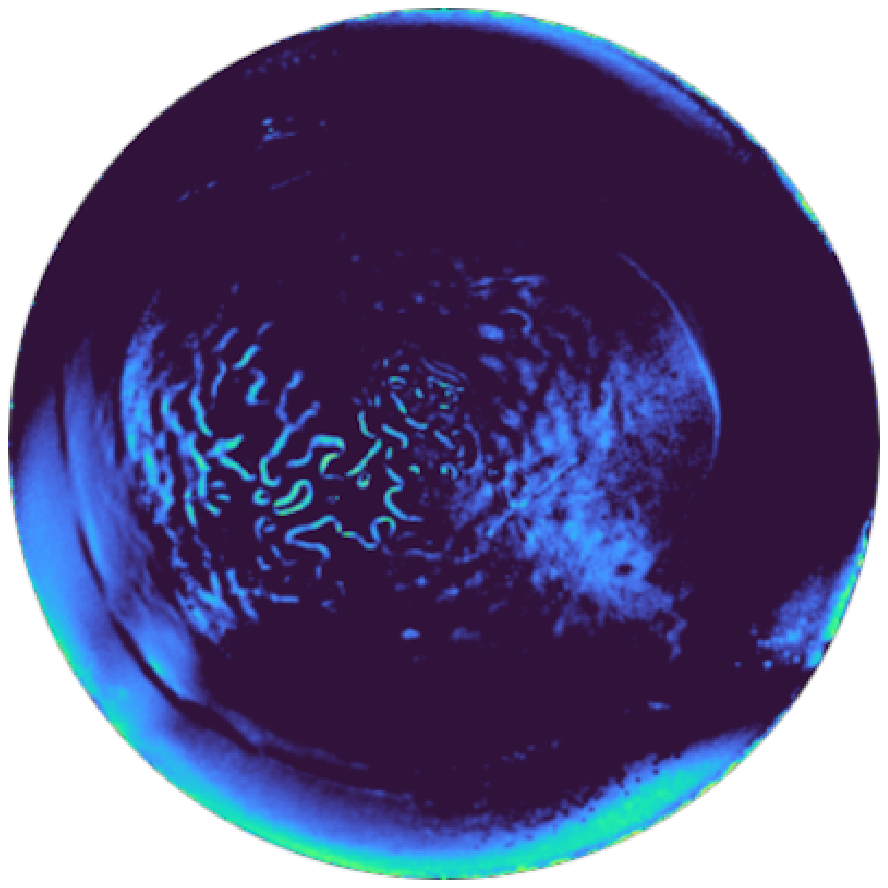}
&\includegraphics[height=\imh]{images/shapevalidation/acetal-original/blank.eps}
&
\\

& & & $12.46$ & $11.25$ & $11.12$ & & \\

&\includegraphics[height=\imh]{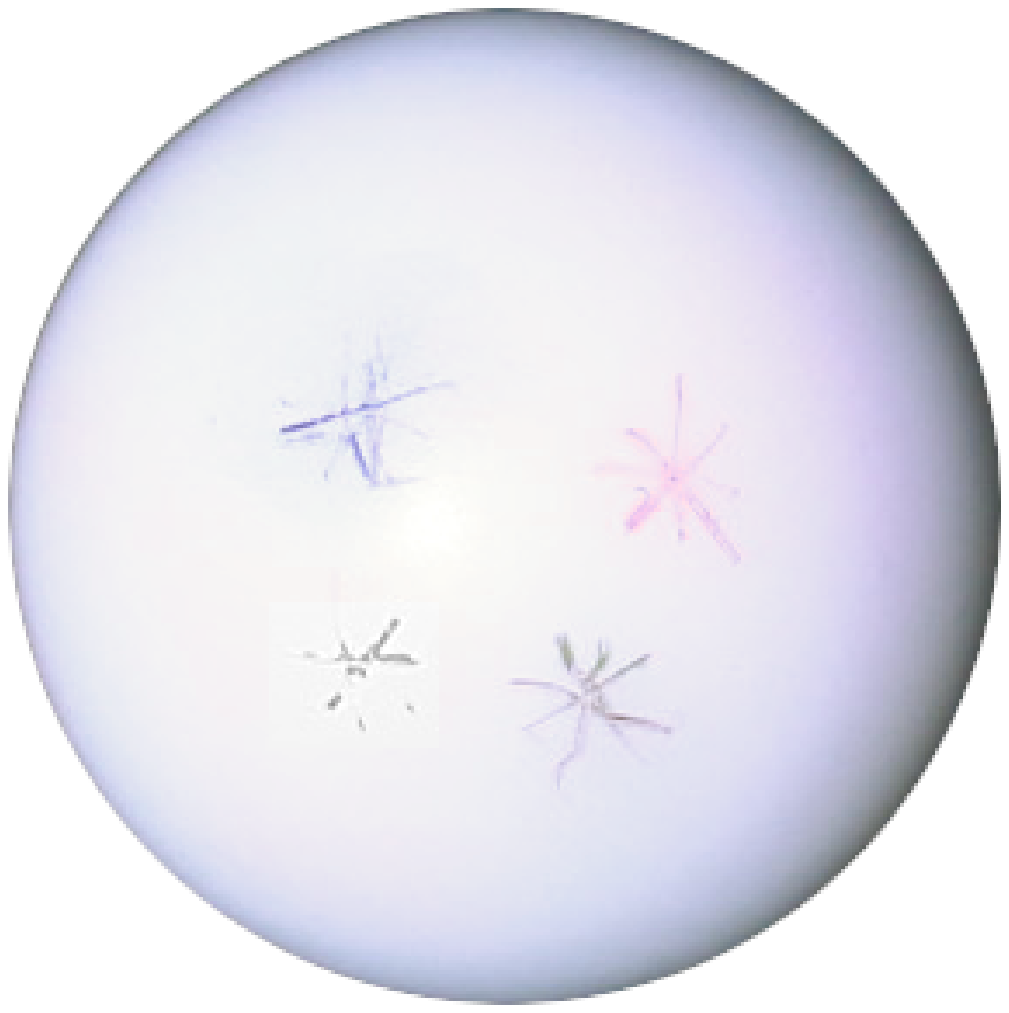}
&\includegraphics[height=\imh]{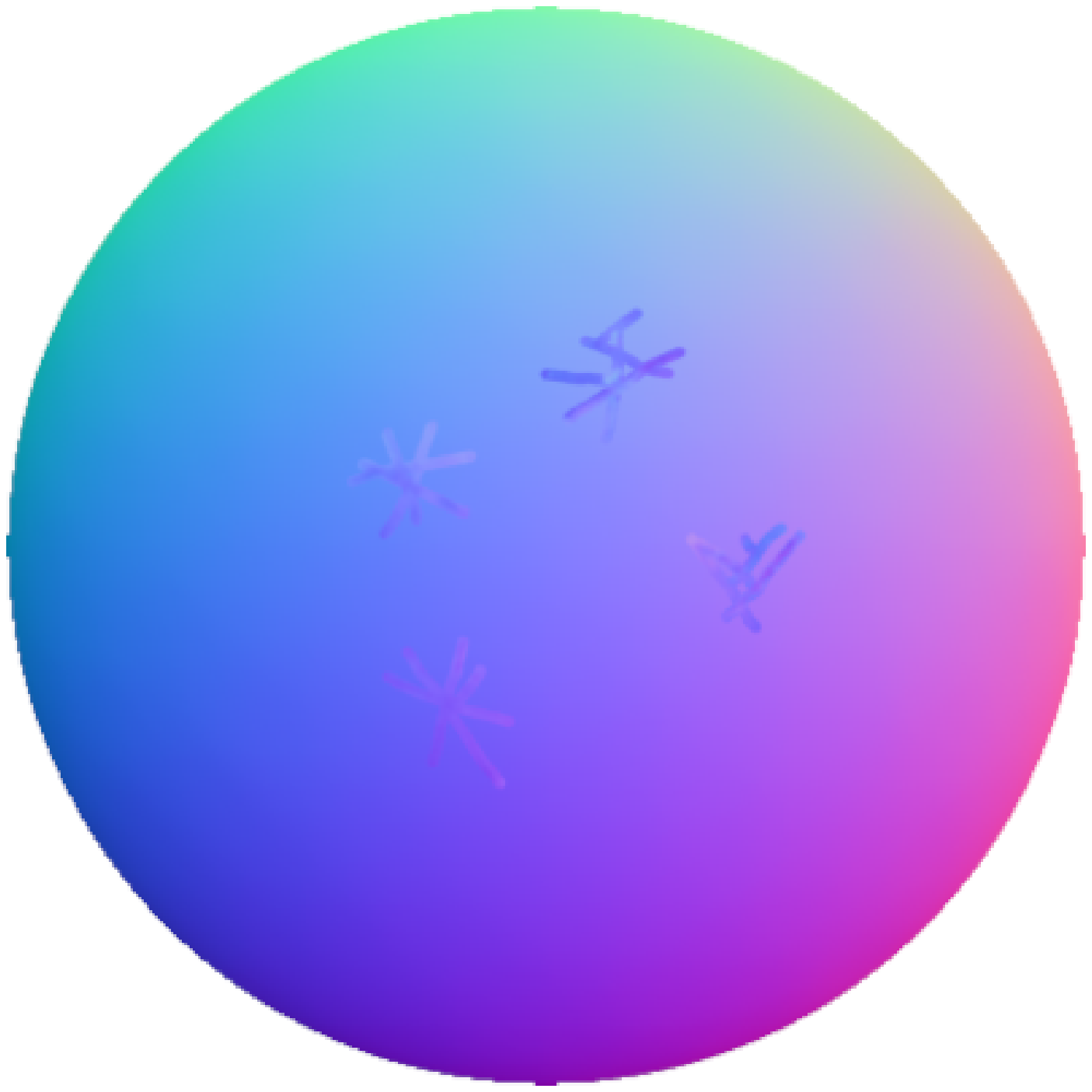}
&\includegraphics[height=\imh]{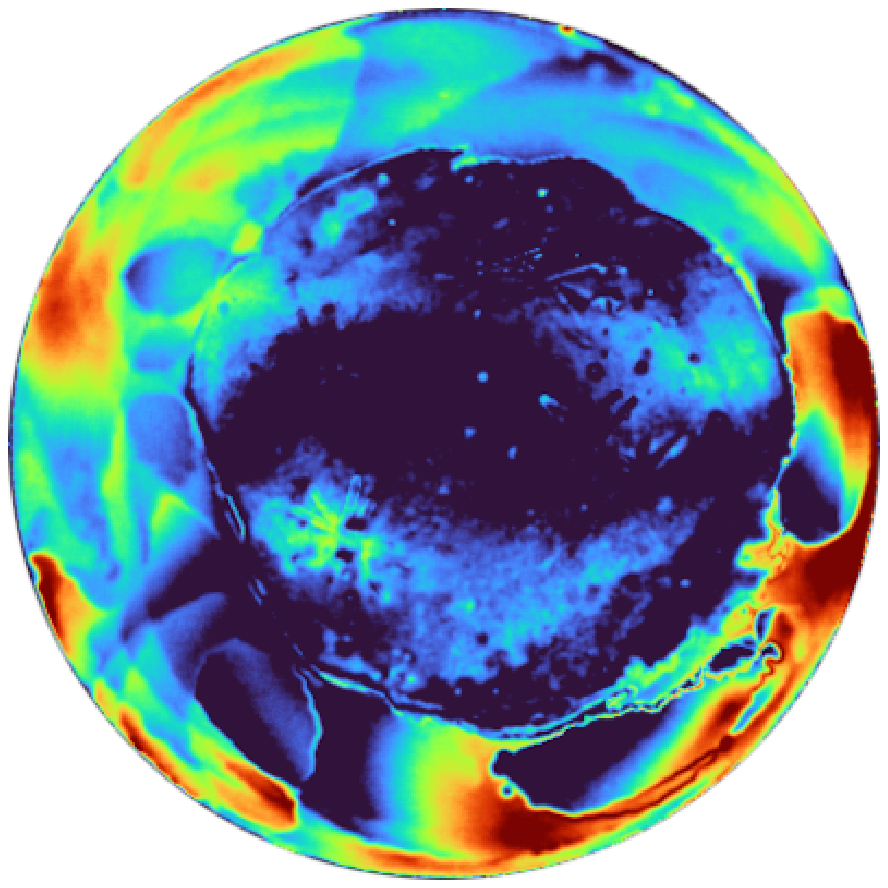}
&\includegraphics[height=\imh]{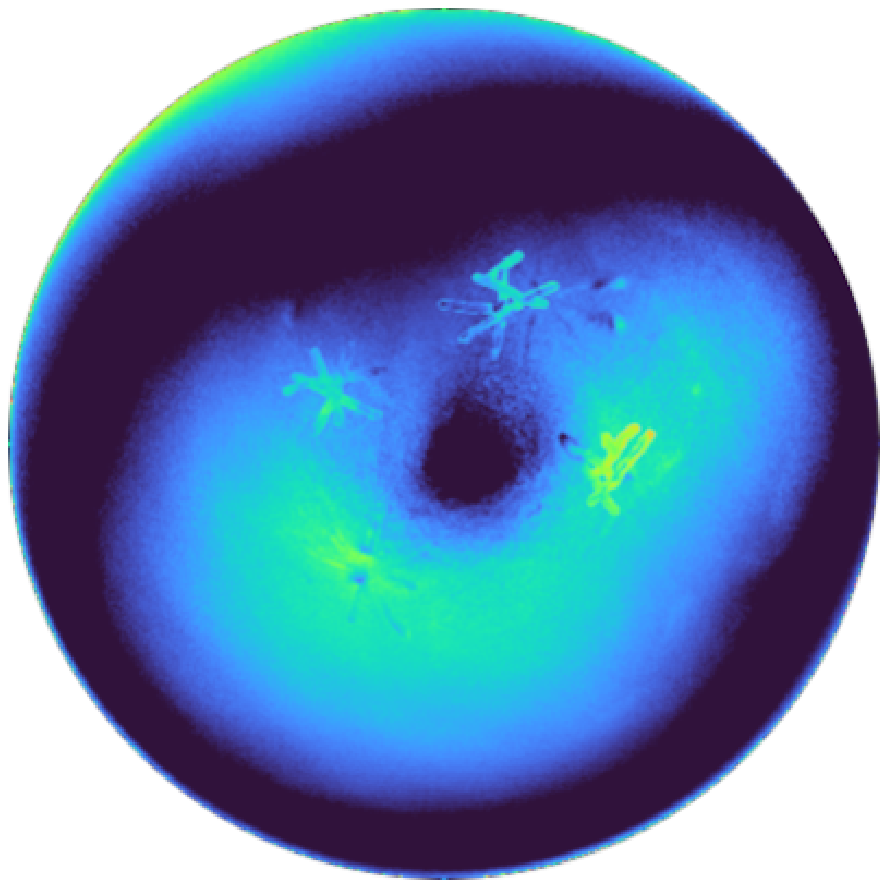}
&\includegraphics[height=\imh]{images/shapevalidation/acetal-original/blank.eps}
&\includegraphics[height=\imh]{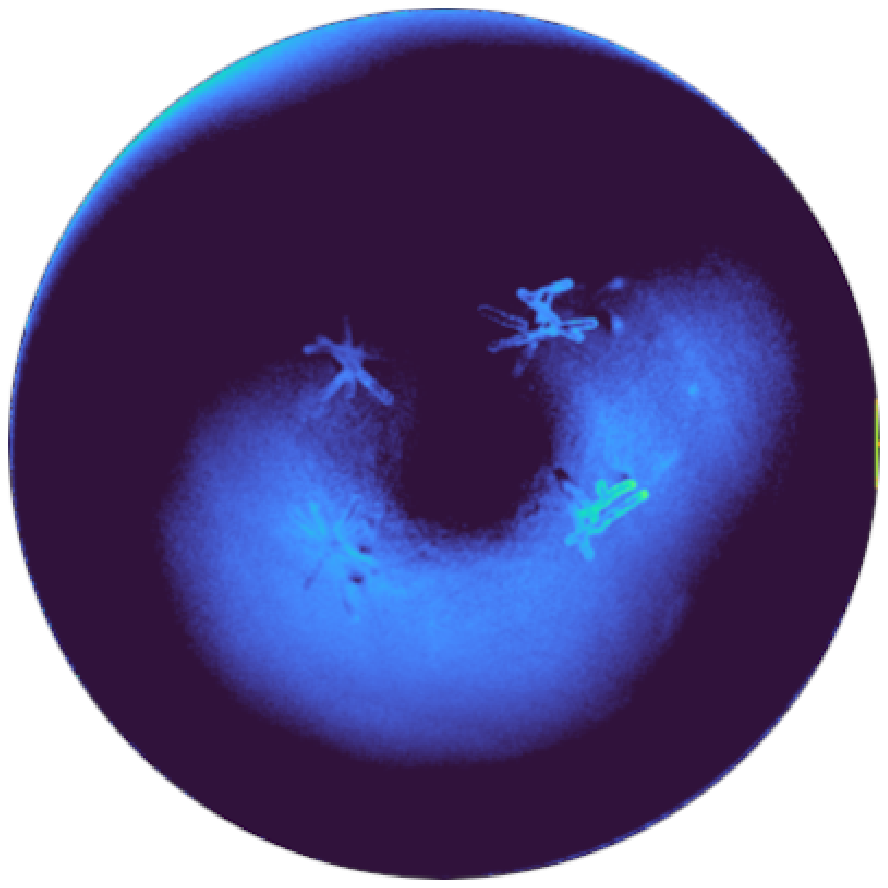}
&
\\

\vspace{8pt}

& & & $13.42$ & $12.06$ & & $12.07$ & \\

\multirow{5}{*}{\rotatebox{90}{\centering $acetal_{green}$}}
&\includegraphics[height=\imh]{images/shapevalidation/acetal-green/blank.eps}
&\includegraphics[height=\imh]{images/shapevalidation/acetal-green/blank.eps}
&\includegraphics[height=\imh]{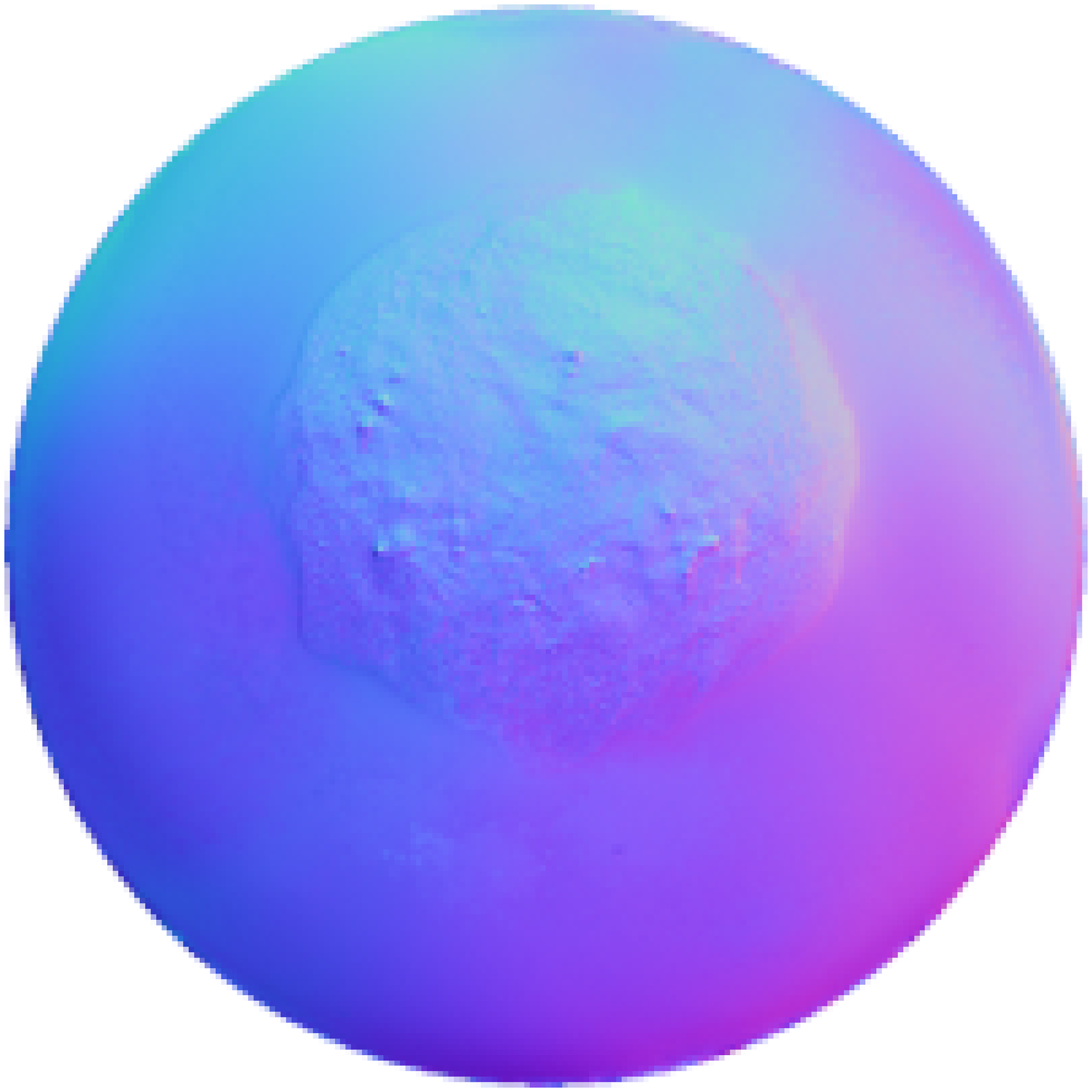}
&\includegraphics[height=\imh]{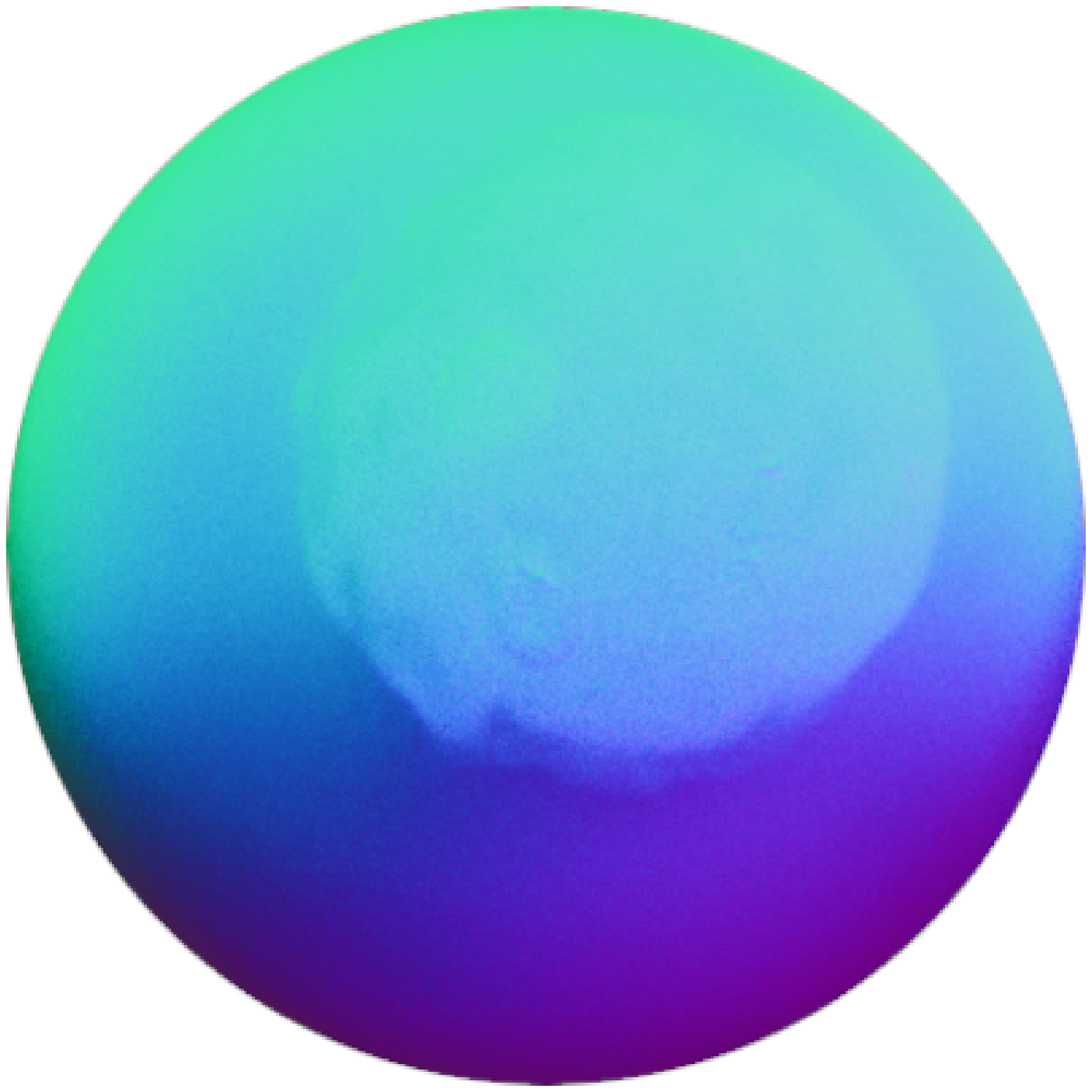}
&\includegraphics[height=\imh]{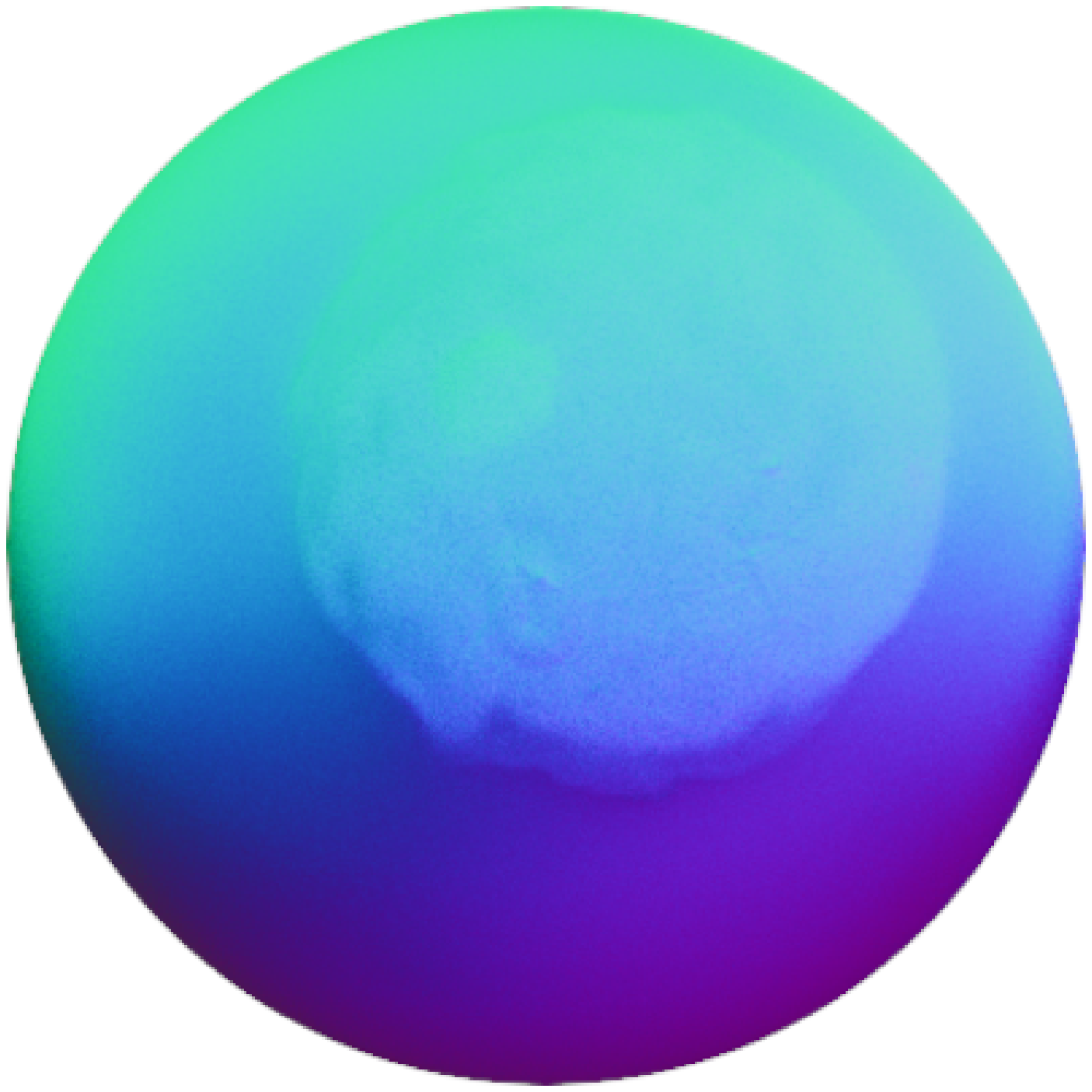}
&\includegraphics[height=\imh]{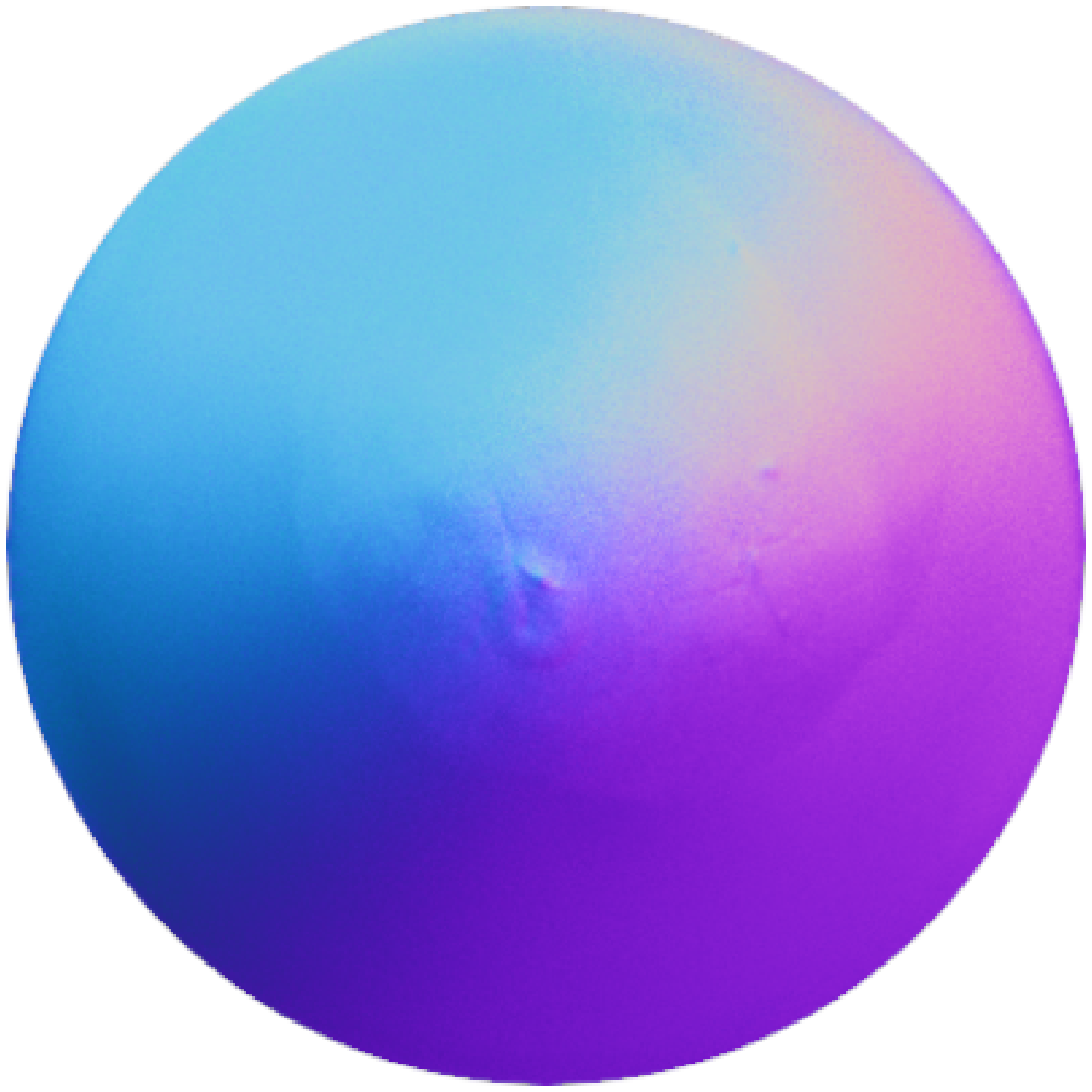}
&\multirow{5}{*}{\includegraphics[height=0.8in]{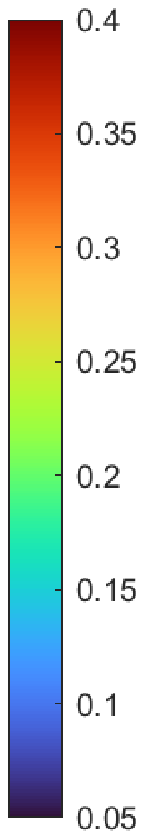}}
\\

&\includegraphics[height=\imh]{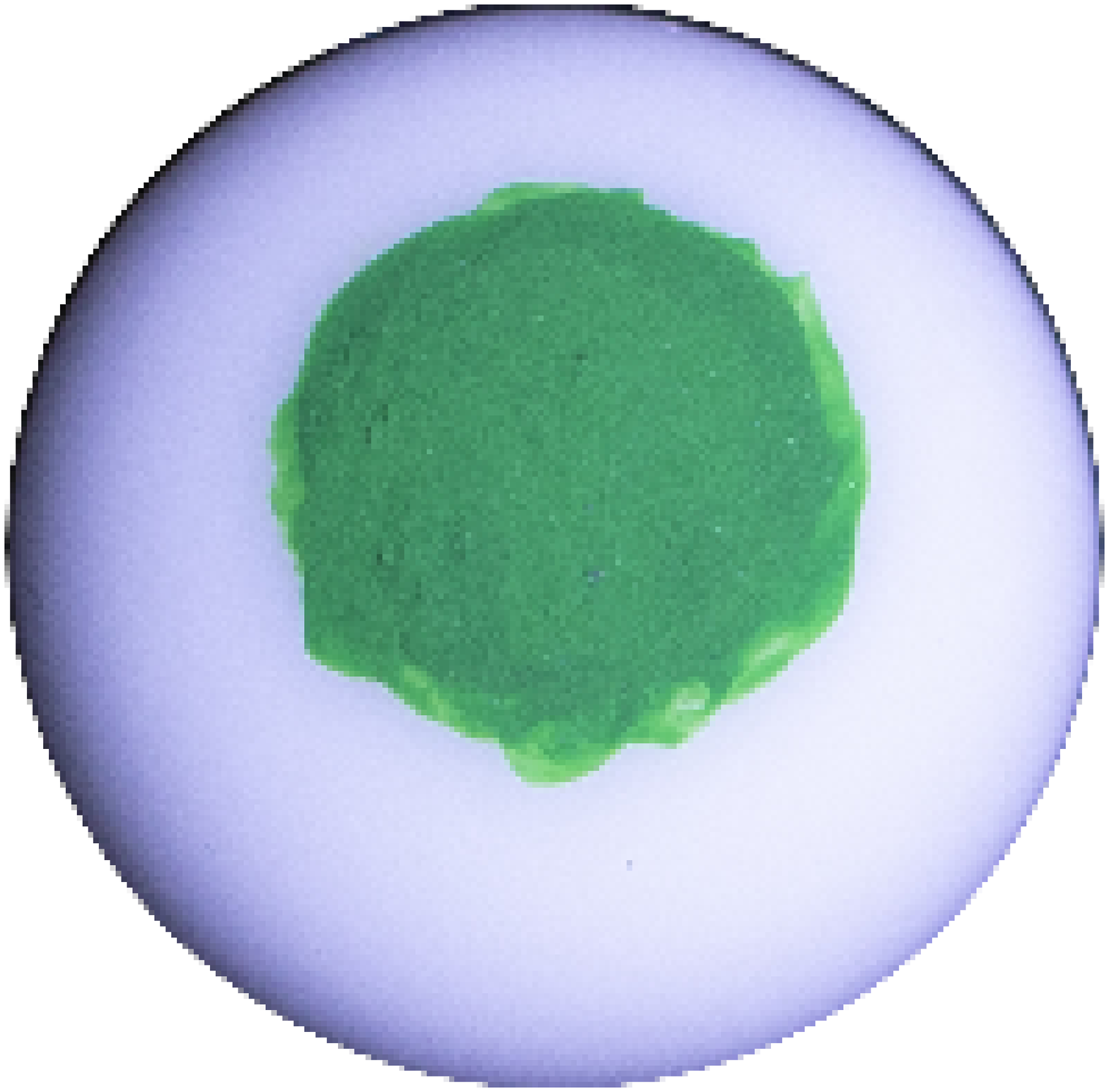}
&\includegraphics[height=\imh]{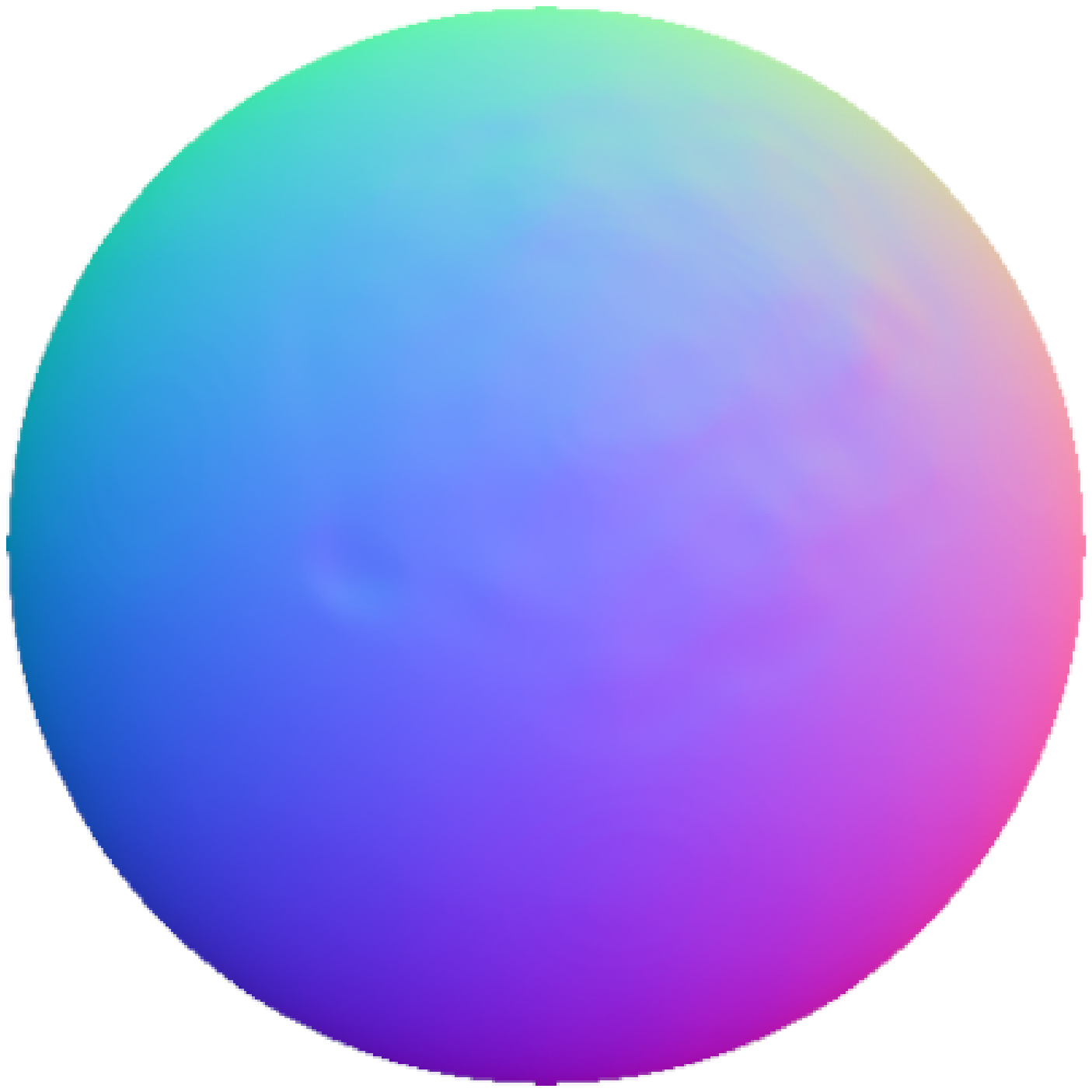}
&\includegraphics[height=\imh]{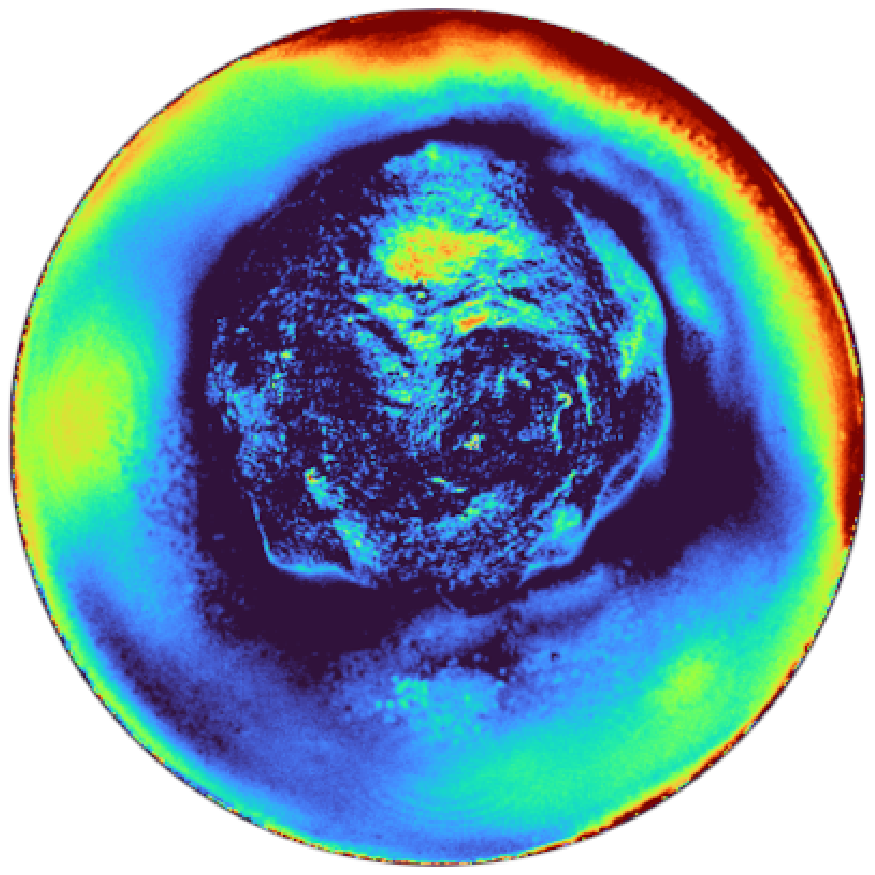}
&\includegraphics[height=\imh]{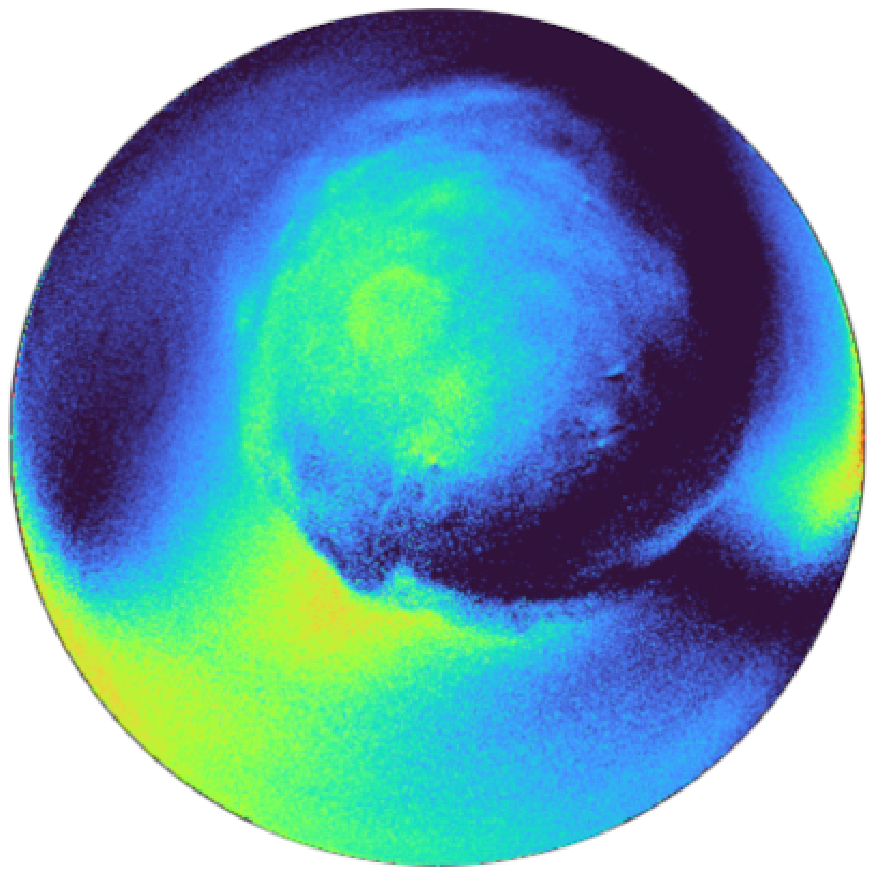}
&\includegraphics[height=\imh]{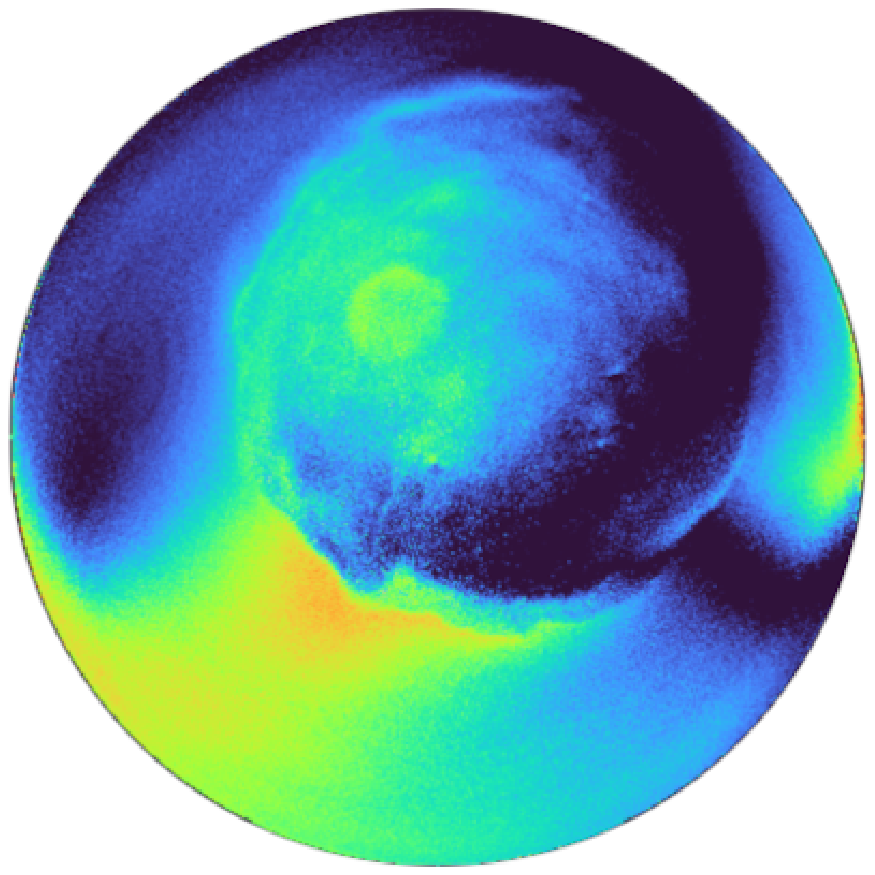}
&\includegraphics[height=\imh]{images/shapevalidation/acetal-green/blank.eps}
&
\\

& & & $8.19$ & $16.13$ & $17.07$ & & \\

&\includegraphics[height=\imh]{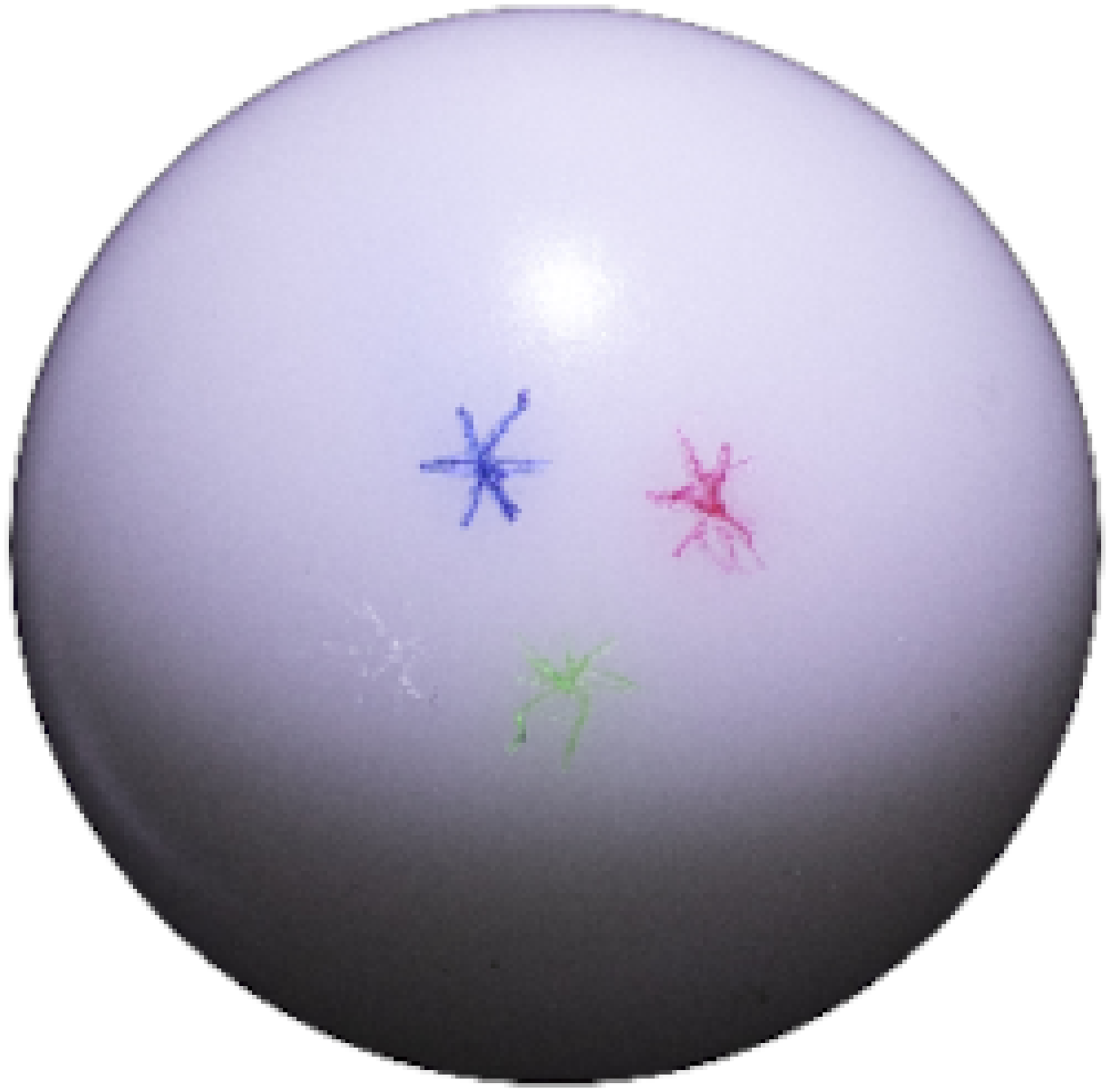}
&\includegraphics[height=\imh]{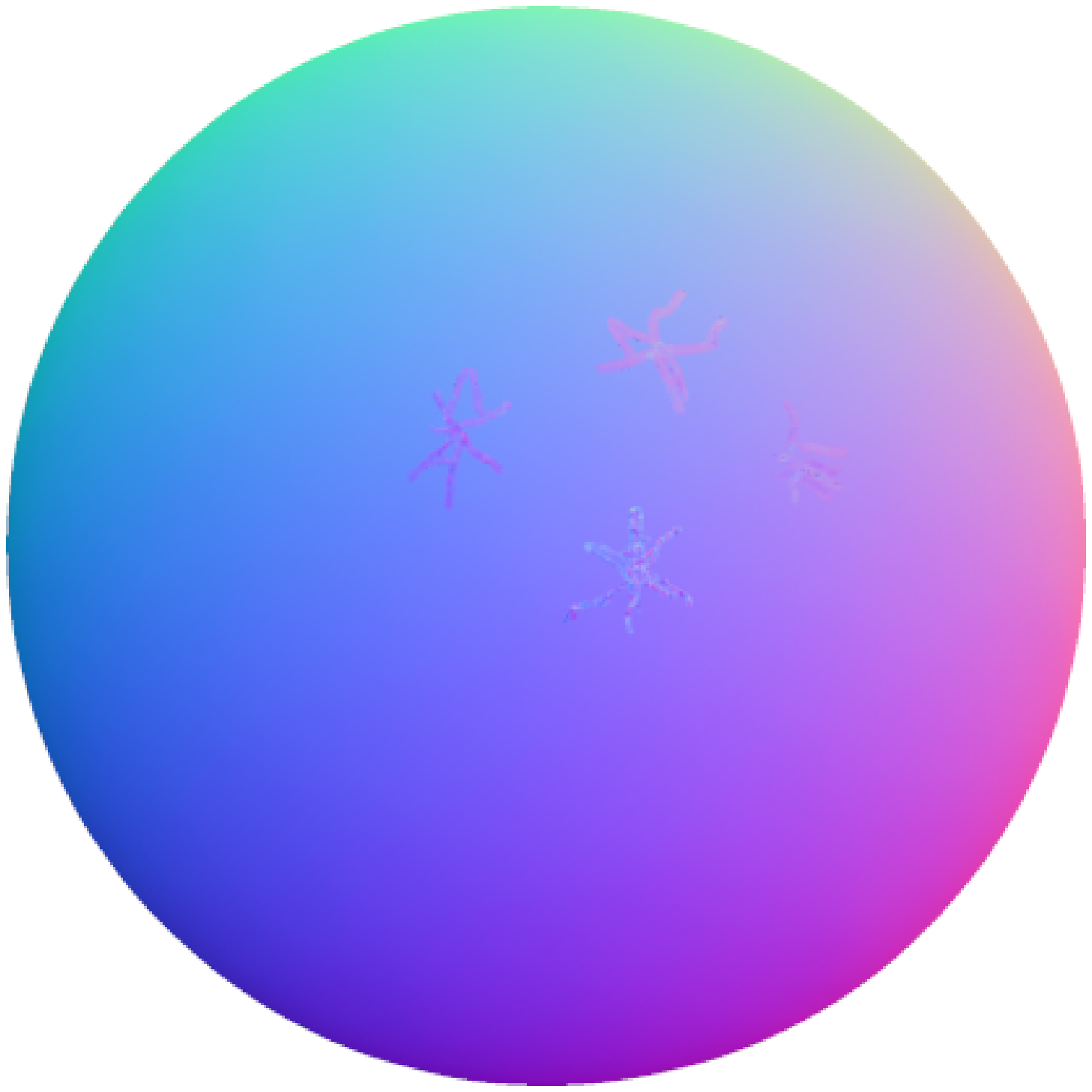}
&\includegraphics[height=\imh]{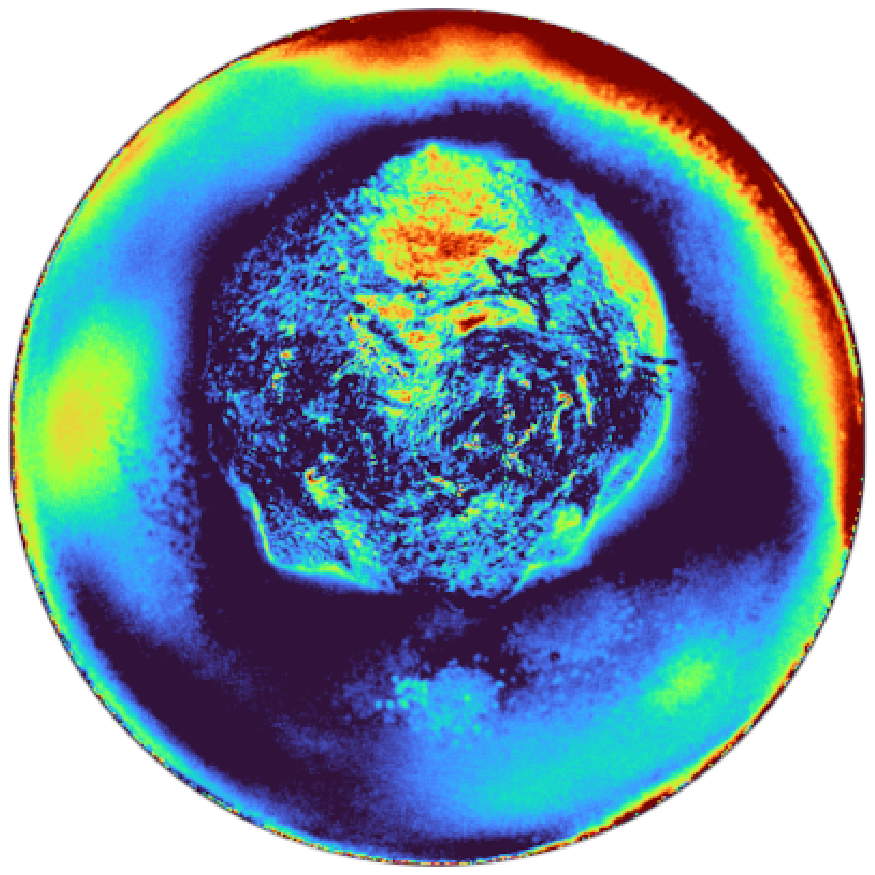}
&\includegraphics[height=\imh]{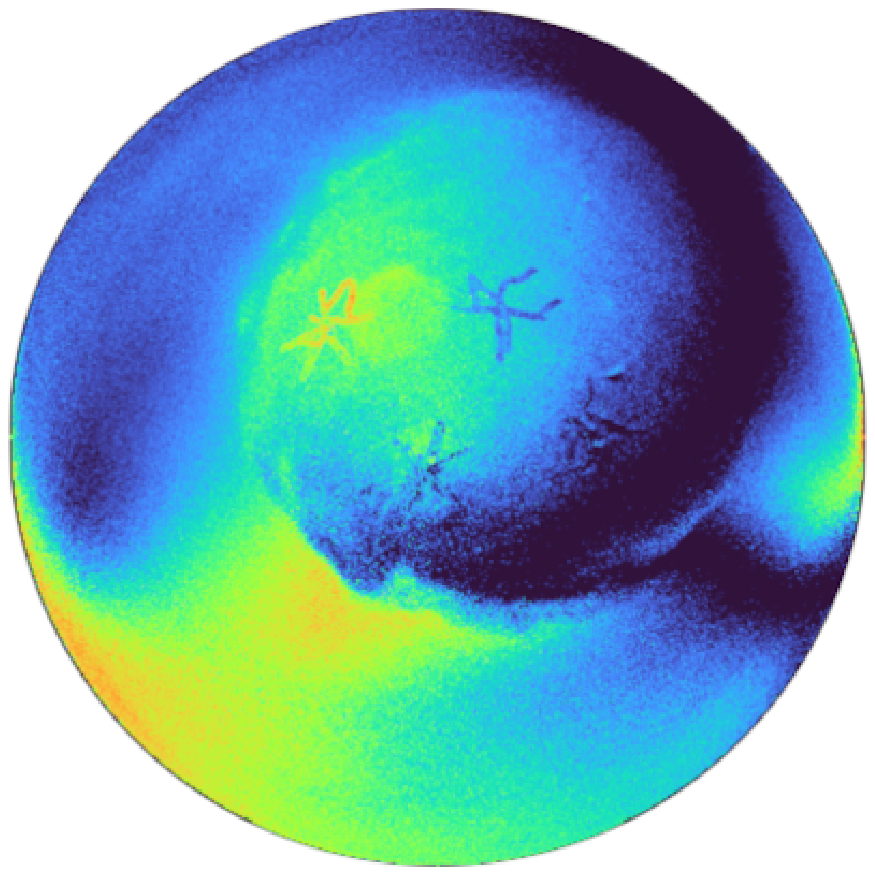}
&\includegraphics[height=\imh]{images/shapevalidation/acetal-green/blank.eps}
&\includegraphics[height=\imh]{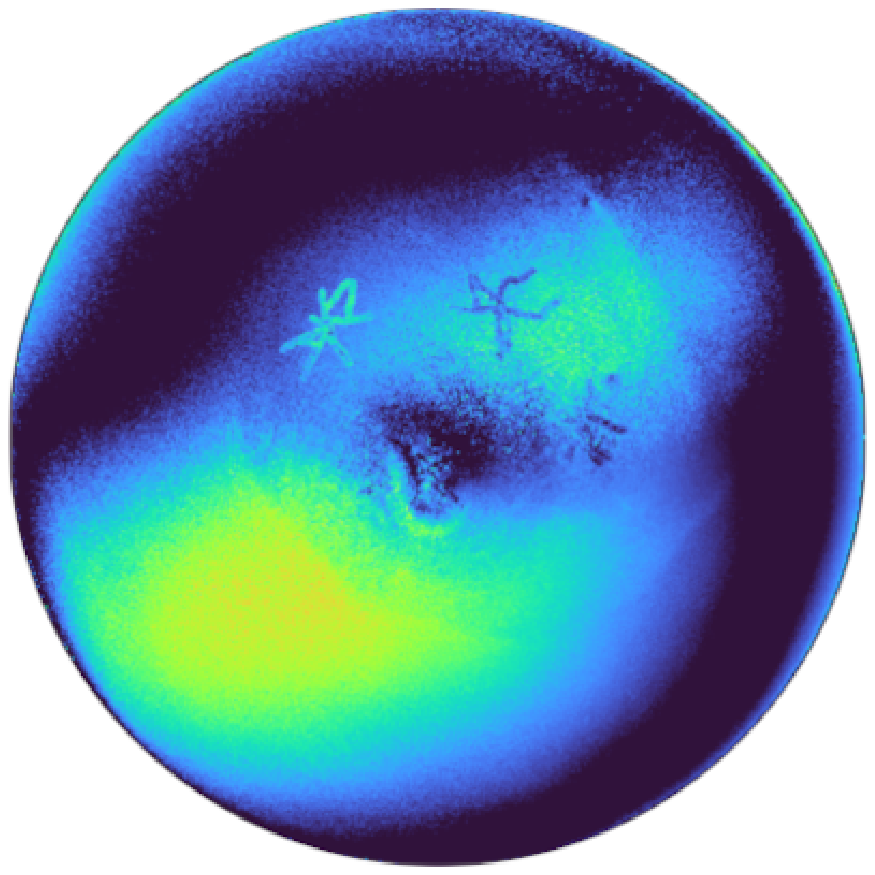}
&
\\

& & & $6.54$ & $17.84$ & & $14.07$ & \\

\end{tabular}
\setlength\abovecaptionskip{-0.3\baselineskip}
\setlength{\belowcaptionskip}{-17pt} 
\caption{ Normal Validation: We compare our computed normals of the top paint layer, and the recovered sub-surface layer with ground truth \hbox{3-D} geometry of the spheres before and after applying the top paint layer. The numbers record average angular error. The heat maps visualize the error in normal differences with blue temperature indicating accuracy and red temperature indicating error. For comparison, we show error with both top and bottom layers for traditional photometric stereo and matrix ranking.}
~\label{fig:normalvalidation-IR}
\end{figure}

\section{Multispectral Normal Reconstruction}
\label{sec:acquisition}

An object's appearance depends on material composition and the wavelength of light hitting the surface. Visible light (VIS $400-700nm$) grazing the surface accentuates surface detail. Longer near-infrared wavelengths (NIR $700nm-3.0\mu m$) penetrate top layers revealing subsurface detail. Bispectral reflectance occurs when short ultraviolet wavelengths (NUV $300-400nm$) are absorbed while longer ones are emitted (fluorescence). In this section, we compute normals using a variant of photometric stereo that captures this wavelength-dependent behavior of light. 

Photometric stereo~\cite{Woodham80} assumes a diffuse object, and uses the relationship  between observed intensity $e$, the normal $n$ and light direction $l$ at point $p$ ($e = n \cdot l$), to compute per pixel normals. Using observations of $p$ lit from at least three directions, we solve an overdetermined system of equations to recover $n$. 

Traditional photometric stereo will not work for our analysis of layered materials as it primarily recovers the top surface shape, and is prone to artifacts from interreflections in non-diffuse objects~\cite{Wu2010}.  Multispectral extensions aim to improve quality in top surface normals by combining the most diffuse normals from different spectral wavelengths~\cite{Takatani2013,Sato2012,nam2016}, and by removing self-emitted radiance for effective interreflection removal~\cite{NamK14}. However, these strategies are problematic when longer wavelengths penetrate beneath the surface to reveal underlying geometry that is different. Solutions that merge normal maps, or otherwise combine reflectance data across wavelengths contain inaccuracies from subsurface scattering at surfaces with varying shape at different material layers.

\subsection{Approach}
 
Our goal is a practical, computationally inexpensive method to extract shape from at least one subsurface layer with only slight modifications to a standard multispectral photometric stereo set-up. First, we record the observed radiance at each wavelength for $37$ light orientations for three exposure values $\mathbf{S}_{\lambda_{Ev}}$. Unlike other methods, we treat $\mathbf{S}_{\lambda_{Ev}}$ as a separate shape layer, and use information from multiple exposure observations, $\{EV_{0}, EV_{1}, EV_{2}\}$ to limit optimizations to a single spectral band. We detect highlights at intensity peaks that spread outward from $EV_{0}$ to $EV_{2}$ (Figure~\ref{fig:highlight}). We examine $\mathbf{S}_{\lambda_{Ev}}$ in the frequency domain as spatial analysis of highlights is challenging with monotone images.  We use Welch's averaged, modified periodogram~\cite{welch67}  to locate frequencies where the magnitude squared coherence ($\mathbf{C}_{xy}$)  between $Ev_0$  and $Ev_1$ (likewise $Ev_0$  and $Ev_2$), is lower than the $50^{th}$ percentile, $th_{ev}$  (divergence for intensity shifts is expected). For two signals, $\mathbf{C}_{xy}$ is a function of their power spectral densities and cross power spectral density. Values range from $0$ to perfect coherence at $1$. Specular-free images  (SF) for highlight detection~\cite{tan2005}  have been generated from scene priors without specularities (flash and no flash images) or by substituting color components.  Without loss of geometric structure, we compute $s_h$, the SF image as the difference in pixel intensities with $s_o$, a highlight-free image at the same tilt angle.  Logarithmic differentiation applied to $s_o$ and $s_h$ extracts pure diffuse reflection. Specular highlights are reduced in $s_h$, by iteratively shifting maximum pixel brightness to that of a more diffuse neighboring pixel. Typical exposure times in seconds are blue and green - $0.2$, $0.5$, $0.7$, red - $1.8$, $2.0$, $2.4$, and $nir$ - $0.8$, $1$, $1.4$, with $th_{ev}$ typically $0.13$.  Additional preprocessing time, image size constraints for FFT, and shadow interference are limitations. Additional exposures are discarded, and per layer  normals are computed as:

\begin{equation}
e_{\lambda}\left(\lambda, x\right) = D\left(\lambda\right)\rho\left(\lambda, x\right)L\left(\lambda\right)n\left(x\right)^{T}l
\label{equ:radiance}
\end{equation}

\noindent where $D\left(\lambda\right)$ is the spectral sensitivity of the camera, $\rho$ is the spectral reflectance of a point, and $L\left(\lambda\right)$ are the spectrum of the light. 

Our capture system includes a radiometric CCD imager with an internal motorized filter wheel~\cite{minkim2011} for aligned multispectral imaging,  essential for accurate per pixel comparison across wavelengths. We also work with affordable visible, and IR modified DSLR cameras (prevalent in our application space). The IR camera is optimized to focus at $720nm$. External lens filters permit imaging at $830nm$ and $1000nm$, but require a manual focus shift, and vignette compensation.  To capture macro detail on small and large samples, we use two light sizes, based on effective attenuation angle and the number of LEDs. Figure~\ref{fig:transcurves} summarizes our filters, lights and sensors. See additional information in supplemental materials. Micro imaging is discussed later in this section.  

\subsection{Validation}

\noindent\textbf{Experimental Setup:} We quantitatively evaluate the accuracy of our normals in Figure~\ref{fig:normalvalidation-IR}. Using our CCD imager, we compute photometric stereo at r, g, b, near-infrared ($nir$), and rgb combined ($vis$) channels on spheres with etched grooves under paint layers. We compare our normals ($\mathbf{n}_{nir}$  and $\mathbf{n}_{vis}$) with ground truth geometry; \hbox{3-D} scans of the grooves ($gt_{bottom}$) and paint layer ($gt_{top}$). Our spheres, labeled  by color, include $ptfe_{red}$, a  near diffuse $\frac{3}{4}$ inch diameter polytetrafluoroethylene (PTFE), and  two one-inch polyoxymethylene  (acetal) spheres,  $acetal_{red}$ and $acetal_{green}$ that exhibit specular reflectance. Opacity, thickness, absorption, transmission, and light wavelength~\cite{Burleigh2016} effect optical properties of paint. We chose tempera paint for its opacity and natural composition, like our data. Several paint layers ensure grooves are not discernable in visible light. We do not use synthetic datasets in our analysis because we require a near-infrared source, but include a spectralon sphere as a base-line standard.

Experimental results show that \textbf{we accurately recover shape beneath the paint layer}. We compare $\mathbf{n}_{nir}$ with $gt_{bottom}$, and $\mathbf{n}_{vis}$ with $gt_{top}$, and use heat maps $\mathbf{H}$ to visualize the error as angular differences. The numbers show average error. Heat maps $\mathbf{H}_{(ptfe_{red}, gt_{bottom})}$, $\mathbf{H}_{(acetal_{red}, gt_{bottom})}$, $\mathbf{H}_{(acetal_{green}, gt_{bottom})}$,\\
\mbox{confirm} that we recover all four sets of grooves, and the overall spherical geometry accurately (dark blue, zoom to see details). Average error is lower in $\mathbf{H}_{(ptfe_{red}, gt_{top})}$  and $\mathbf{H}_{(ptfe_{red}, gt_{bottom})}$ because the diffuse surface has less interreflection artifacts than the reflective acetal examples. Most of our error is due to interreflections in unpainted regions of acetal spheres. Although matrix ranking~\cite{Takatani2013} consistently selected the most diffuse spectra per segment,  it did not accurately recover shape in $gt_{top}$ or $gt_{bottom}$ for $ptfe_{red}$ or $acetal_{green}$ because it combined shape across different material layers (reddish yellow regions). This method had nearly the same performance as our approach (error $12.056$ and  $12.067$ respectively)  for  $\mathbf{H}_{(acetal_{red}, gt_{bottom})}$ because, due to the simplicity of our data, the normal map had primarily near-infrared normals. Grooves appear in both normal layers for $acetal_{red}$ which transmits red ($660nm$) and infrared ($720nm$). Conversely,  $acetal_{green}$ absorbed $vis$ and $nir$, appearing black in our red, blue  and $nir$ filters, and dark grey under the green. In this case, imaging required a high-powered IR light ($315nm$ to $3000nm$) and a $920nm-1080nm$  filter. We did not see grooves below $920nm$ (though one was recorded in the $vis$ green, lowering expected error in $\mathbf{H}_{(acetal_{green}, gt_{bottom})}$). We used a full-spectra light and a $920nm-1080nm$ filter with $ptfe_{red}$ which had lower $nir$ reflectance than acetal. Traditional photometric stereo had higher average errors from interreflection and subsurface scattering ($12.46$, $8.19$,  for $acetal_{red}$ and $acetal_{green}$ respectively), but low error for $pfte_{red}$  ($7.51$) and spectralon ($5.31$).   

\noindent{\textbf{Limitations}} Our method is effective for naturally occurring materials that exhibit near-infrared transmittance. Synthetic polymer-based paint with complex structure and high reflectance hide lower level shape (see supplemental examples). Although we investigate separate channels in the visible range, we use $\mathbf{n}_{vis}$ for our rendering, and handle $\mathbf{n}_{nir}$ as a separate layer.  

\subsection{Object-Scale Shape}	

\begin{figure}[t]
\centering
\includegraphics[width=0.7\hsize]{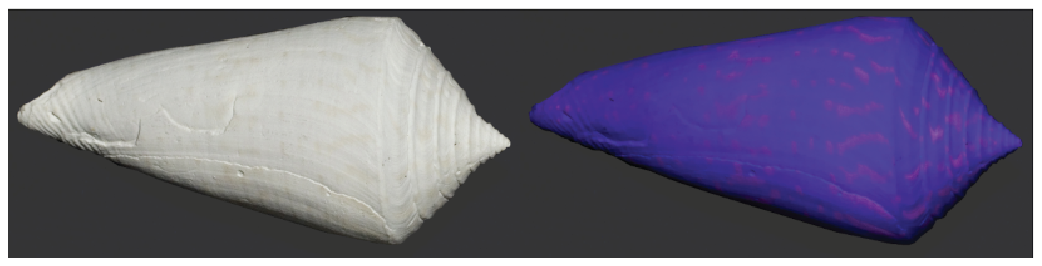}\\
\includegraphics[width=0.7\hsize]{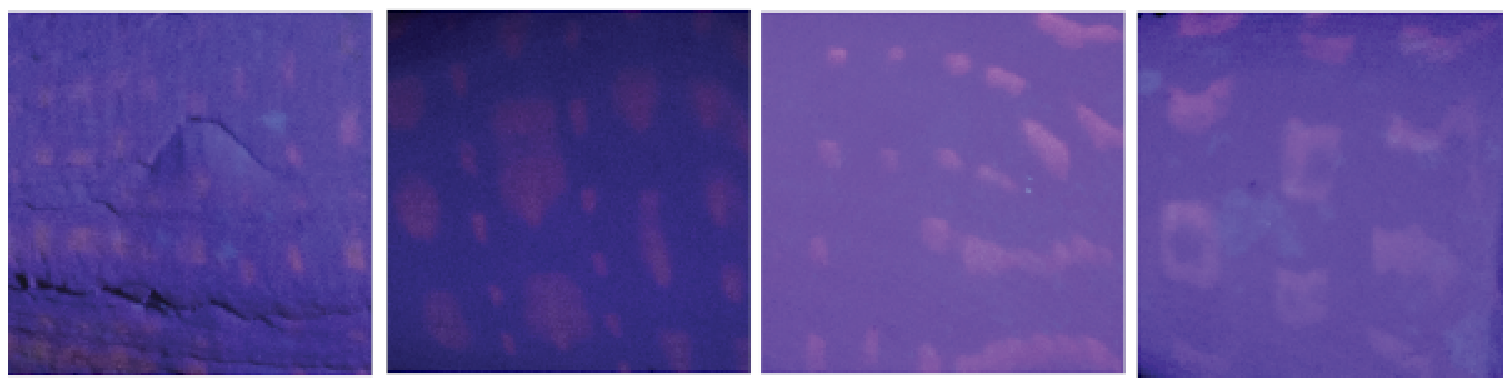}\\
\setlength{\belowcaptionskip}{-10pt} 
\setlength\abovecaptionskip{2pt}
\caption{\label{fig:uvandsamples}%
Top: UV fluorescence (right) of a colorless shell fossil reveals color patterns not distinguishable under visible light (Left). These unique identifiers are used for species classification. Bottom: (left to right) Conus delisserti No. AMNH-10695, Conus delisserti No. UF-60317 FLMNH, Conus delisserti No. UF11729 FLMNH, Conus spurius No. UF-113870-803 FLMNH.}
\end{figure}

Thus far, we have focused on visible and near-infrared normals. Bispectral normals from emissive rather than reflective  wavelengths approximate shape better when there are specularities (~\cite{Sato2012,Treibitz2012}). Bispectral shape recovery is challenging for purely fluorescent~\cite{Glassner1995AMF} objects with changing material composition (Figure~\ref{fig:uvandsamples}). Even objects with relatively constant emission spectra are difficult when material concentration varies~\cite{wiley2011, zhang2011}. Hybrid deep learning approaches combine near and far approximation lights to capture spatially varying reflectance~\cite{santo2020deep}, but ignore effects of light penetration depth. Large volumes of data are required, and simple visible-only datasets are limiting.  GPU memory constraints and patch-based operations reduce output quality.  Layered material simulations focus on photo realism ~\cite{Yamaguchi2019}. 

\begin{figure}[t]
\centering
\includegraphics[width=0.9\hsize]{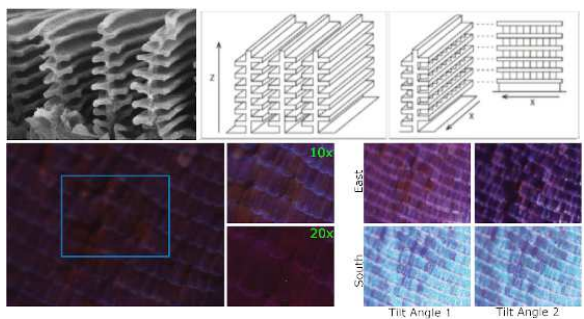}
\setlength\abovecaptionskip{2pt}
\setlength{\belowcaptionskip}{-13pt} 
\caption{\label{fig:micro}%
Layered materials in a Morphidae blue butterfly wing: (top left) Scanning electron microscopy image. (top right)  Lamellae multilayer and microribs. (bottom left) Object scale features at two magnifications. (bottom right) Color varies with orientation but not tilt angle. West and east for two tilt angles. Morpho sulkowsly butterfly, McGuire Center for Lepidoptera and Biodiversity, FLMNH.} 
\end{figure}

We incorporate bispectral shape and luminance in our shading algorithms to enhance object-scale features and structural differences between  materials. Combinations of thin translucent and colorful opaque layers in butterfly wings~\cite{Tabata1996} cause hue to vary with light orientation but remain consistent with changes in tilt angle (Figure~\ref{fig:micro} bottom row). We approximate the emittance spectra from observed bispectral reflectance in micro-images, using a standard Olympus micro camera, common to museums and life science research labs. We image $15$ light orientations,  limited by the closeness objectives ($5x$, $10x$ and $20x$) to the object. We then compute normals using the visible emission spectra (r, g and b) from ultraviolet excitation ($365 nm$), $\mathbf{n}_{bispectral}$, which we combine with $\mathbf{n}_{vis}$, only at fluorescing pixels. Though bispectral shape recovery for composite materials is beyond scope, the results ($\mathbf{n}_{combined}$)  capture more structural detail than current NPR approaches. We now discuss our datasets and layered normals in the context of our guiding principles.

\begin{figure}[h]
\centering
\includegraphics[width=0.9\hsize]{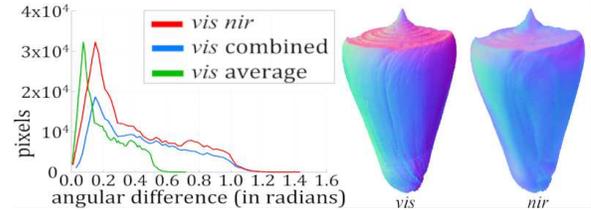}
\setlength\abovecaptionskip{2pt}
\setlength{\belowcaptionskip}{-12pt}
\caption{\label{fig:normals}%
(left) Comparing multispectral normals: angular difference between visible(middle) and near-infrared (right).}
\end{figure}

\begin{figure}[h]
\centering
\def\imh{0.7in}
\setlength{\tabcolsep}{0.2pt}
\begin{tabular}{ccc}

$\lambda = 400-700nm$ & $\lambda = 720nm$ & $\lambda = 830nm$\\ 

\includegraphics[height=\imh]{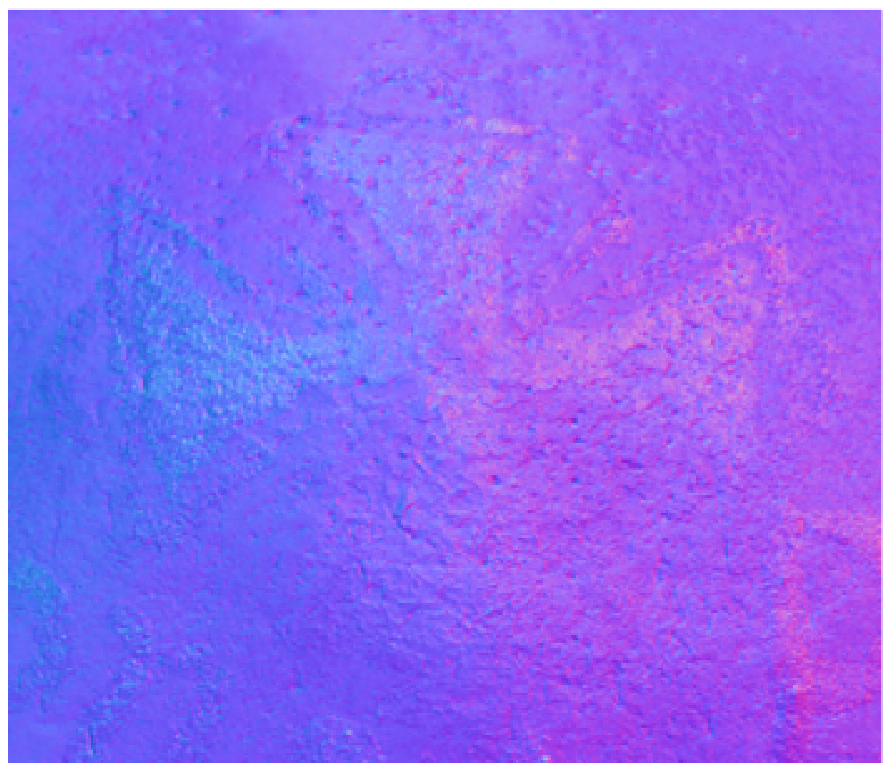}
&\includegraphics[height=\imh]{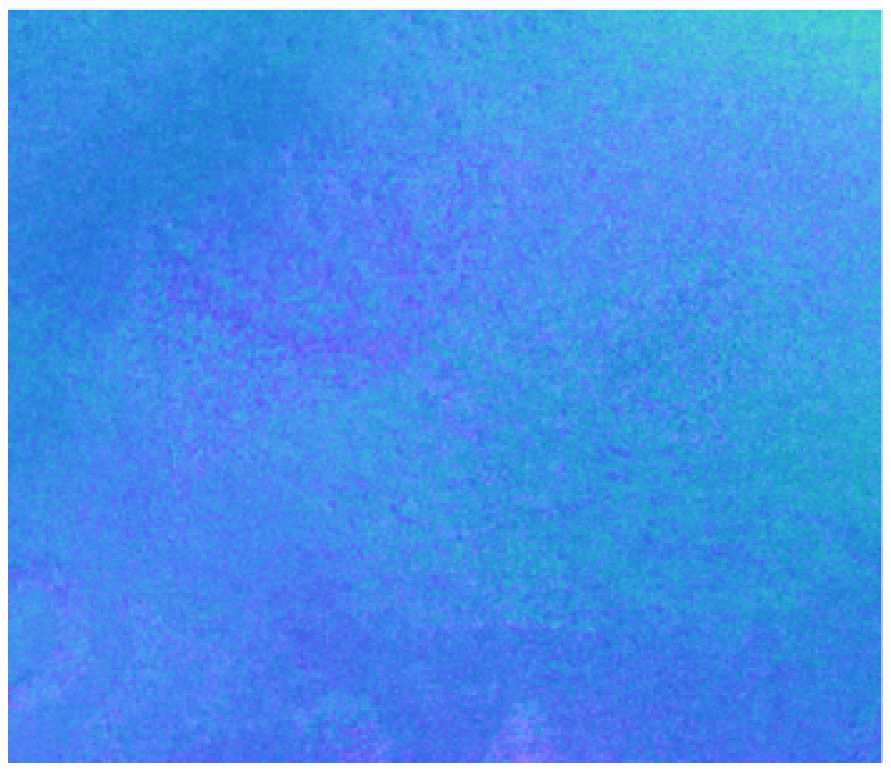}
&\includegraphics[height=\imh]{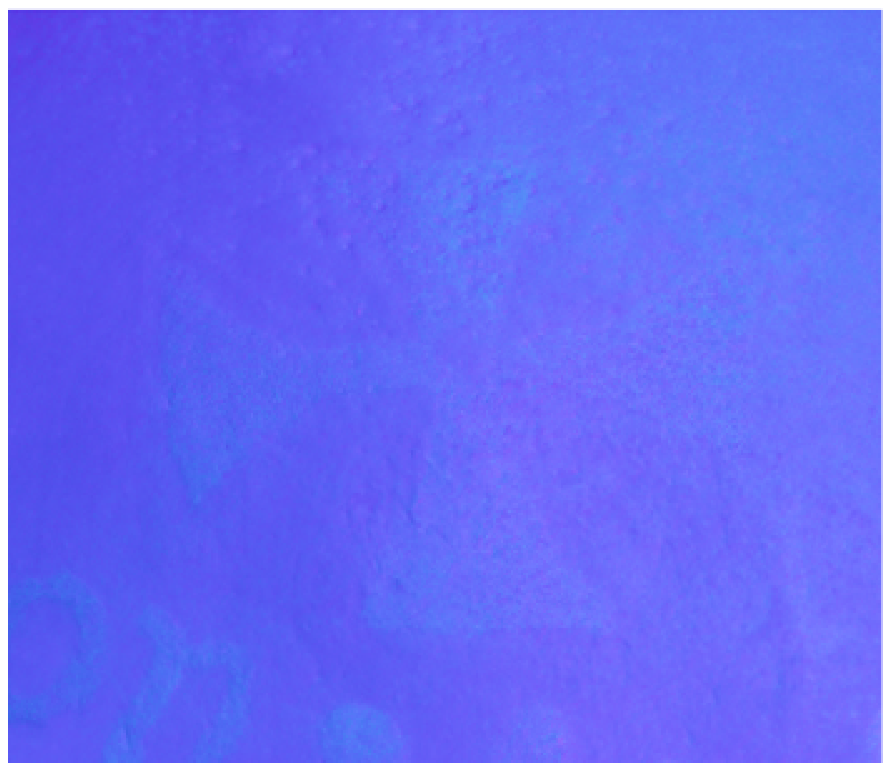}
\\

\includegraphics[height=\imh]{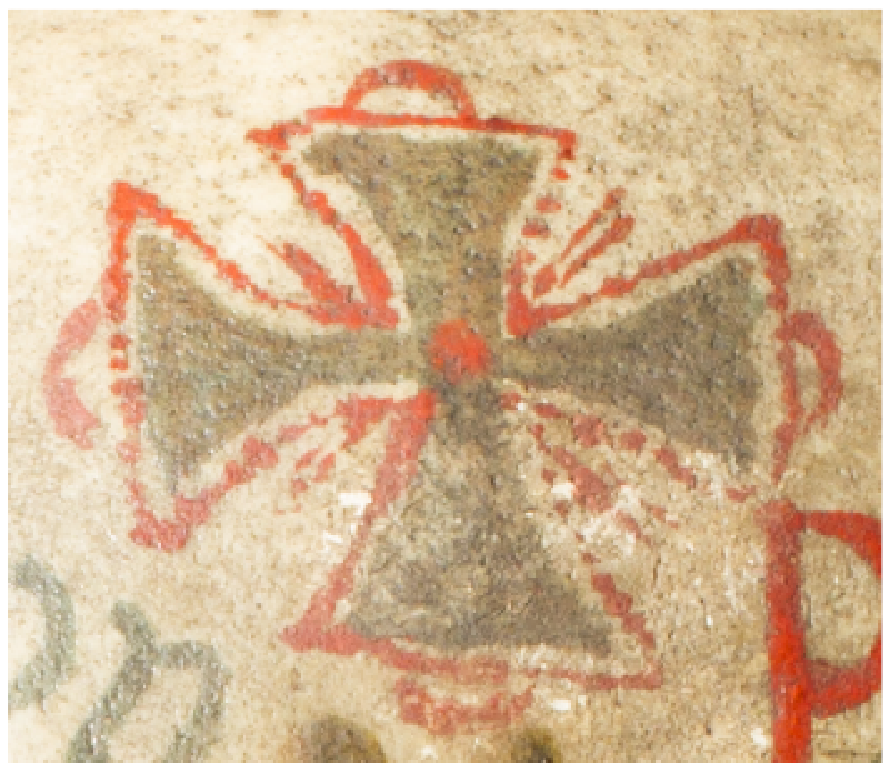}
&\includegraphics[height=\imh]{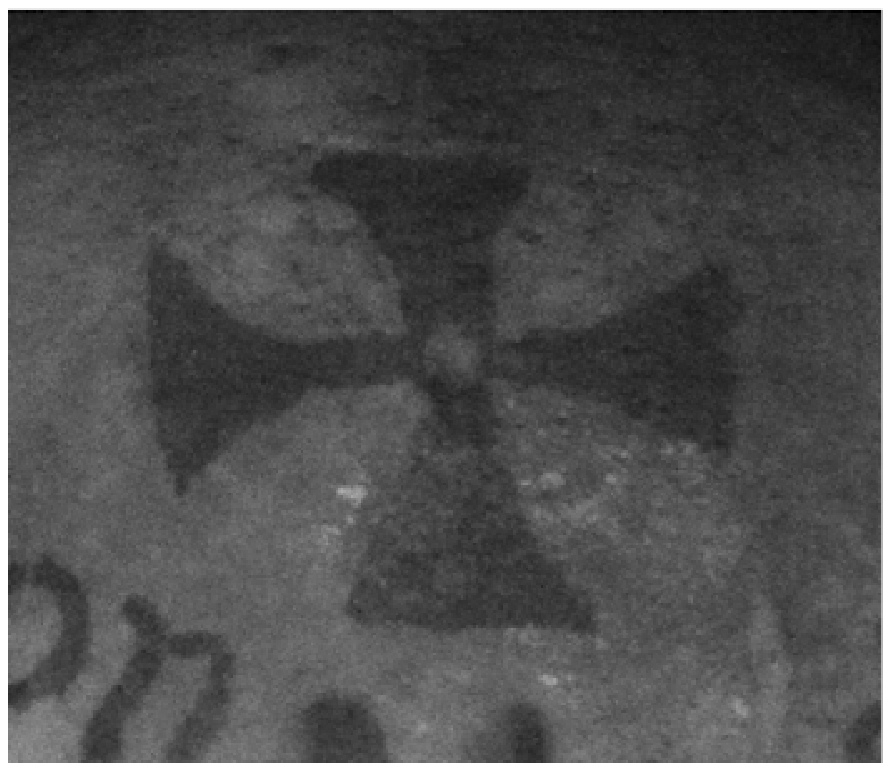}
&\includegraphics[height=\imh]{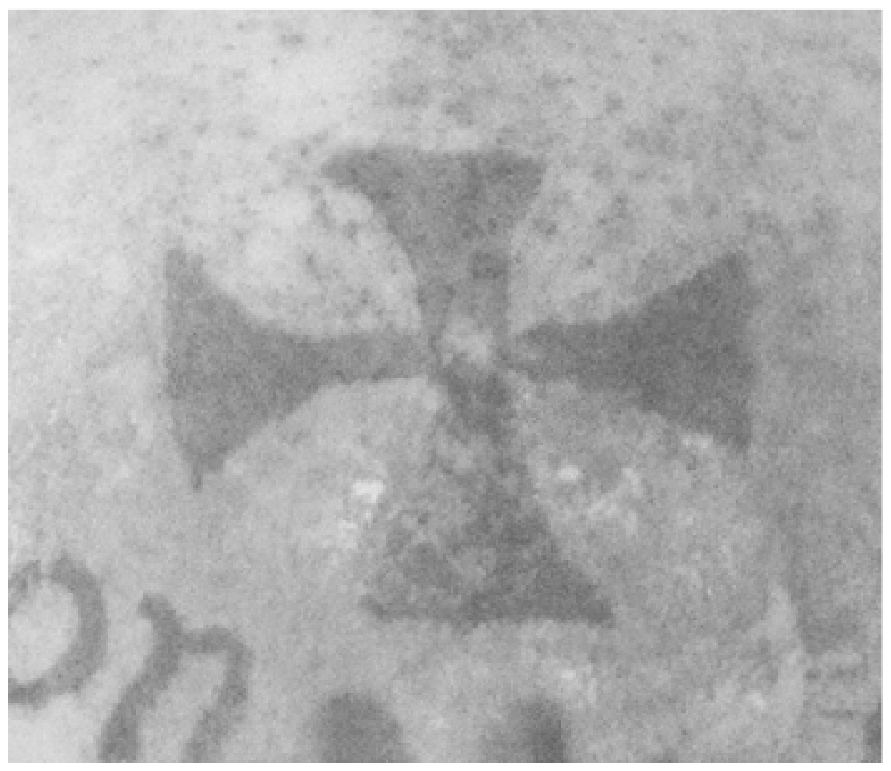}
\\

\end{tabular}
\setlength\belowcaptionskip{2pt} 
\caption{Multispectral Normals. Painted cranium, No. VL/122, courtesy of the Division of Anthropology, American Museum of Natural History.}
\label{fig:multispectralcross}
\end{figure}

\textbf{Principle~\ref{itm:filtering}} recommends minimal smoothing to preserve longer wavelengths that penetrate deeper, and shorter wavelengths that encode shape details and discontinuities. Figure~\ref{fig:normals} (\emph{middle and right}) shows recovered shape at different segment layers for a shell fossil after smoothing.

\textbf{Principle~\ref{itm:contrastenhancement}} emphasizes differences in shape which we recover at multiple wavelengths. Figure~\ref{fig:normals}  \emph{left} shows the shape differences between layers for the normals shown to the right, which we incorporate in our contrast enhancement techniques. In Figure~\ref{fig:multispectralcross}, reflectance from the red pigment produces shape in the visible normal map while the near-infrared map captures curvature features

\begin{figure}[h]
\centering
\includegraphics[width=1.0\hsize]{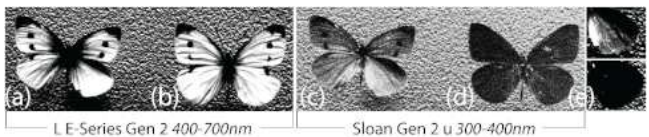}
\setlength\abovecaptionskip{-0.7\baselineskip}
\setlength{\belowcaptionskip}{-10pt} 
\caption{\label{fig:uvcabbagewhite}%
UV reflected imaging for analyzing bio-materials. The female (a) and male (b) cabbage white butterfly appear equally bright under a clear L E-Series Gen 2 filter. Under the same visible lighting and view point, UV absorbing pteridins cause the male (d) to appear black under a Sloan Gen $2$ u $300-400nm$ filter, while the female (c) appears a lighter grey. Pieris rapae, McGuire Center for Lepidoptera and Biodiversity, FLMNH.}
\end{figure}

\noindent of the underlying bone. Note how the red letter \emph{P} fades as the near-infrared wavelength increases from $720nm$ to $830nm$.

\textbf{Principle~\ref{itm:uvpatterndetection}} leverages bispectral imaging~\cite{Sato2012} to record fluorescence and residual color patterns (Figure~\ref{fig:uvandsamples} top). We use observed fluorescence as emitted red, green and blue intensities under ultraviolet excitation.

\textbf{Principle~\ref{itm:microdetail}} incorporates shape variations at different magnifications which are known to influence object-scale reflectance (Figure~\ref{fig:micro} \emph{bottom left}). Butterfly wings are complex layered structures with ridges, lamellae and a lattice of microribs (Figure~\ref{fig:micro} \emph{top}) that produce complex reflectance properties. Figure~\ref{fig:micro} \emph{bottom right} shows how chrominance changes with orientation but not tilt angle.  Reflected UV imaging in Figure~\ref{fig:uvcabbagewhite} shows how the structure and composition of a \emph{cabbage white} butterfly wing influences the absorption and emission spectra.



\section{Multispectral Data Processing}
\label{sec:msprocessing}

\subsection{Filtering}
Joint bilateral filtering removes residual capture artifacts. When contrast modulation operators are applied to near-infrared intensity values, they often produce values that range close to zero. Following Principle~\ref{itm:filtering}, we apply minimal filtering to prevent loss of this information (typically $9$ passes using a $3$ pixel width filter with domain, range and normal kernel widths of $0.25$, $5$, and $1$ respectively).

\begin{figure}[h]
\centering
\includegraphics[width=1.0\hsize]{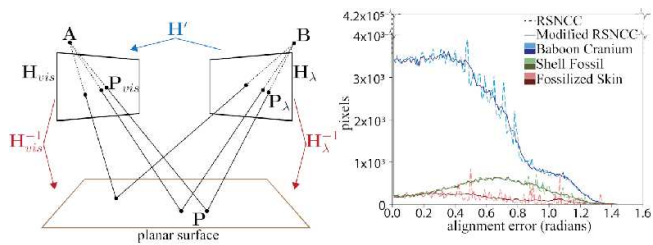}
\setlength{\belowcaptionskip}{-10pt} 
\setlength\abovecaptionskip{-0.7\baselineskip}
\caption{\label{fig:align}%
(left) The affine homography matrix $H^{\prime}$ transforms point $P_{\lambda}$ with the intrinsic parameters $B$ to $P_{vis}$, the corresponding point in the reference frame with the intrinsic parameters $A$. (right) Comparing RSNCC and modified RSNCC performance.}
\end{figure}

\subsection{Multispectral Multimodal Registration}

Multispectral, multimodal registration is required for sub-pixel accuracy when comparing information across spectral bands from different devices.  We modify a Robust Selective Normalized Cross Correlation (RSNCC)~\cite{Shen2014} algorithm (Appendix~\ref{append:RSNCC}). The inputs are visible and near-infrared intensity/normal map pairs. We first generate a composite image $\digamma_{\lambda}$ at each wavelength using the $80^\text{th}$ percentile intensity at each pixel across the multilight stack (from normal computation) to remove shadows that skew results. We use RSNCC to register each $\digamma_{\lambda}$ to a reference visible image $I_{\mathbf{vis}}$. Dense pixel correspondences are used to generate a global and local alignment transformation between each $\digamma_{\lambda}$ and $I_{\mathbf{vis}}$.

Next, normals from different spectra must be aligned. RSNCC cannot be applied to normals without introducing shape distortions as it computes similarities between image-based color and gradient features. The three regularization terms in the cost function $\mathnormal{E}_{\mathbf{w}}$ (Equation~\ref{equ:local}) use convolution operations that over-smooth normal vectors. Large variations in normals across spectral bands make them unsuitable as a similarity measure. We  iteratively transform color and geometry without disassociating per-pixel normals from corresponding pixels in the intensity map. We use the homography matrix $\mathnormal{H}$ from the global alignment of $\digamma_{\mathbf{\lambda}}$ with $I_{\mathbf{vis}}$ to perform a coarse rigid body alignment that shifts the pixel coordinate location of each normal in the image frame. We do not use the RSNCC cubic bilinear interpolation to shift and blend pixel values but shift normal positions directly. Assuming one camera is fixed, we use the homography relationship between the two images (Figure~\ref{fig:align})to transform the position $\mathnormal{P}_{\mathbf{\lambda}} \left(u, v, 1\right)$ to $\mathnormal{P}_{\mathbf{vis}} \left(u, v, 1\right)$.  $\mathnormal{H}$ is not ideal when computed on small regions or a broad range of focal lengths. Thus, we use an affine homography transformation $\mathnormal{H}^{\prime}$ by modifying  the last row of $\mathnormal{H}$ to $\{0, 0, 0, 1\}$.  Finally, we apply the transformation from the localization phase. We adjust the process by filtering on $\frac{n_x}{n_z},\frac{n_y}{n_z},\frac{nz}{n_z}$ to prevent foreshortening effects and limit discontinuities from data loss which occur when pixel positions are shifted~\cite{rgbn07}. The cost functions we choose are more effective (than SIFT and mutual information) because they are resistant to noise, intensity variations, gradient reversal and structural inconsistencies between near-infrared and visible images. Figure~\ref{fig:align}\emph{right} plots the error between ground truth near-infrared normals from aligned imaging with our CCD  (with motorized filter wheel) an infrared DSLR camera aligned using global and local transformations from traditional RSNCC and our modified RSNCC. The reference $I_{\mathbf{vis}}$ stores the combined r, g, b CCD images. Modified RSNCC (solid line) is more accurate than traditional RSNCC (dotted line) for different specimens captured with different intrinsic and extrinsic camera parameters.

\subsection{Near-Infrared Shape Analysis}
\label{sec:contrast}
We introduce two shape analysis methods, a dynamic light dependent approach and a static light independent approach, that identify where to incorporate near-infrared features in stylization. Building upon Principle~\ref{itm:contrastenhancement}, both algorithms operate on visible and near-infrared normals, $\bm{n}_{vis}$ and $\bm{n}_{nir}$, to generate \emph{near-infrared enhancement maps}, $\mathbf{C}$, per-pixel weights that quantify the amount of contribution from each spectral wavelength.

\textbf{Dynamic Near-Infrared Enhancement} uses contrast modulation operators from signal processing to compare wavelength dependent shape functions. We compute $\mathbf{C}$ dynamically by comparing the local contrast in Lambertian shape computed on $\bm{n}_{vis}$ and $\bm{n}_{nir}$:

\begin{equation}
m_{\mathsf{g}} = \frac{\mathsf{g}_{max} - \mathsf{g}_{min}}{\mathsf{g}_{max} + \mathsf{g}_{min}} 
\label{equ:modulation}
\end{equation}

\noindent where $\mathsf{g}(n, \lambda)$ is a luminance function that depends on wavelength and surface normal orientation. Lambertian map $\bm{\chi}_{nir}$ stores $\bm{n}^i \cdot l$ where $\bm{n}^i$ is the normal at pixel $i$ in $\bm{n}_{nir}$ and $\bm{l}$ is the light direction. We choose $\bm{l}$ interactively using grazing angles until features appear clearly. This is analogous to the \emph{raking light} used to exaggerate local details in art analysis. We compute another Lambertian map $\bm{\chi}_{vis}$ using $\bm{n}_{vis}$ and the same $\bm{l}$. Finally, a \emph{Michelson} contrast modulation operator, $\mathnormal{m}$, is applied to each Lambertian map:

\begin{equation}
m^i\left(\bm{\chi}, \lambda, r\right)   = \frac{\bm{L}_{max} - \bm{L}_{min}}{\bm{L}_{max} + \bm{L}_{min}} 
\label{equ:contrastmodulation}
\end{equation}

\noindent where $\bm{\chi}$ is the Lambertian map at wavelength $\bm{\lambda}$ and $\bm{L}_{min}$ and $\bm{L}_{max}$ are the minimum and maximum luminance over a pixel neighborhood $\bm{r}$.  The operator $m$ identifies variations in shape by measuring the relationship between the spread and the sum of $\bm{L}_{min}$ and $\bm{L}_{max}$ over localized regions in $\bm{\chi}$. Small $r$ values produce sharper contrast while larger values produce smooth variations. The results are two per-pixel contrast maps $m_{vis}$ and $m_{nir}$. Equation~\ref{equ:contrastoperation} computes the contrast difference $\bm{\varphi}$ between $m_{vis}$ and $m_{nir}$ at pixel $i$ to encode shape differences in $\bm{\chi}_{vis}$ and $\bm{\chi}_{nir}$:

\begin{equation}
\bm{\varphi}^i = \abs{m_{vis}^i  - m_{nir}^i}
\label{equ:contrastoperation}
\end{equation}

\begin{equation}
\mathbf{C}^i = \begin{cases}
		    \bm{\varphi}_{th}^i,     &if \hspace{4pt} \mathnormal{m}_{nir}^i > \mathnormal{m}_{vis}^i.\\
                0,              &otherwise.                            \end{cases}
\label{equ:contrastmap}
\end{equation}

We clamp $\bm{\varphi}^i$ below a threshold $th$ ($0.1$) to zero as relative spectral shape differences are negligible.  Using thresholded differences, $\bm{\varphi}_{th}^i$, the value $\mathbf{C}^i$ stored at the $i^{th}$ pixel in the final enhancement map is a weight on the interval $[0,1]$ that satisfies Equation~\ref{equ:contrastmap}. Luminance varies with wavelength. Thus, lighter pixel values indicate changes in shape relative to the average at each spectra.

\begin{figure}[ht]
\centering
\includegraphics[width=0.8\hsize]{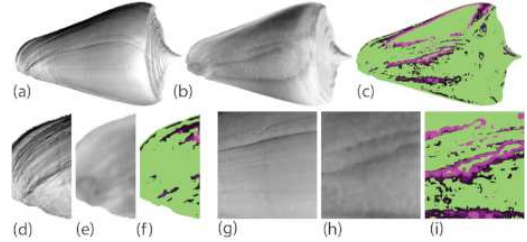}
\setlength{\belowcaptionskip}{-10pt} 
\caption{\label{fig:contrastcloseup} Lambertian shape computed on visible (a) and near-infrared normals (b) for the same light create the dynamic near-infrared enhancement map (c). Closeups d-f and  h-i compare (a), (b) and (c) left to right.}\end{figure}

\begin{figure}[h]
\centering
\def\imh{0.5in}
\setlength{\tabcolsep}{0.1pt}
\begin{tabular}{ccc}

 \small{$\bm{l}\left(-0.83,-0.1,0.56\right)$} 
&\small{$\bm{l}\left(0.51,-0.68,0.53\right)$}
&
\\

  \includegraphics[height=\imh]{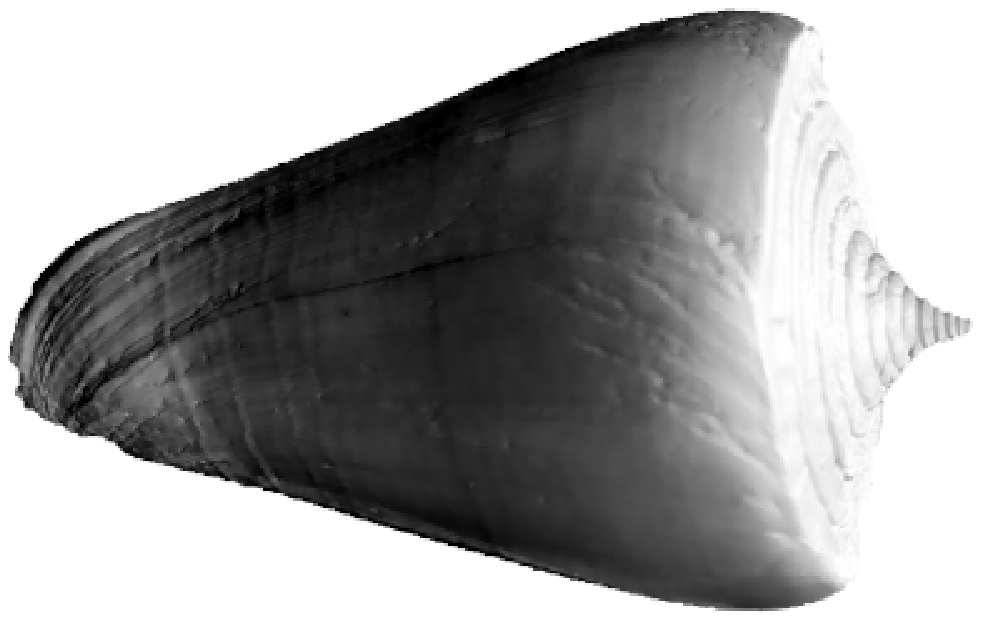}
& \includegraphics[height=\imh] {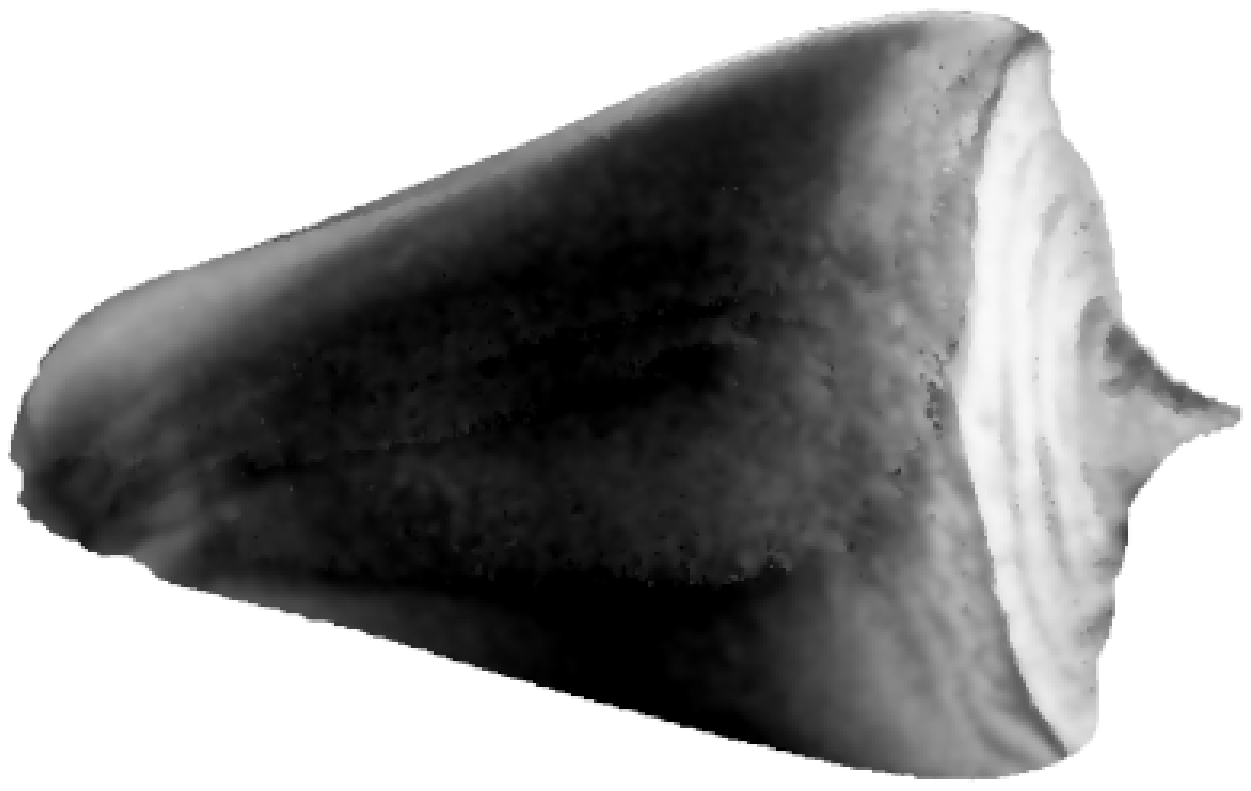}
& \includegraphics[height=\imh] {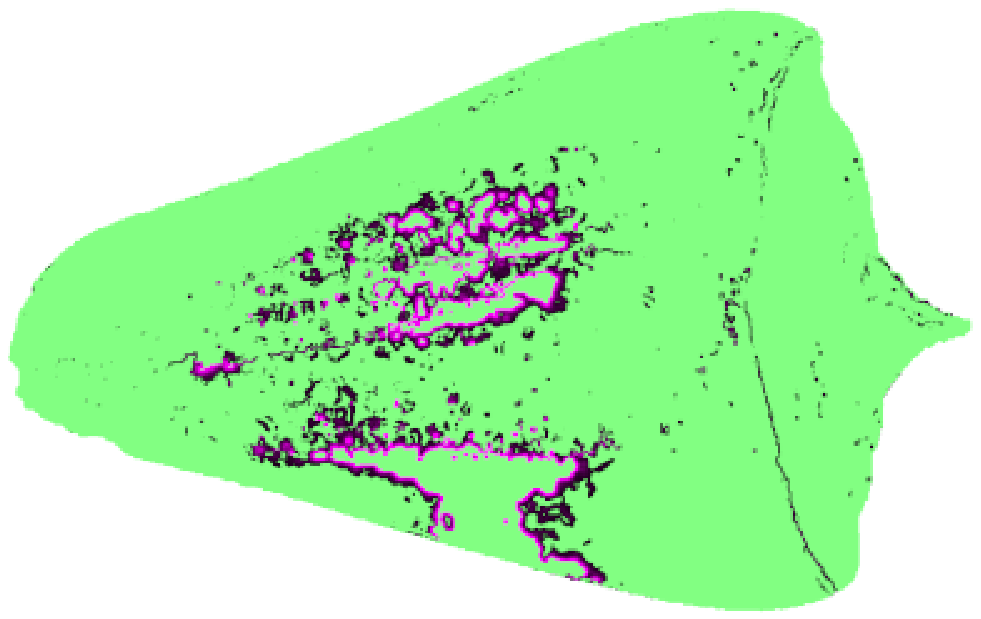}
\\

$\bm{\chi}_{vis}$ & $\bm{\chi}_{nir}$ & $\mathbf{C}$\\

  \includegraphics[height=\imh]{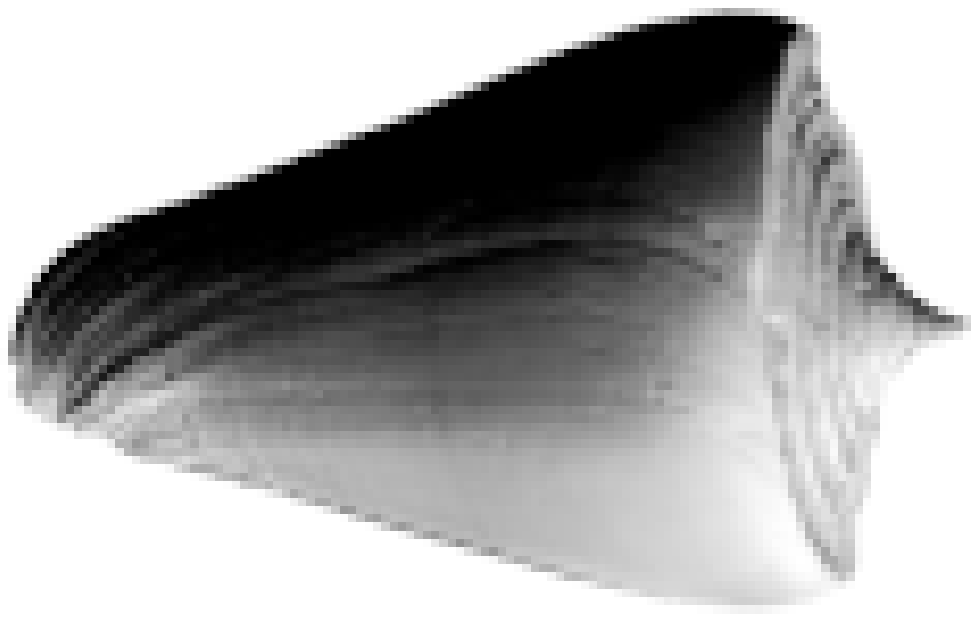}
& \includegraphics[height=\imh] {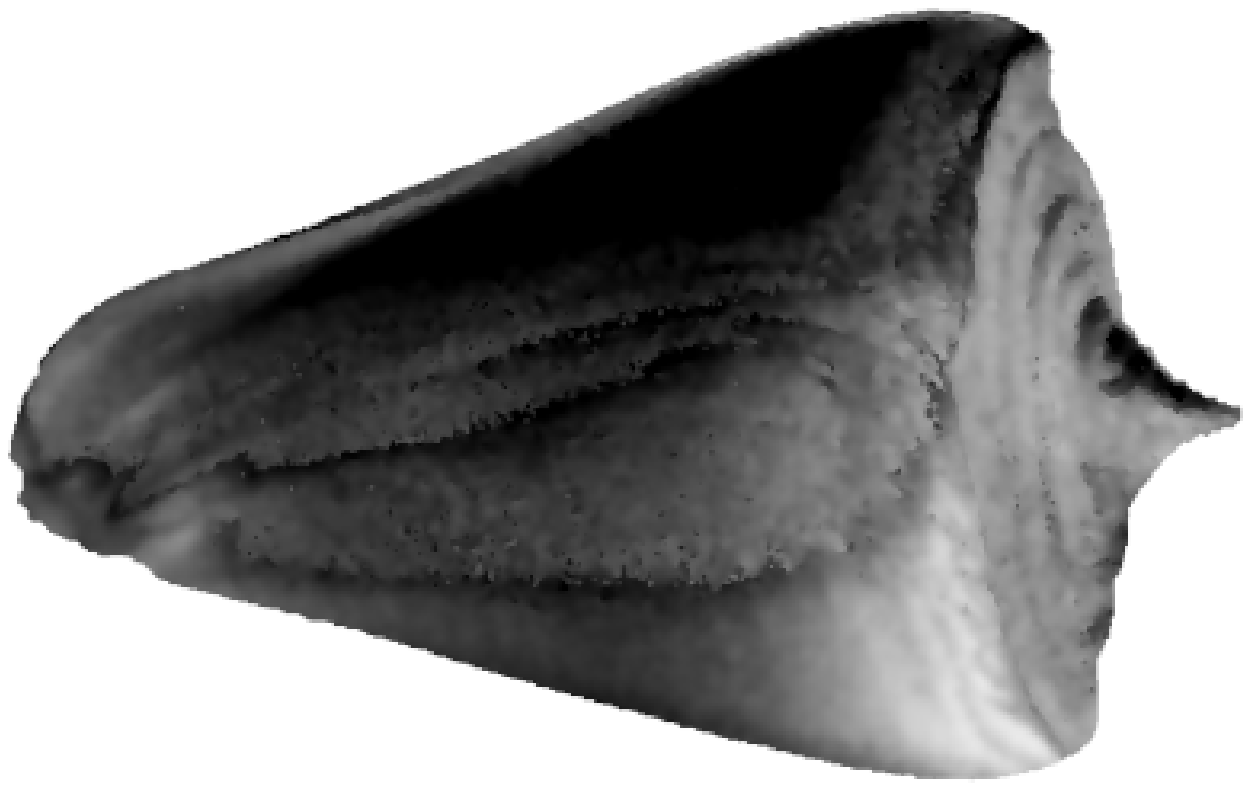}
& \includegraphics[height=\imh] {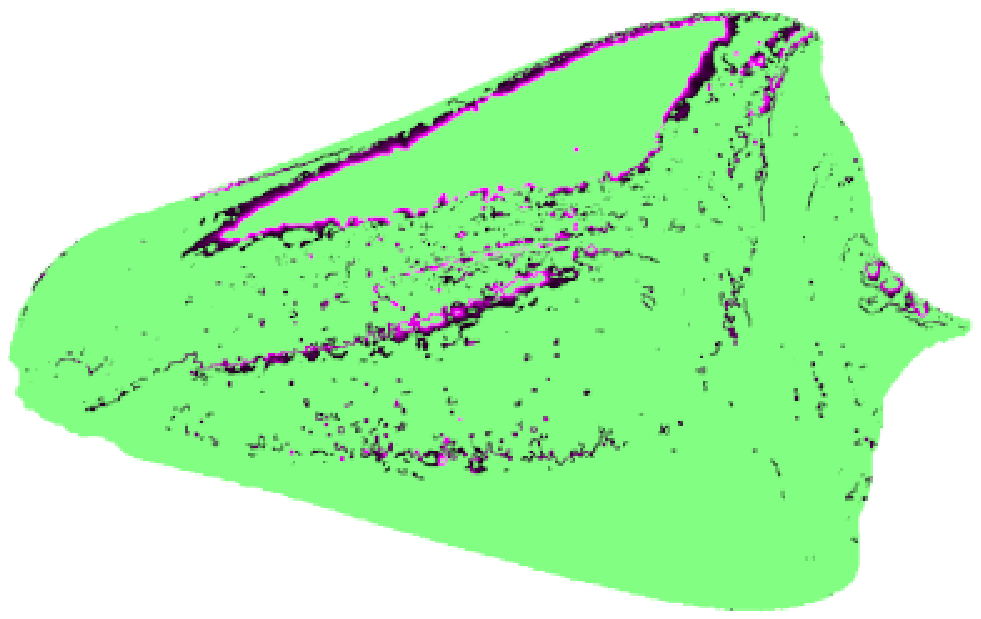}\\

$\bm{\chi}_{vis}$ & $\bm{\chi}_{nir}$ &  $\mathbf{C}$\\

\end{tabular}
\setlength{\belowcaptionskip}{-10pt} 
\caption{$\bm{l}$ is positioned to increase contrast in near-infrared (bottom) rather than visible shape (top) In both examples, we examine the top left features.}
\label{fig:testsnirfirst}
\end{figure}

We chose \emph{Michelson} contrast because it detects multiscale features in both bright and dark regions by examining differences in maximum and minimum intensity within a region rather than differences in foreground and background intensity (\emph{Weber contrast}) or standard deviation from mean intensity (\emph{RMS}). Lambertian shape in $\bm{n}_{nir}$ (Figure~\ref{fig:contrastcloseup} b) reveals structures not apparent in $\bm{n}_{vis}$ (Figure~\ref{fig:contrastcloseup}a). Close-ups 
(Figure~\ref{fig:contrastcloseup} d-f and g-i) compare details at each spectra with corresponding enhancement maps  (shown left to right). Regions for enhancement (magenta) are brighter for higher contribution weights. Only regions where shape in $\bm{n}_{nir}$  does not appear in $\bm{n}_{vis}$ are captured because we identify enhancement where $\mathnormal{m}_{nir}$ is greater than $\mathnormal{m}_{vis}$. Other formulations transfer unwanted visible shape to $\mathbf{C}$.  Using $\bm{\chi}_{nir}$  to position $\mathbf{l}$ is also key (Figure~\ref{fig:testsnirfirst}).

\textbf{Static Near-Infrared Enhancement} leverages relationships from differential geometry to compute the full near-infrared feature set. A linear shape operator $S$ is applied at each wavelength to compare local curvature using derivatives of surface normals in $\bm{n}_{vis}$ and $\bm{n}_{nir}$ with respect to direction $v$. Operator $S$ is a Weingarten Map:

\begin{equation}
S = \left( EG - F^2\right)^{-1} \mat{eG - fF}{fG - gF}{fE - eF}{gE -  fF}
\label{eqn:weingarten}
\end{equation}

\noindent where $E$, $F$, and $G$ are coefficients of the first fundamental form $\I$ and $e$, $f$ and $g$ are coefficients of the second fundamental form $\II$

\noindent and  $S$ is defined at each tangent plane $\left(u, v\right)$. We extract algebraic invariants of $\I$ and $\II$~\cite{Ohtake04}  including normal curvature, principle curvatures and mean curvature ($k_n$, $k_1$ and $k_2$, and $H$ respectively). After smoothing with foreshortening correction (scales normals by $\frac{1}{z}$), a Sobel operator is applied to $\bm{n}_{vis}$ and $\bm{n}_{nir}$ to approximate $\I$ and $\II$ and compute curvature maps $\bm{\mathcal{K}}_{vis}$ and $\bm{\mathcal{K}}_{nir}$. Per pixel curvature differences $\bm{\varrho}$ determine regions for near-infrared enhancement:

\begin{equation}
\bm{\varrho}^i = \begin{cases}
		    \Bigl| |\bm{\mathcal{K}}_{vis} | - |\bm{\mathcal{K}}_{nir} | \Bigr|, &if \hspace{4pt} \bm{\mathcal{K}}_{nir} > \bm{\mathcal{K}}_{vis}.\\
                0,              &otherwise.                            \end{cases}
\label{equ:Kdiff}
\end{equation}

\begin{equation}
\mathbf{C}^i_{\mathcal{K}} = \frac{\bm{\varrho}^i_{th}}{\bm{\eta}}
\label{equ:kenhance}
\end{equation}

\begin{equation}
\bm{\eta} = \abs{\bm{\mathcal{K}}^i_{vis}} + \abs{\bm{\mathcal{K}}^i_{nir}}
\label{equ:knorm}
\end{equation}

\noindent The thresholded difference at pixel $i$, $\bm{\varrho}^i_{th}$ ($th = 0.02$), is used to compute the enhancement value $\mathbf{C}^i_{\mathcal{K}}$ (Equations~\ref{equ:kenhance} and~\ref{equ:knorm}). Normalization $\bm{\eta}$ is applied to generate the final contribution weights. Weighting different shape types has negligible effect. We only emphasize regions where near-infrared curvature is greater (beyond a threshold) regardless of orientation as regions of similar curvature already appear in visible renderings. Thus, when comparing concave and convex shapes, near-infrared shape is accentuated if one curves more than the other, and any intrinsic function of $k_1$ and $k_2$ may be used despite differing signs (directions). Normal curvature produces more consistent results than methods like mean curvature, which loses detail after normalization.

\begin{figure}[t]
\centering
\def\imh{0.7in}
\setlength{\tabcolsep}{0.1pt}
\begin{tabular}{ccc}

\includegraphics[height=\imh]{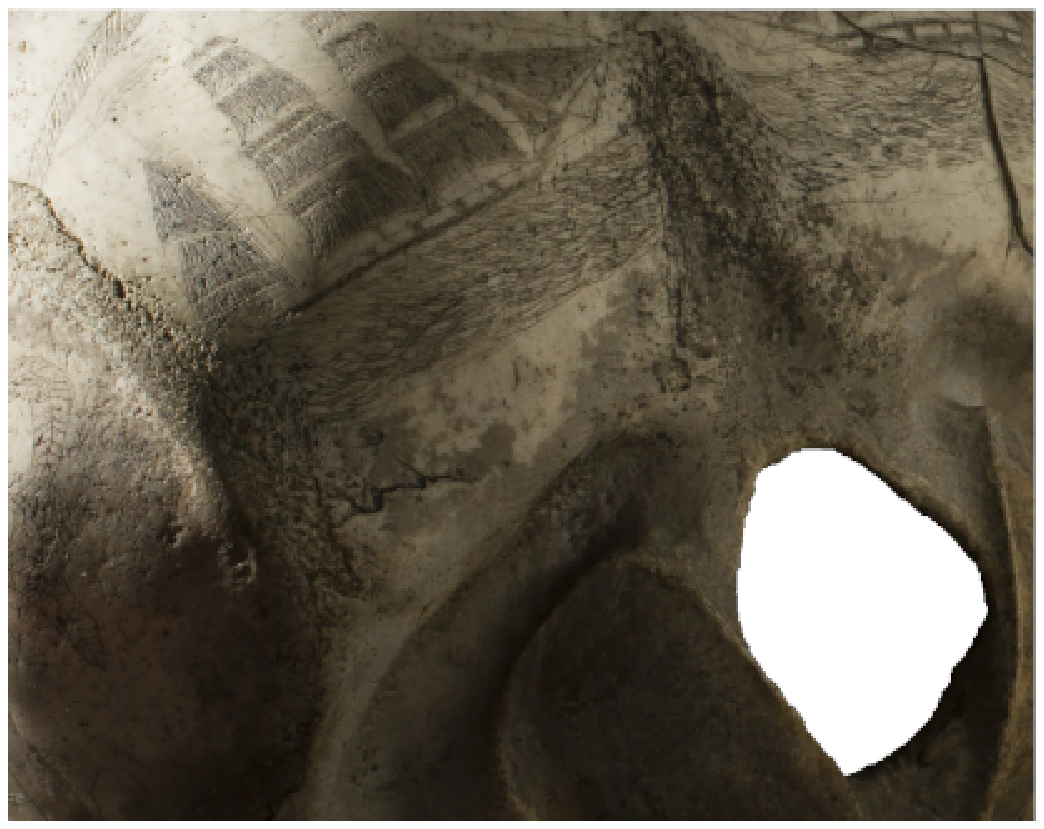}
& \includegraphics[height=\imh]{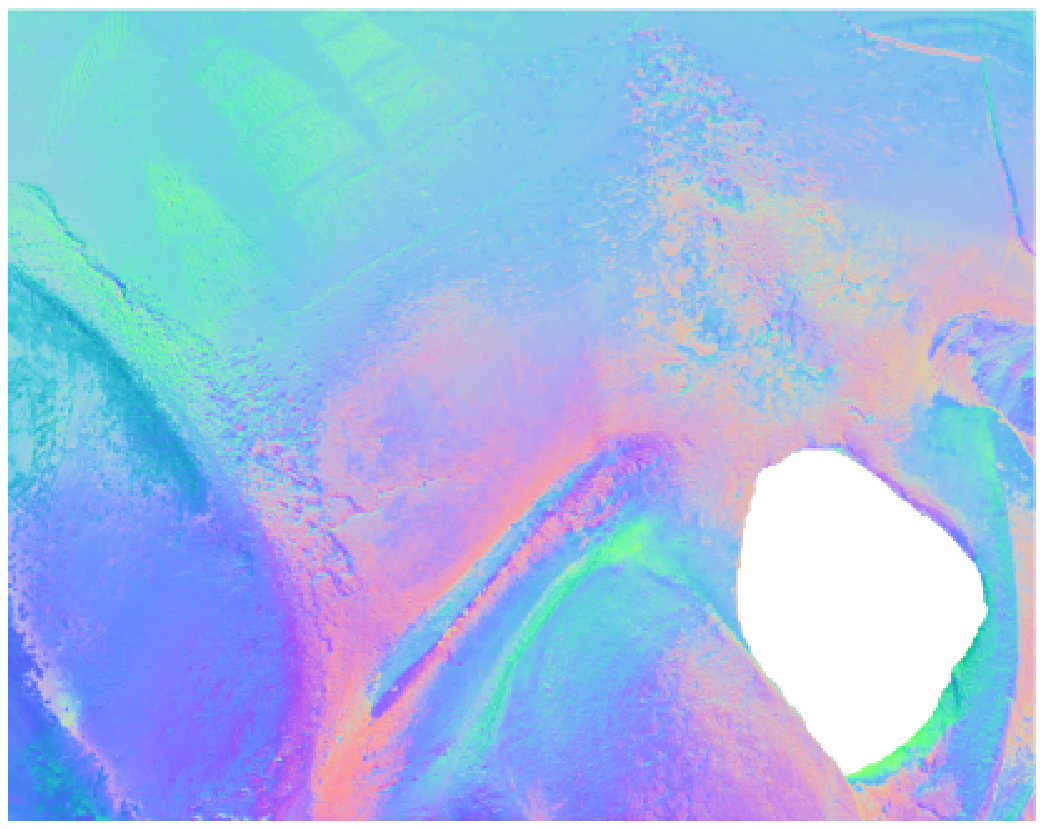}
&\includegraphics[height=\imh]{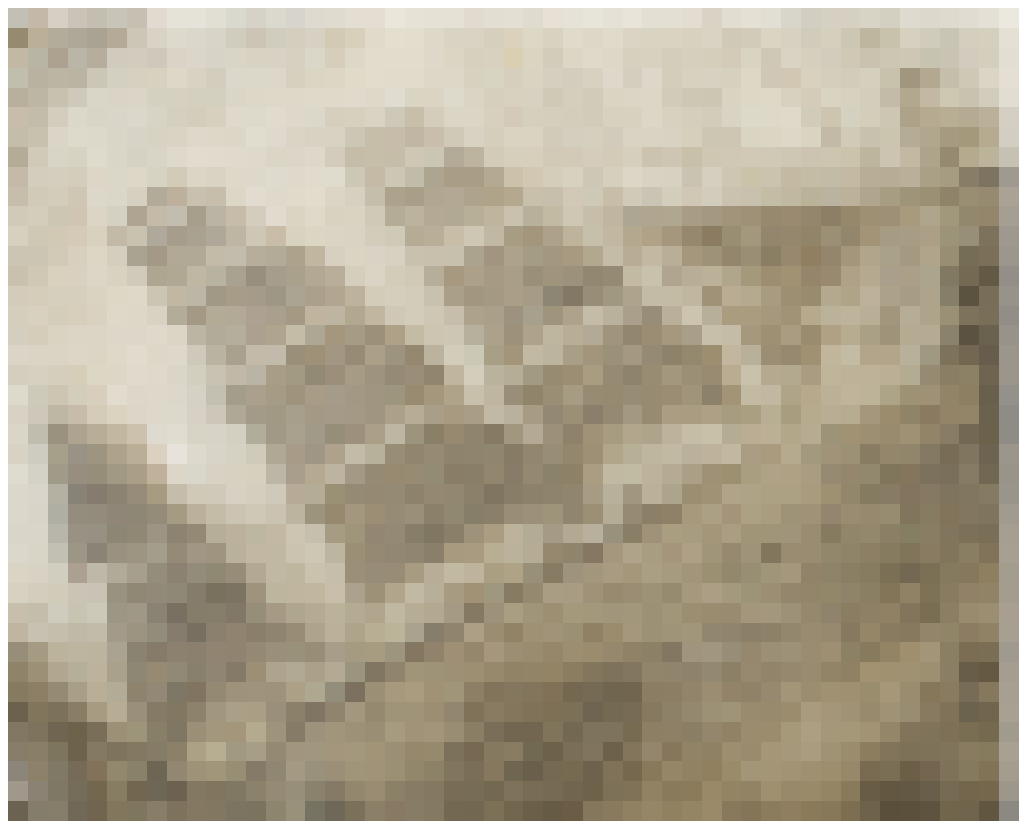}
\\

$vis$ & $\bm{n}_{vis}$ & scrimshaw ship\\

\includegraphics[height=\imh]{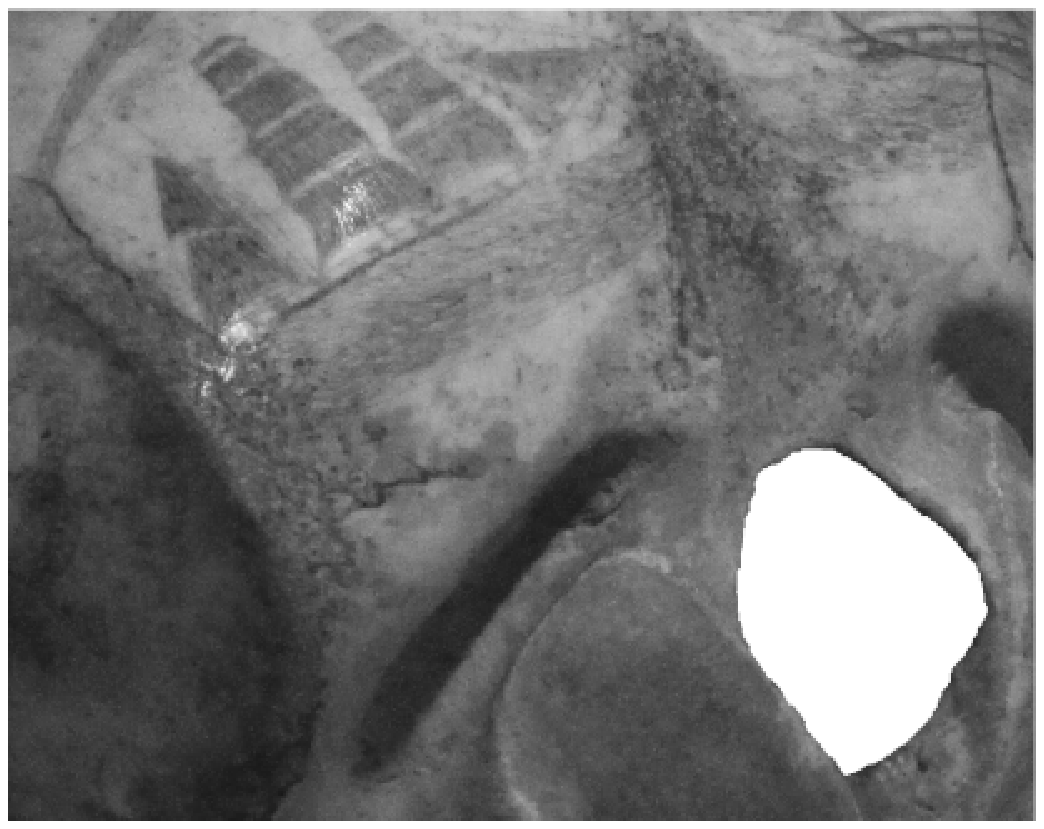}
& \includegraphics[height=\imh]{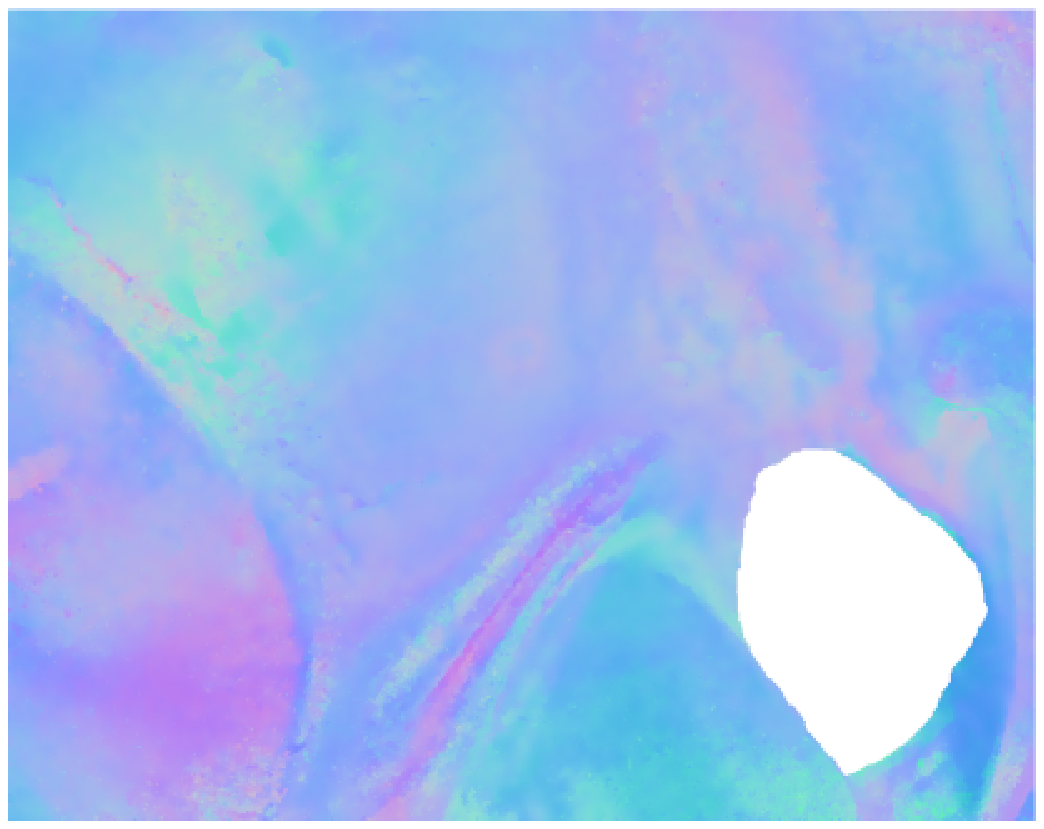}
&\includegraphics[height=\imh]{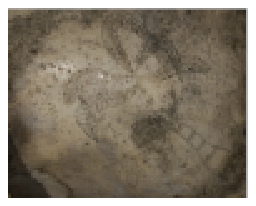}
\\

$nir$ & $\bm{n}_{nir}$ & scrimshaw tree\\

\end{tabular}
\setlength{\belowcaptionskip}{-10pt} 
\caption{Multispectral normal maps capture biological and artistic features. Whale cranium with scrimshaw, Department of Mammalogy, AMNH.}
\label{fig:contrastlightdata}
\end{figure}

\begin{figure}[h]
\centering
\def\imh{0.6in}
\setlength{\tabcolsep}{0.3pt}
\begin{tabular}{cccccc}

&\small{$\bm{l}\left(-0.92,0.35,0.19\right)$} 
&\small{$\bm{l}\left(0.96,0,0.28\right)$}
&        
& 
& 
\\

\rotatebox{90}{$\bm{\chi}_{vis}$}
& \includegraphics[height=\imh]{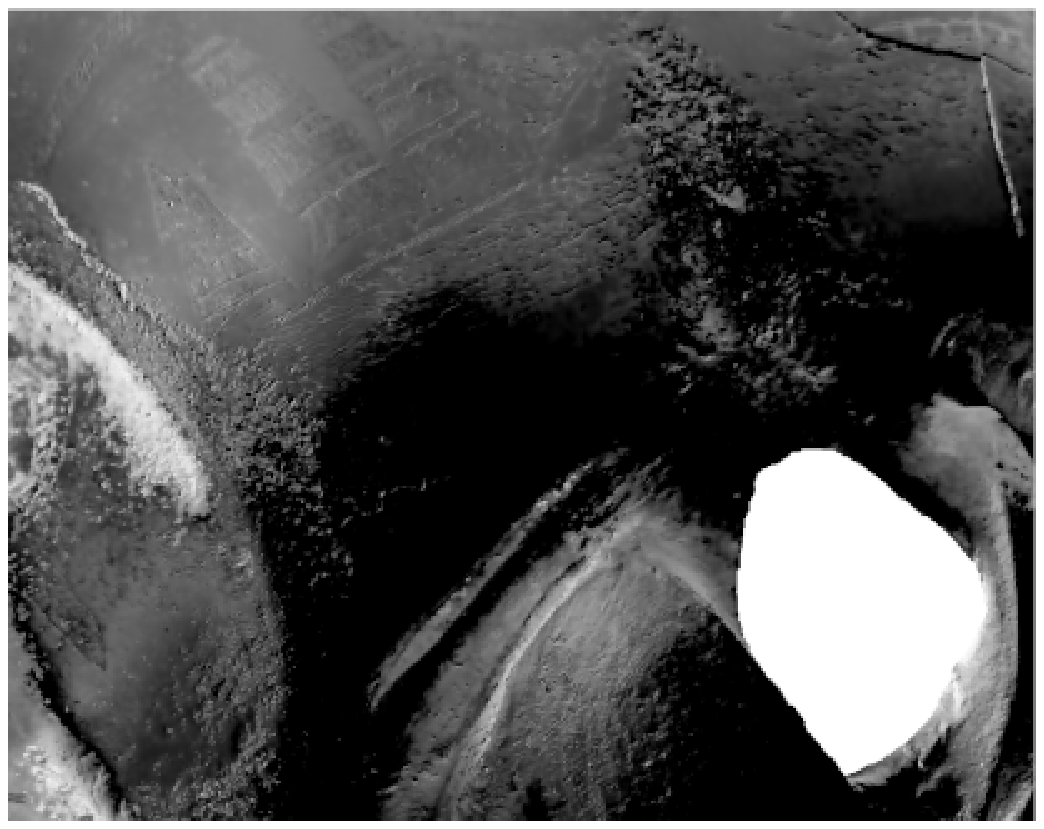}
& \includegraphics[height=\imh]{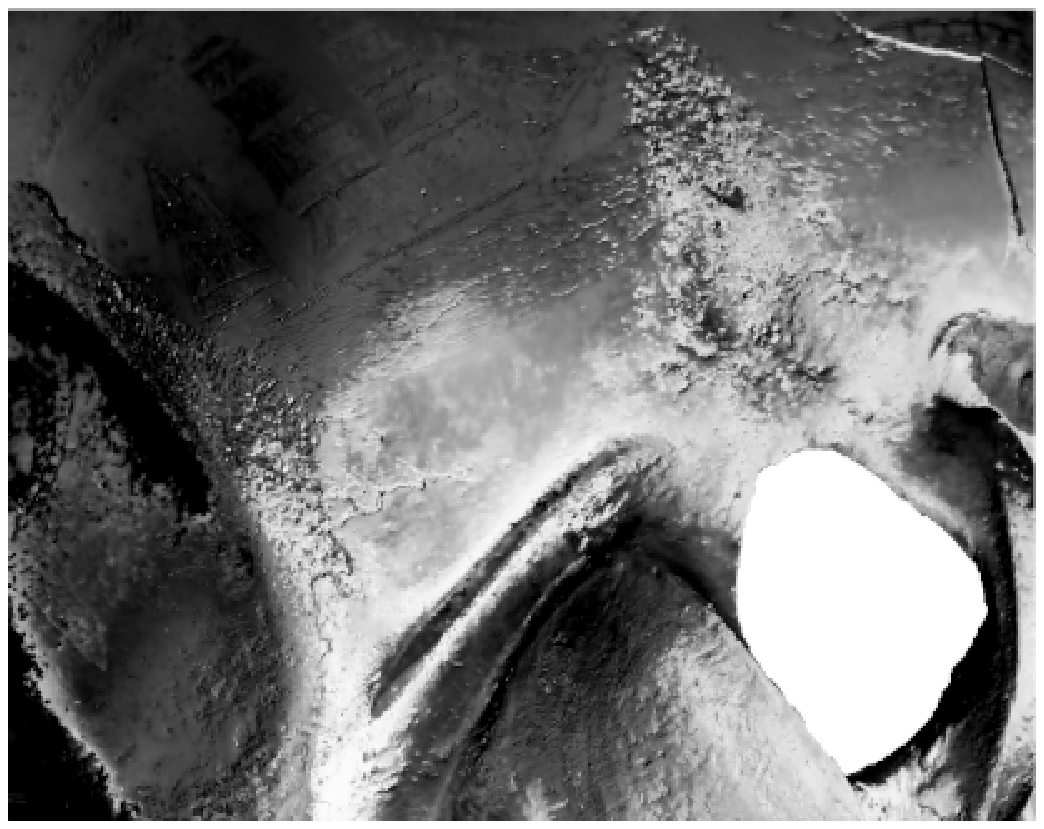}
& \phantom{\rotatebox{90}{$\bm{\mathcal{K}}_{vis}$}}       
&\rotatebox{90}{$\bm{\mathcal{K}}_{vis}$}
&\includegraphics[height=\imh] {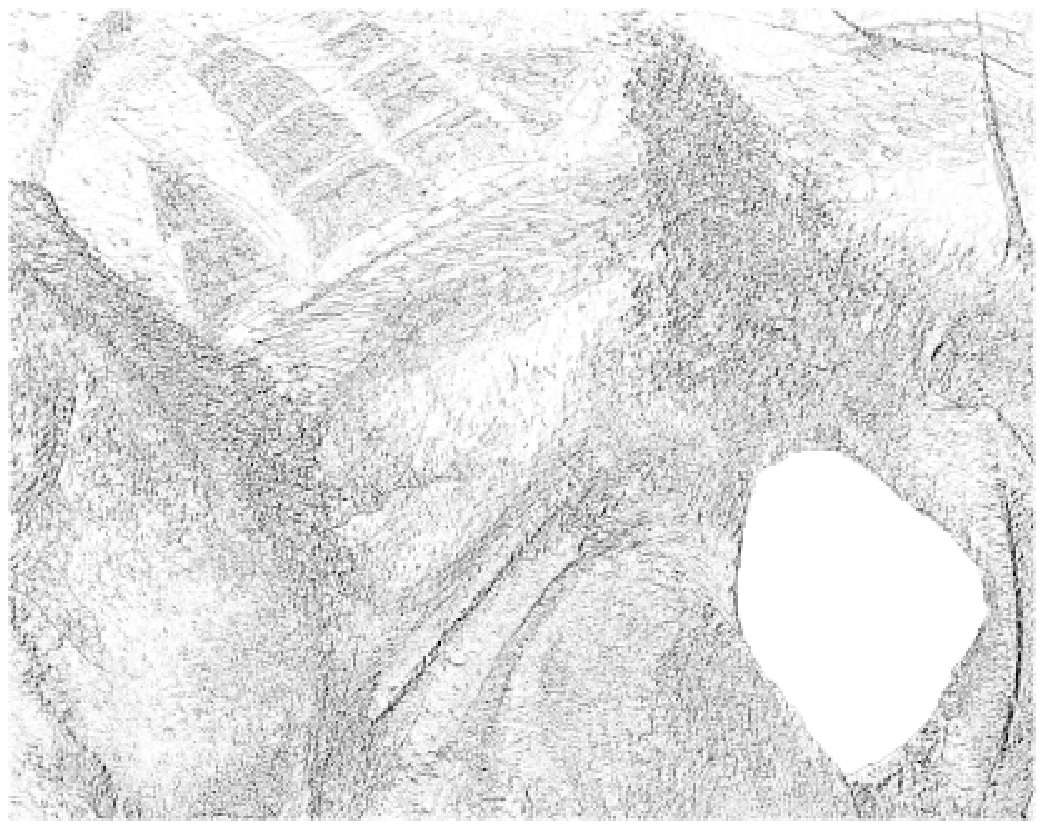}
\\

\rotatebox{90}{$\bm{\chi}_{nir}$}
& \includegraphics[height=\imh]{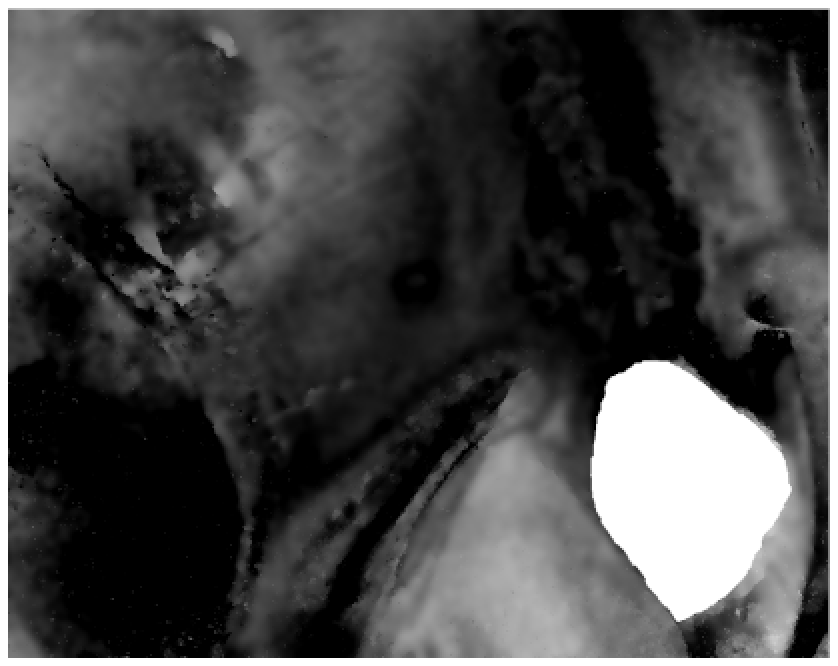}
& \includegraphics[height=\imh]{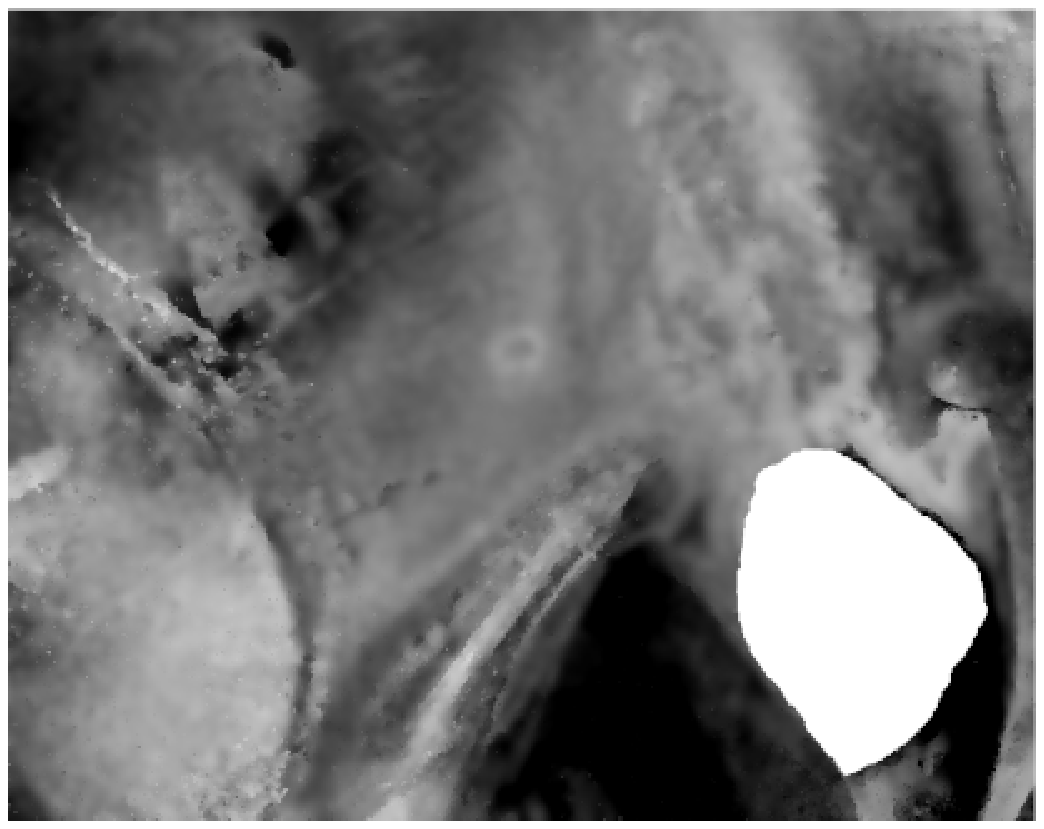}
& \phantom{\rotatebox{90}{$\bm{\mathcal{K}}_{nir}$}}       
&\rotatebox{90}{$\bm{\mathcal{K}}_{nir}$}
&\includegraphics[height=\imh] {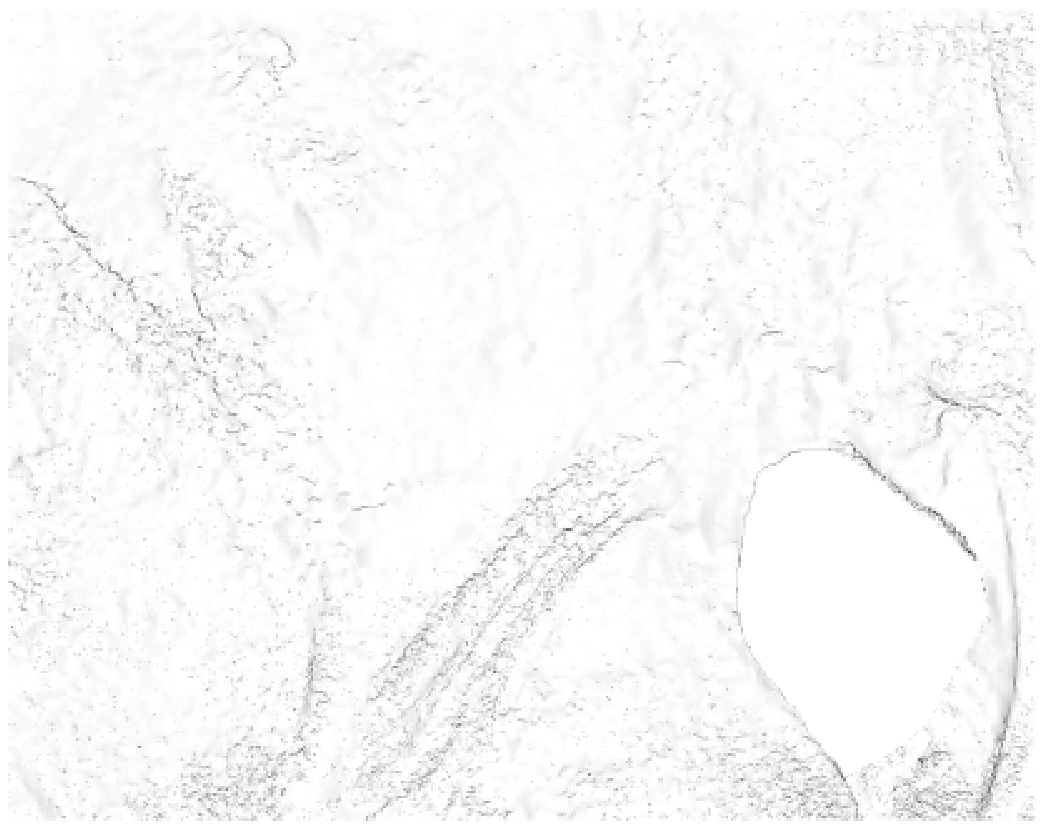}
\\

\rotatebox{90}{$\mathbf{C}$}
& \includegraphics[height=\imh]{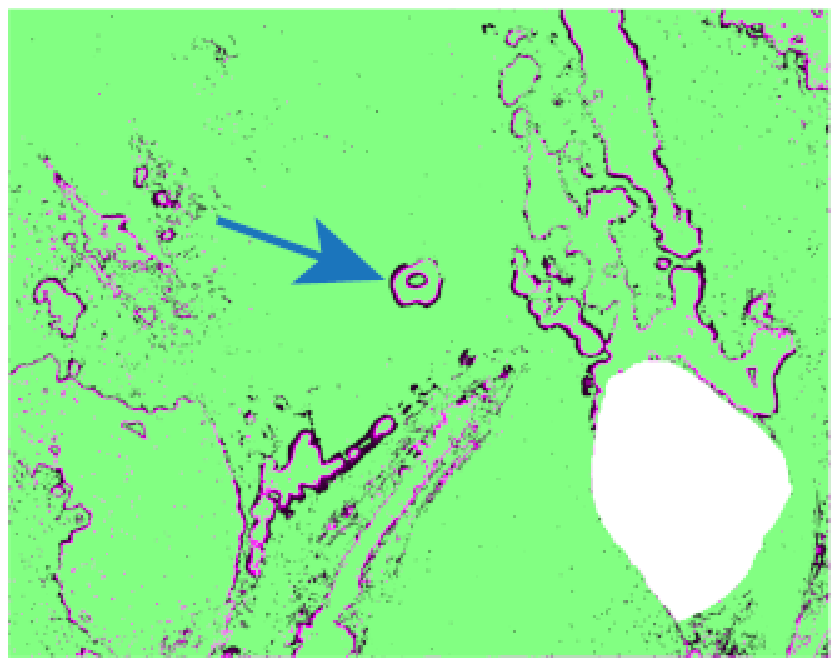}
& \includegraphics[height=\imh]{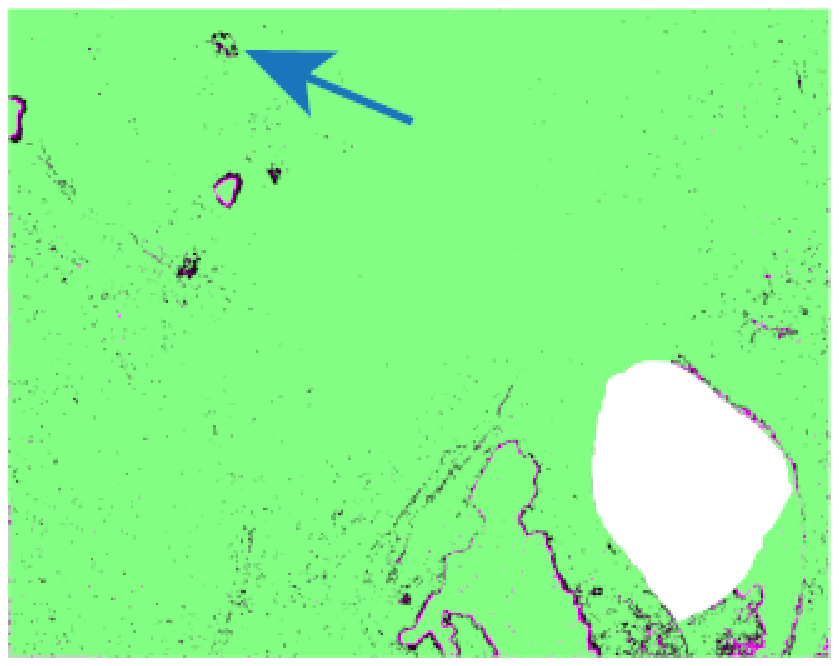}
& \phantom{\rotatebox{90}{$\mathbf{C}_{\mathcal{K}}$}}      
&\rotatebox{90}{$\mathbf{C}_{\mathcal{K}}$}
&\includegraphics[height=\imh] {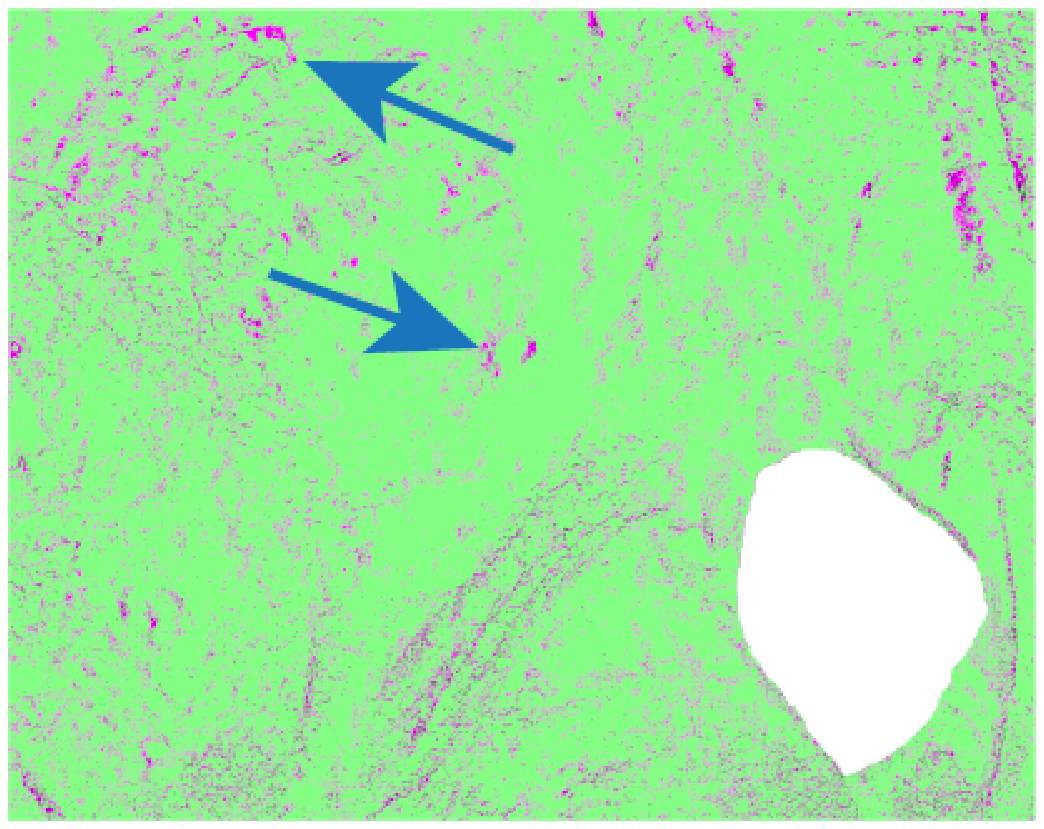}
\\

& Dynamic 
& Dynamic 
&      
& 
& Static
\\

\end{tabular}
\setlength{\belowcaptionskip}{-10pt} 
\caption{Near-infrared enhancement maps: (bottom left and bottom center) Dynamic feature selection. (bottom right) Static curvatures encode the complete feature set. Both circular features (blue arrows) are detected. Inputs $\bm{n}_{vis}$ and $\bm{n}_{nir}$ shown in Figure~\ref{fig:contrastlightdata}}
\label{fig:contrastlight}
\end{figure}

Static enhancement eliminates false enhancement readings from dynamic lights that illuminate regions in either $\bm{\chi}_{vis}$ or $\bm{\chi}_{nir}$. Static enhancement reveals all features (both circular features appear in Figure~\ref{fig:contrastlight} bottom right) while dynamic enhancement reveals different features at different light orientations (good for feature selection, Figure~\ref{fig:contrastlight} bottom left and middle). Static enhancement produces continuous, smoothly varying differences suitable for shading while dynamic enhancement emphasizes discrete shapes.

\section{Multispectral Stylization}
\label{sec:shading}

Building upon principles and tools introduced in Sections~\ref{sec:principles} and~\ref{sec:msprocessing}, we present multispectral NPR stylization techniques that express the shape, composition and structure of biological materials.

\subsection {Spectral Band Shading}
\label{sec:spectralband}

Spectral band shading emphasizes multiscale near-infrared shape across multiple wavelengths for each surface orientation. The \emph{spectral band control} enabled by the integration of near-infrared enhancement maps (Section ~\ref{sec:msprocessing}) distinguishes Spectral Band Shading from other multiscale NPR enhancement algorithms~\cite{Fattal2007,rgbn07,Rusinkiewicz06}. We develop strategies for simulating narrow and broad band transmission filters that suppress visible content while accentuating near-infrared wavelengths. Our methods combine spectral wavelengths (horizontally) at each scale before combining (vertically) across frequencies.

We start with two stacks of smoothed normals in the visible and near-infrared ($\bm{n}_{vis}$ and  $\bm{n}_{nir}$) and near-infrared enhancement maps $\mathbf{C}$ computed at each smoothing level $\bm{\mathcal{L}}_x$. The shading contribution $\bm{e}$ at pixel $i$ on  $\bm{\mathcal{L}}_x$ from each spectra is:

\begin{equation}
\bm{e}_{vis}^i = \left(1 - \mathbf{C}^i\right)\left(\bm{n}_{vis} \cdot \bm{l}_{vis}^{contrast}\right)
\label{equ:evis}
\end{equation}

\begin{equation}
\bm{e}_{nir}^i = \mathbf{C}^i \left(\bm{n}_{nir} \cdot \bm{l}_{nir}^{contrast}\right)
\label{equ:enir}
\end{equation}

\begin{equation}
\bm{e}^i = a\left(\bm{e}_{vis}^i + \bm{e}_{nir}^i\right)  
\label{equ:sbsblend}
\end{equation}

We use weight $\mathbf{C}^i$ to combine spectral contributions before applying a shape enhancement term $a$ (typically 35). For the remainder of the paper, we refer to near-infrared enhancement maps in equations as $\mathbf{C}$ and distinguish dynamic from static in the main body of the text. In spectral band shading, $\mathbf{C}^i$  can be dynamic or static depending  on the light direction choice $\bm{l}_{\lambda}^{contrast}$. We discuss application specific options for $\bm{l}_{\lambda}^{contrast}$ later in
this section. For now, assume $\bm{l}_{\lambda}^{contrast}$ is  $\bm{l}$ used to generate $\mathbf{C}^i$ with dynamic lighting at each level.

\begin{figure}[h]
\centering
\def\imh{0.6in}
\setlength{\tabcolsep}{0.1pt}
\begin{tabular}{ccc}

 $\bm{r}=5$ 
&$\bm{r}=13$ 
&$\bm{r}=25$ 
\\

  \includegraphics[height=\imh]{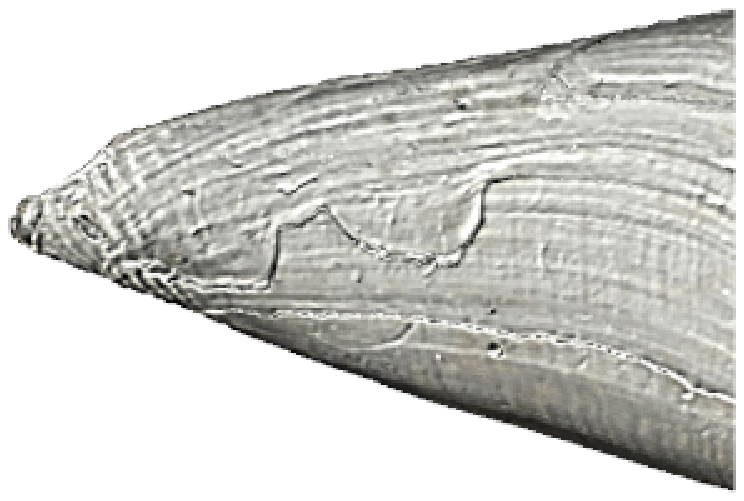}
& \includegraphics[height=\imh]{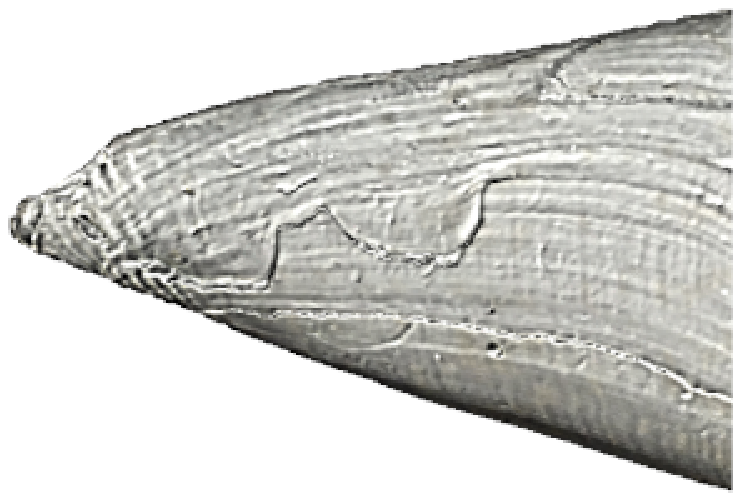}
& \includegraphics[height=\imh]{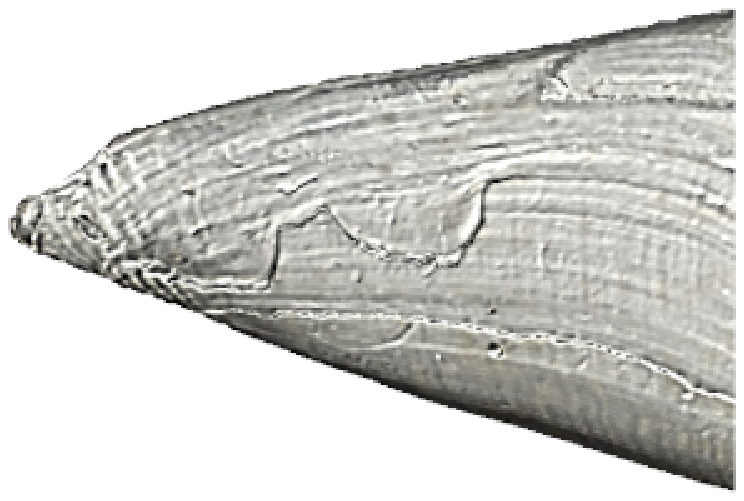}
\\


\end{tabular}
\setlength{\belowcaptionskip}{-16pt} 
\setlength\abovecaptionskip{-0.2\baselineskip}
\caption{ Narrow modulation windows reveal fine detail
while large windows produce smoother course detail. }
\label{fig:testswindow}
\end{figure}

\begin{figure}[h]
\centering
\includegraphics[width=0.6\hsize]{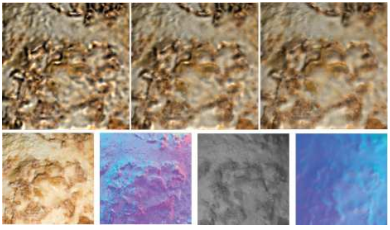}
\setlength{\belowcaptionskip}{-8pt} 
\setlength\abovecaptionskip{2pt}
\caption{\label{fig:csbsresult2}%
Dynamic Spectral band shading for a single focus light. Frequency enhancement. top: (left) $f = -1$ Weights shifted toward sharper levels. (middle) $f = 0$ Equal weights. (right) $f = 1$ Weights shifted to smoother levels. bottom: (left to right) Visible color and normals. Near-infrared spectra and normals. Primate cranium, DPC No. 11835, Archaeolemur, Field No. 92-M-257, Fossil Primates Division, Duke University Lemur Center.}
\end{figure}

\begin{figure*}[h]
\centering
\def\imh{0.9in}
\setlength{\tabcolsep}{0.4pt}

\begin{tabular}{cccccccc}

  $vis$  
& $nir$ 
& Combined 
& Dynamic Focus 
& Combined 
& Dynamic Multilight 
& Combined 
& Static Principle\\

\includegraphics[height=\imh]{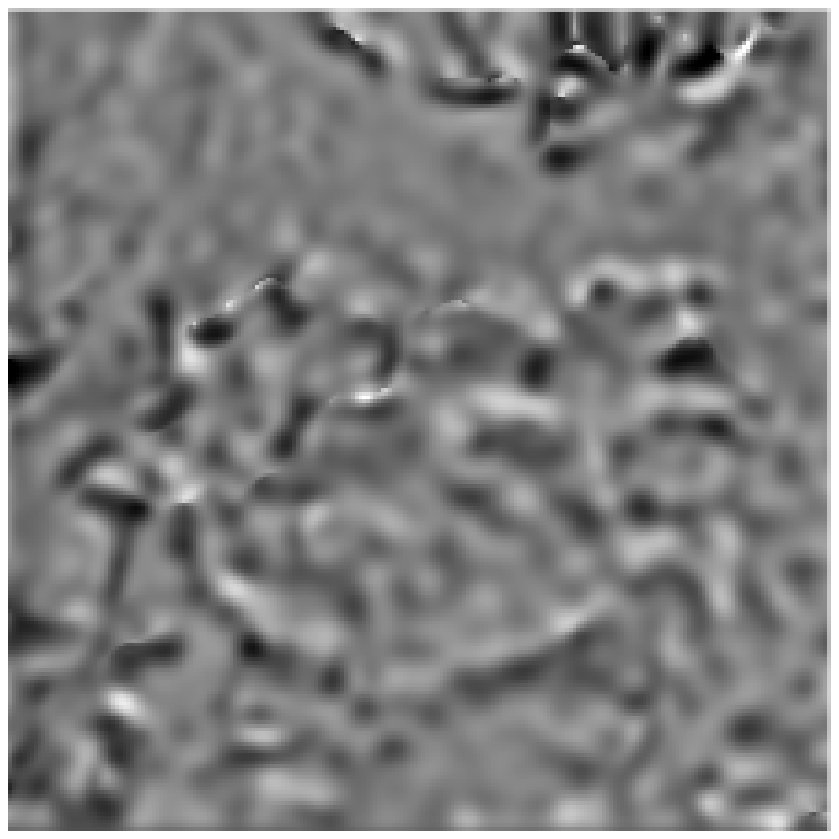}

&\includegraphics[height=\imh]{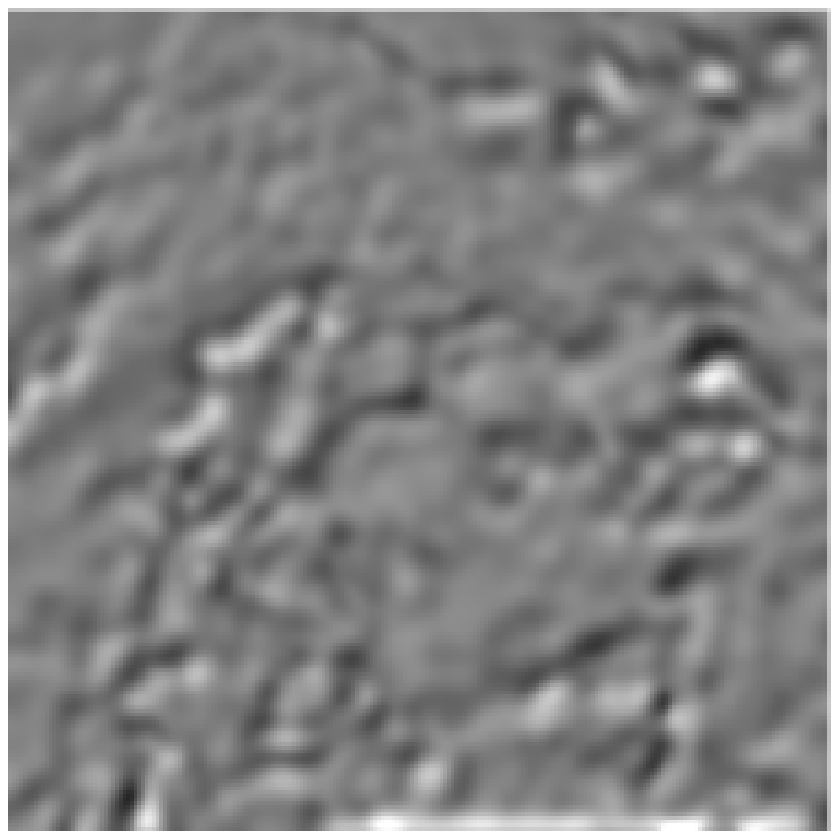}

&\includegraphics[height=\imh]{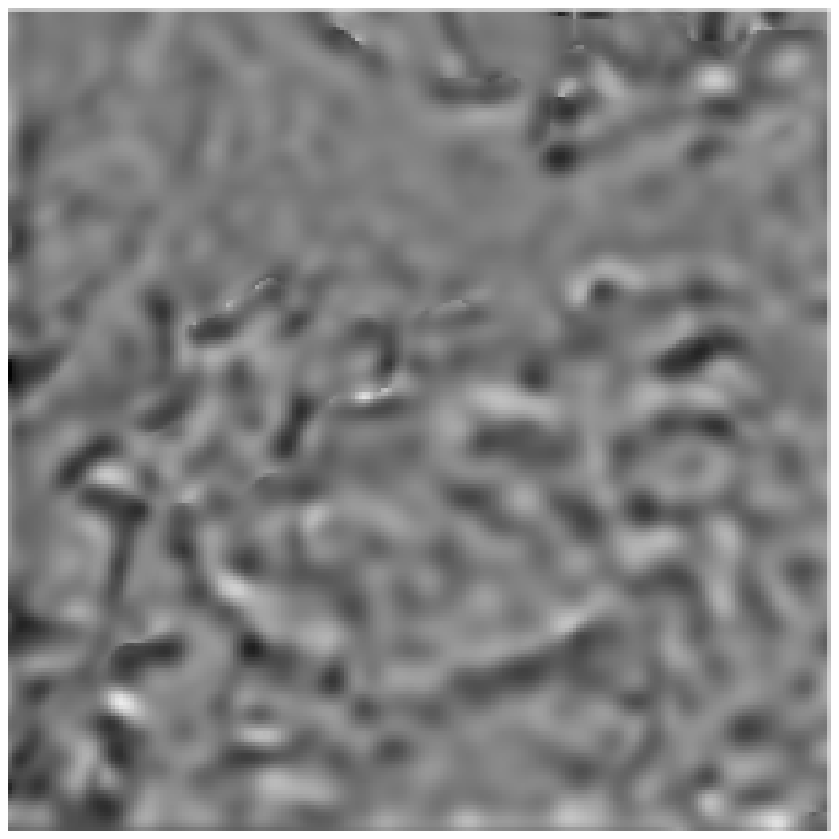}

&\includegraphics[height=\imh]{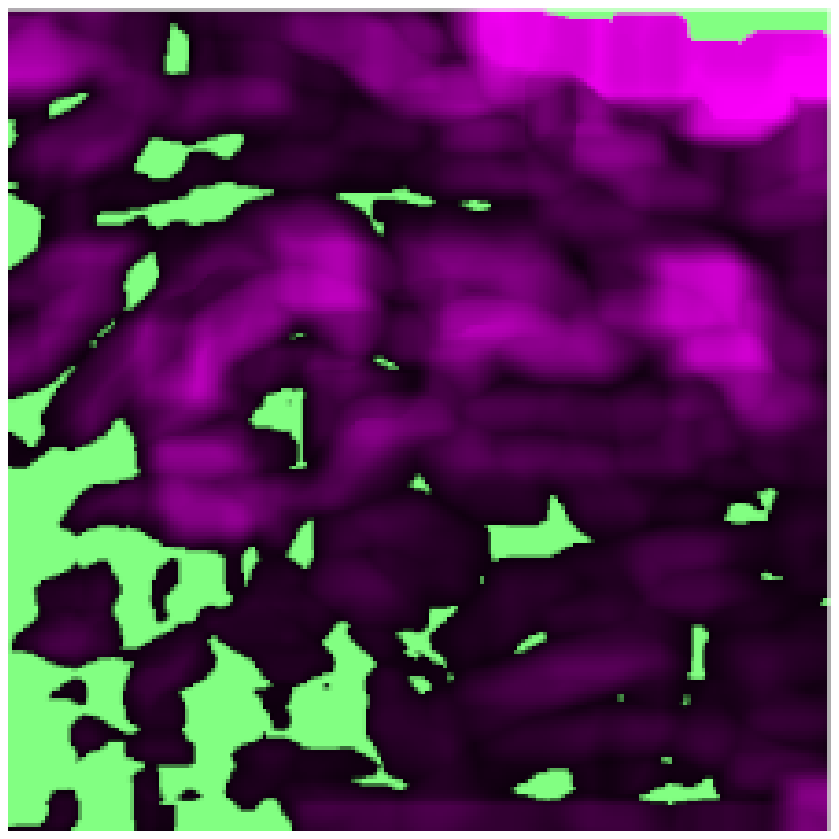}

&\includegraphics[height=\imh]{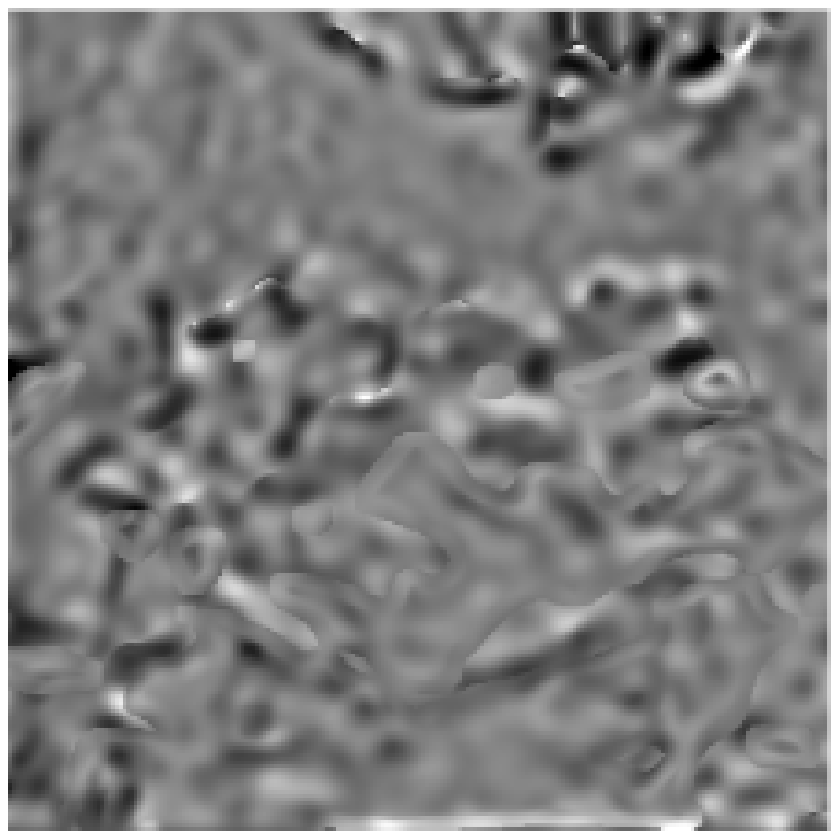}

&\includegraphics[height=\imh]{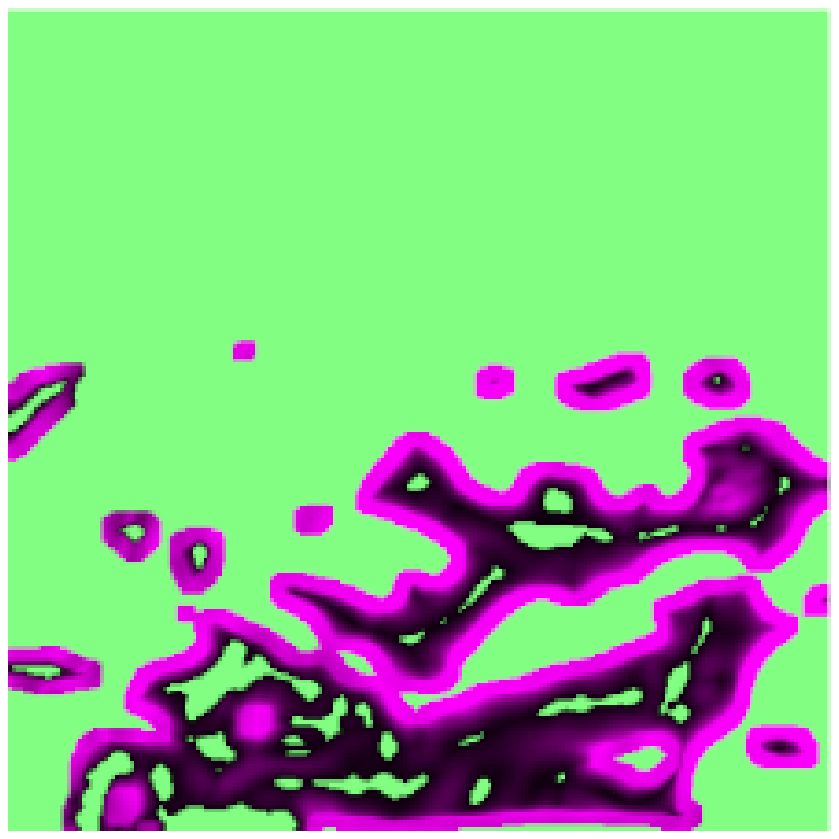}

&\includegraphics[height=\imh]{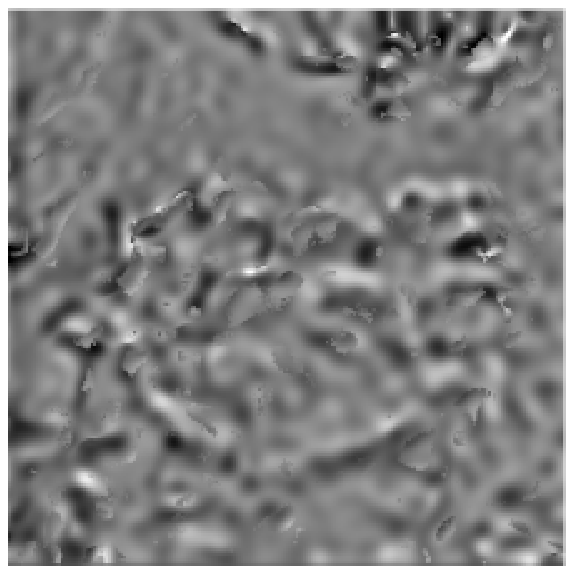}

&\includegraphics[height=\imh]{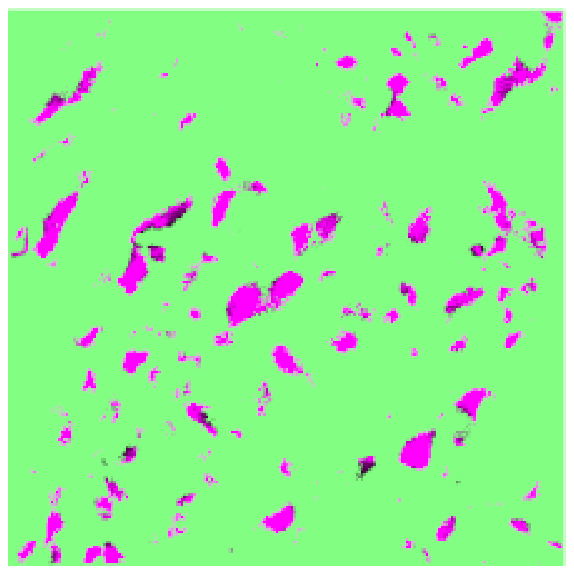}

\end{tabular}
\setlength\abovecaptionskip{-0.3\baselineskip}
\setlength{\belowcaptionskip}{-6pt} 
\caption{Spectral band shading at level $2$ of for visible, near-infrared, and combined near-infrared enhancement. In each example $a = 25$, $f = -1$, $\bm{l}^{global}\left(0.37,0.47,0.801374\right). $Dynamic Focus: $\bm{l}^{contrast}\left(0.73, 0.05, 0.68\right)$, $\bm{r} = 15$, $th = 0.112$. Dynamic Multilight: $\bm{r} = 3$, $th = 0.112$. Static: $\bm{r} = 3$, $th_{\mathcal{K}} = 0.2$. DPC No. 11835, Archaeolemur, Field No. 92-M-257, Fossil Primates Division, Duke University Lemur Center.}
\label{fig:csbsirlevelsOne}
\end{figure*}

\begin{figure}[h]
\centering
\def\imh{0.725in}
\setlength{\tabcolsep}{0.35pt}
\includegraphics[height=\imh]{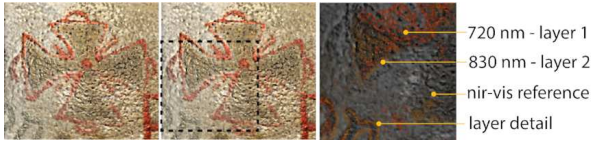}
\setlength{\belowcaptionskip}{-12pt} 
\setlength\abovecaptionskip{2pt}
\caption{Spectal Band Shading: Rendering layered effects. Wavelengths: (a) Competing. (b, c) Dominating. (right) Wavelengths with layer colormap. Painted cranium, AMNH Anthropology Catalog Number VL/122, Courtesy of Division of Anthropology, American Museum of Natural History.}
\label{fig:sbsspectralcontrol}
\end{figure}

Next, we incorporate multiple wavelengths using two methods. The first is analogous to a narrow bandpass filter where a specific wavelength dominates (Figure~\ref{fig:sbsspectralcontrol} b). For each pixel, we compute a per-wavelength contrast modulation $m_{\lambda}  \in   {m_{\lambda_1}, m_{\lambda_2} \cdot \cdot \cdot m_{\lambda_q}}$ and choose the near-infrared wavelength  $p_{\lambda}  \in   {p_{\lambda_1}, p_{\lambda_2} \cdot \cdot \cdot p_{\lambda_q}}$ at the current pixel, to be the spectra with the highest modulation value. We use this value to compute the contrast $\varphi$ at this pixel. Thus, contrast differences shift between wavelengths from pixel to pixel as each wavelength competes for its  contribution. When $\mathbf{C}$ is dynamic,  the weight $\mathbf{C}^i$ is:

\begin{equation}
\mathbf{C}^i = \begin{cases}
		    \bm{\varphi}_{th}^i,     &if \hspace{4pt} \mathnormal{m}_{\lambda}^i > \mathnormal{m}_{vis}^i.\\
                0,              &otherwise.                            \end{cases}
\label{equ:contrastmapmulti}
\end{equation}

\noindent where:

\begin{equation}
\mathnormal{m}_{\lambda}^i  = max(m_{\lambda_1}, m_{\lambda_1} \cdot \cdot \cdot m_{\lambda_x})
\label{equ:maxm}
\end{equation}

\noindent When $\mathbf{C}$ is static, weights are computed using the wavelength with the maximum curvature value in $\bm{\mathcal{K}}_{\lambda}$:

\begin{equation}
\bm{\varrho}^i = \begin{cases}
		    \Bigl| |\bm{\mathcal{K}}_{vis} | - |\bm{\mathcal{K}}_{\lambda} | \Bigr|, &if \hspace{4pt} \bm{\mathcal{K}}_{nir} > \bm{\mathcal{K}}_{vis}.\\
                0,              &otherwise.                            \end{cases}
\label{equ:Kdiffmulti}
\end{equation}

\noindent where:

\begin{equation}
\bm{\mathcal{K}}_{\lambda}^i  = max(\bm{\mathcal{K}}_{\lambda_1}, \bm{\mathcal{K}}_{\lambda_1} \cdot \cdot \cdot \bm{\mathcal{K}}_{\lambda_x})
\label{equ:maxk}
\end{equation}

The second method combines shape contributions from different near-infrared wavelengths at each level linearly so that they perform like a broad band filter (Figure~\ref{fig:sbsspectralcontrol} a). We compute $m$ and $\mathbf{C}$ for each wavelength. We then normalize the per-pixel contributions in each $\mathbf{C}^i$ by the number of wavelengths. We compute Equation~\ref{equ:sbsblend} for each wavelength. For both approaches, the final pixel shading is combined vertically across the smoothed stack weighted by $\omega_{x}$, the normalized kernel widths:

\begin{equation}
\bm{e}_{final}^i = \frac{1}{2} + \frac{1}{2} \sum\limits_{x=0}^{\mathcal{L}}
\omega_{x}\bm{e}_{\mathcal{L}}^i \label{equ:irsum}
\end{equation}

We shade with soft toon clamped from $[0, 1]$. Smoothing consists of a base layer lit by a global light in the upper left for flat objects (lower for curved objects). The pixel width of our smoothing window increases by powers of two. A geometric series is used to increase the width of our Gaussian kernels $\bm{\sigma}^g_{vis}$  and $\bm{\sigma}^g_{nir}$ (which are consistent across spectral bands at each level). A  base layer plus five levels of smoothing are sufficient for our datasets. Results are shown in Figure~\ref{fig:csbsresult2}. We set $\sigma^c_{vis}$, $\sigma^c_{nir}$, $\sigma^n_{vis}$ and $\sigma^n_{nir}$ to $10$.

Following Principle~\ref{itm:measureddata}, we use measured parameters for stylization controls that illustrate the power of our algorithm. 

\paragraph{\textbf{Shape:}} The analysis in Section~\ref{sec:msprocessing} provides the foundation for our shape enhancement effects. Smoothing and sharpening near-infrared detail is achieved by widening and narrowing $\bm{r}$, the contrast modulation window from Equation~\ref{equ:contrastmodulation} (Figure~\ref{fig:testswindow}). Figure~\ref{fig:csbsresult2} shows how the frequency term $f$ controls sharpness by pushing near-infrared enhancement up and down the normal stack. Selective enhancement at any number of wavelengths exploits robust shape features that persist at longer wavelengths that are less prone to noise. Recall that details from $\bm{\chi}_{nir}$ (which are pre filtered for noise removal) will smooth away while significant information remains  in $\bm{\chi}_{vis}$. Using a weighting term when combining visible and near-infrared information also eliminates undesirable artifacts and blends the two spectra with more gradual transitions for a smooth shading result.

\begin{figure}[h]
\centering
\def\imh{0.625in}
\setlength{\tabcolsep}{0.1pt}
\begin{tabular}{cccc}

\includegraphics[height=\imh]{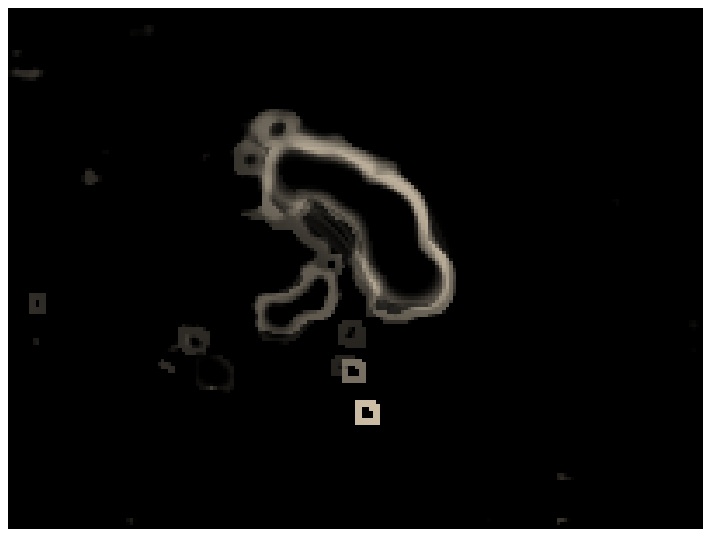}
& \includegraphics[height=\imh]{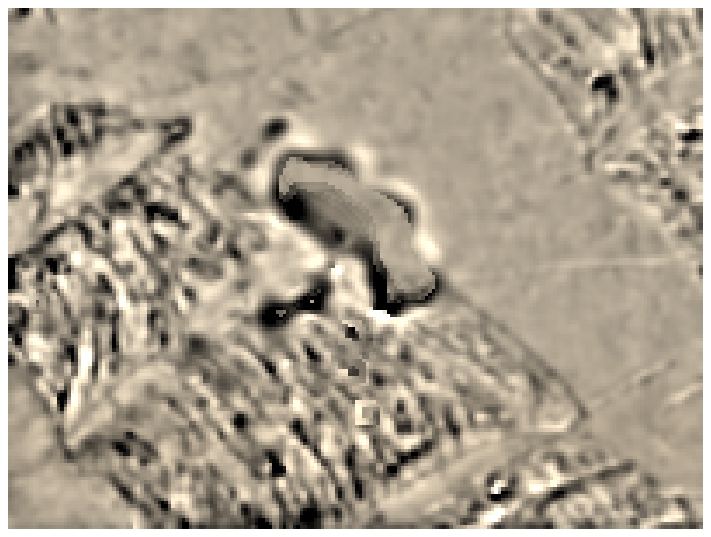}
& \includegraphics[height=\imh]{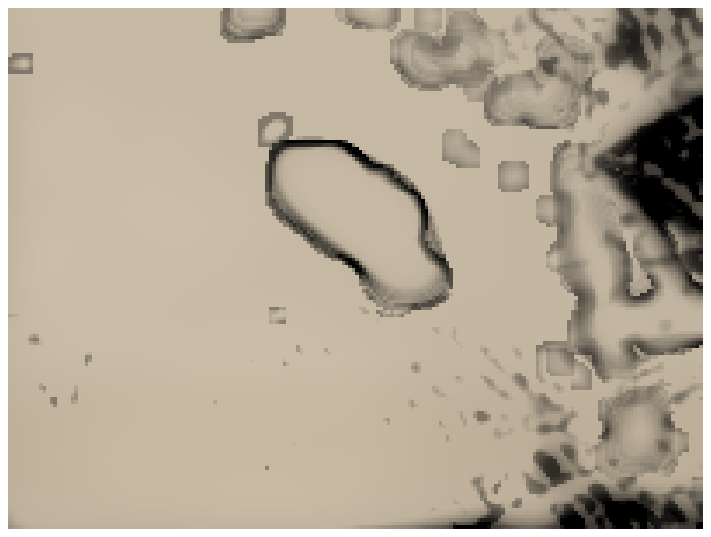}
&\includegraphics[height=\imh]{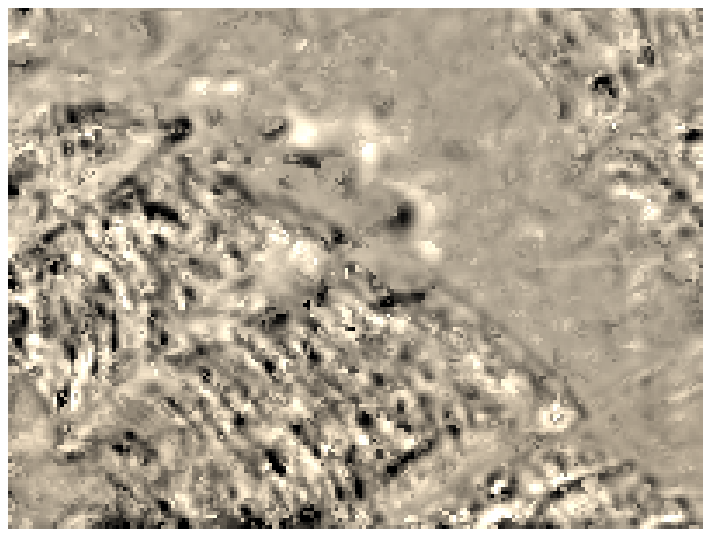}
\\

(a) & (b) & (c) & (d)\\

\end{tabular}
\setlength\abovecaptionskip{-0.1\baselineskip}
\setlength{\belowcaptionskip}{-14pt} 
\caption{(a) Dynamic Enhancement Map Light (b) Dynamic Multilight (c) Dynamic Focused Light (d) Static Principle Light.}
\label{fig:shadinglights}
\end{figure}

\paragraph{\textbf{Lights:}} The options for $\bm{l}_{\lambda}^{contrast}$  for dynamic spectral band shading are shown in Figure~\ref{fig:shadinglights} a, b and c. Enhancement map lights emphasize near-infrared detail where there are greater shape variations in $\bm{n}_{nir}$ as both $\bm{l}_{vis}^{contrast}$ and $\bm{l}_{nir}^{contrast}$ are positioned 
at $\bm{l}$, the light used to generate $\mathbf{C}$ at each level. The multilight option maximizes shape from both spectra by setting $\bm{l}_{vis}^{contrast}$ and $\bm{l}_{nir}^{contrast}$ to a light projected into the tangent plane of the next smoothed normal  in $\bm{n}_{vis}$ and $\bm{n}_{nir}$ respectively. The system generates the single focus light to maximize contrast in a user chosen focus region by positioning $\bm{l}^{contrast}_{focus}$ to maximize the near-infrared contrast in $\bm{\chi}_{nir}$ for persistent features across the normal stack. The static principal light uses pre-computed per-pixel curvature values to place  $\bm{l}_{\lambda}^{contrast}$ along per-pixel principal directions in each spectral band at each level. This accentuates near-infrared detail where the subsurface layer has sharply varying curvature. Figure~\ref{fig:csbsirlevelsOne} shows how three light options effectively combine multispectral shape information at smoothing level $2$.

\paragraph{\textbf{Color:}} Layered rendering effects are inspired by our pilot study, where we observed multispectral HDR contrast enhancement that suppressed visible content while enhancing subtle near infrared detail in crime scene photos. In Figure~\ref{fig:sbsspectralcontrol}, we visualize signatures at specific wavelengths. A multispectral blending function shifts between visible color to near-infrared intensity across wavelengths by adjusting a near-infrared blend function from $0-1$. This intensity map is used in spectral band shading to create a background reference image. Next, each wavelength is associated with a color from a user specified palette and rendered on top of the reference. Weights (Equations~\ref{equ:contrastmapmulti} and~\ref{equ:Kdiffmulti}) control the visual contribution  of each wavelength in the foreground.

\subsection {Multispectral Curvature and Line Drawing}

\begin{figure}[h]
\centering
\includegraphics[width=1.0\hsize]{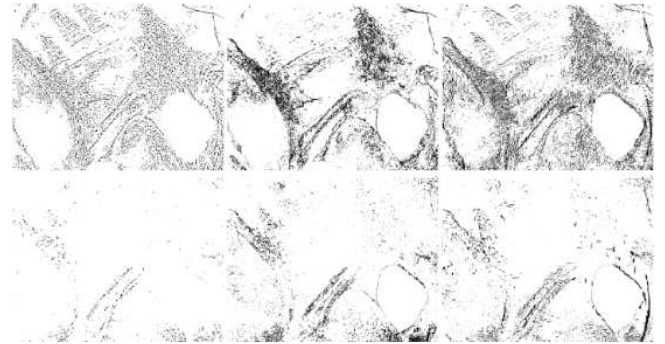}
\setlength{\belowcaptionskip}{-22pt} 
\setlength\abovecaptionskip{-0.7\baselineskip}
\caption{\label{fig:lines_whalecrania}%
Multispectral line extraction in the visible (top) and near-infrared
 (bottom) for three line types: (left) Suggestive contours (middle) Discontinuity lines (right) Normal curvature $k_1$ Whale cranium with scrimshaw, Department of Mammalogy, AMNH.}
\end{figure}

We combine spectral band shading with multiscale multispectral curvature shading~\cite{rgbn07} to simulate per pixel lighting. The algorithm substitutes the diffuse component with curvature values (Section~\ref{sec:contrast}) computed on $\bm{n}_{vis}$  and $\bm{n}_{nir}$.   The wide range of curvature values in our specimens make it difficult to see subtle details as high curvature regions are pushed to white while small curvature regions are gray. Near-infrared curvature values are more readily apparent than visible ones. To mitigate this, we scale curvature values by a user defined parameter $Q$. We double $Q$ when working with visible spectra. The scaled curvature values are then combined across spectra using enhancement maps. We clamp the shaded output between $[-1, 0]$ to darken valleys and $[0, 1]$ to brighten ridges. The contribution at each level is a function of $\sigma$, the width of the domain filter. We also extract shape conveying lines directly from multispectral normals. Suggestive contours  occur as a contour is forming on the horizon~\cite{DeCarlo03}. We create a head-lit image $\mathbf{n}_{\lambda} \cdot v$, search for minima of intensity over a local neighborhood and extract lines where a majority of pixels are darker than the current pixel. Discontinuity lines are computed by searching for locations where neighboring normal orientations are very different and normals are not perpendicular to the view. We also find locations of maximum normal curvature in the principal directions (Section~\ref{sec:msprocessing}). Smoothing and denoising are key for creating aesthetically pleasing results as the variation of parameters for robust line extraction changes across spectra. Figure~\ref{fig:lines_whalecrania} shows curvature shading and multispectral line extraction. The mean filter radius, pixel neighborhood  and threshold for suggestive contours are $3$, $7$ and $80\%$. View and normal thresholds for discontinuity lines are $0.9$ and $0.9$.

\subsection {Bispectral Shading}
\label{sec:bispectral}

We present multiscale shading variants to enhance object-scale anatomical features in layered materials by incorporating bispectral shape, bispectral luminance, and multiple diffuse reflectance maps.

\begin{figure}[ht]
\centering
\includegraphics[width=1.0\hsize]{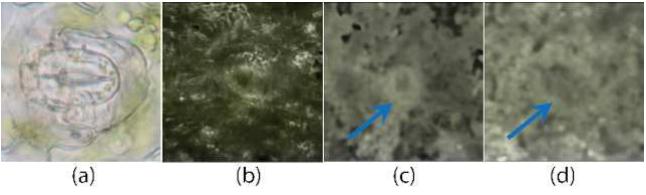}
\setlength\abovecaptionskip{-0.7\baselineskip}
\setlength{\belowcaptionskip}{-12pt} 
\caption{\label{fig:bispectral_stoma_diffuseshading}%
(a) Micro-image of a guard cell made manually from an invasive peal imprint \textcopyright{shinyu/Adobe Stock}. (b) Our diffuse shading using multiscale bilateral filtering on c. (c) The combined visible and bispectral shape map enhances guard cell structure better than the visible-only example (d). Galax urceolata, FLAS263100, University of Florida Herbarium.}
\end{figure}

\paragraph{\textbf{Bispectral shape:}} In Figure~\ref{fig:bispectral_stoma_diffuseshading} b, multiscale bilateral filtering on combined visible and bispectral normals ($\mathbf{n}_{combined}$, Section~\ref{sec:acquisition}) enhances the shape of \emph{guard cells} so that they are easily identified in micro-images. Guard cells (Figure~\ref{fig:bispectral_stoma_diffuseshading}a) are two specialized curved cells that surround stomata, tiny openings found  in the lower epidermis of a leaf underneath a thin cuticle layer. Manual imaging methods use static photographs of peal leaf-imprints in a process which may damage the specimen. Our non-destructive alternative controls levels of  detail in observations of the surface. Shape map $\mathbf{n}_{combined}$ (Figure~\ref{fig:bispectral_stoma_diffuseshading}c) shows more details of the guard cell structure (like the opening), than shape map $\mathbf{n}_{vis}$ (Figure~\ref{fig:bispectral_stoma_diffuseshading}d) due to object-scale detail in $\mathbf{n}_{bilateral}$. Shadows and yellowish color map tones are added for curvature enhancement. 

\begin{figure}[h]
\centering
\includegraphics[width=1.0\hsize]{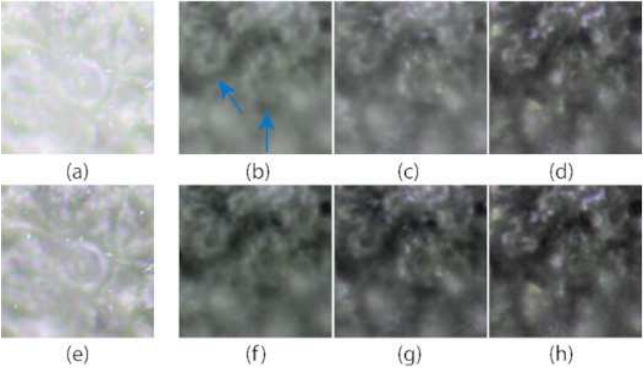}
\setlength\abovecaptionskip{-0.7\baselineskip}
\setlength{\belowcaptionskip}{-14pt} 
\caption{\label{fig:bispectral_stoma}%
(a and e) Traditional static MLIC for depth parameters $\beta = 0.5$ and $\beta = 1.00$ repectively. (b-d, f-h) Our MLIC with bispectral luminance emphasizes anatomical features better. Diffuse reflectance map interpolation for three different light directions for depth parameters $\beta = 0.5$ (b, c, d) and $\beta = 1.00$ (f, g, h). Galax urceolata, FLAS263100, University of Florida Herbarium.}
\end{figure}

\paragraph{\textbf{Bispectral Luminance:}} Our multilight enhancement is unique because it captures features that are only revealed by the interaction of short ultraviolet wavelengths with nanoscale structures (Figure~\ref{fig:uvcabbagewhite}). Multilight enhancement methods~\cite{Fattal2007} (MLIC) separate luminance and chrominance, operating in $YUV$ space. Details ($I_{D}$) from multiscale luminance ($Y$) image decomposition are combined before adding chrominance ($U$ and $V$):
 
\begin{equation}
    I_{Result} = I_{D} + \beta \cdot I_{B}, 0 < \beta \leq 1
\end{equation}

\noindent Base image, $I_{B}$, combines shading across all input at the coarsest scale. User parameter $\beta$ ($[0 – 1]$) controls interpolation. 

We introduce a new structural term $\mathbf{\Gamma}_{bis}$ computed from bilateral filter decomposition of bisectral reflectance  maps captured from all light directions $\mathbf{l}$ and scales $s$. We control the sharpness (apparent \emph{focus}) of shape details, by replacing the visible decomposition term,  $\mathbf{\Gamma}_{vis}$, with $\mathbf{\Gamma}_{bis}$  at pixels in the bispectral luminance map, $Y^{i}_{p, bis}$ where the intensity is greater than a threshold $Y_{th} = 0.25$. These pixels are emitting visible light from uv excitation ($365nm$). We compute $I_{D}$ from difference images of smoothed luminance for each $\mathbf{l}$: 

\begin{equation}
    I_{D}(p) = \sum^{l}_{i=1}\sum^{s}_{j=1}\frac{f_{D}(i, j, p)}{N}
\end{equation}

\noindent where:

\begin{equation}
    f_{D}(i, j, p) = w(i,j,p, \mathbf{\Gamma})o(i,j,p,\mathbf{\Gamma})
\end{equation}

\noindent  and $N$ is a normalized sum of all detail weights.  Weight, $w$ is a per wavelength gradient with high frequency noise removed. The detail function $o$ computes the difference in luminance between levels across all lights and scales. As discussed,  $\mathbf{\Gamma}$ is $\mathbf{\Gamma}_{bis}$ for $Y^{i}_{p, bis} > Y_{th}$, otherwise $\mathbf{\Gamma}_{vis}$. Each pixel $p$ in $f_{B}(i, p)$  is selected directly from $\mathbf{\Gamma}_{bis}$  or $\mathbf{\Gamma}_{vis}$  at the course most level (without detail enhancement).

\begin{equation}
    I_{B}(p) = \sum^{l}_{i=1}w_{B}(\mathbf{l}^{input}, \mathbf{l}^{i})f_{B}(i, p)
\end{equation}

\noindent Unlike prior methods, we compute a linear combination $U$ (and $V$) sampled from more than one reflection map, which we combine with $Y$ before converting to RGB color space. Our bispectral formulation eliminates shadows (a limitation of traditional methods)  across all MLICs in regions with high structural detail (apparent in fluorescence). Light dependent chroma effects are simulated by incorporating multiple visible diffuse reflection maps  when shading (see \emph{Multiple Diffuse Maps}  for further reasoning behind this step):

\begin{equation}
   Y_{Result} = e^{I_{Result}}
\end{equation}

\begin{equation}
    U_{Result} = \sum^{l}_{i=1}w_{B}(\mathbf{l}^{input}, \mathbf{l}^{i})U^{i}_{vis}
\end{equation}

\begin{equation}
    V_{Result} = \sum^{l}_{i=1}w_{B}(\mathbf{l}^{input}, \mathbf{l}^{i})V^{i}_{vis}
\end{equation}

Figure~\ref{fig:bispectral_stoma} compares our method for three light orientations (three right columns) with traditional MLIC enhancement (left column). Increasing $\mathbf{\Gamma}_{bis}$ from $0.5$ to $1.0$ creates an effect like focusing a camera lens. The out-of-focus threshold is approached in Figure~\ref{fig:bispectral_stoma} d and h. The combined effect of $\mathbf{\Gamma}_{bis}$ and $\mathbf{\Gamma}_{vis}$ increases apparent depth of the cell in Figure~\ref{fig:bispectral_stoma} c and g  compared to flatter renderings in Figure~\ref{fig:bispectral_stoma} a and e. Biologists manually count guard cells to determine the number of sets of 
chromosome pairings in plants (~\cite{MISHRA1997689}), an evaluation method in our user study. 

\begin{figure}[h]
\centering
\includegraphics[width=0.7\hsize]{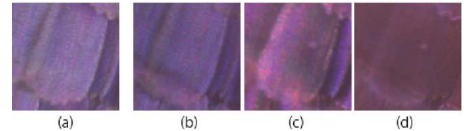}
\setlength\abovecaptionskip{2pt}
\setlength{\belowcaptionskip}{-14pt} 
\caption{\label{fig:bispectral_butterfly}%
Our bispectral MLIC shading emphasizes additional object scale structure in butterfly scales (b-d) when compared to traditional static MLIC enhancement (a) $Y_{vis} = 0.5$ (b-d) $Y_{bis} = 0.5$}
\end{figure}

\begin{figure}[t]
\centering
\includegraphics[width=0.9\hsize]{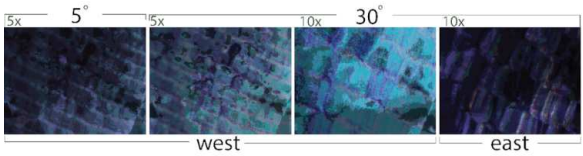}
\setlength{\belowcaptionskip}{-10pt} 
\caption{\label{fig:bispectralshading}%
Toon shading with interpolated diffuse maps. Morphidae blue butterfly wing. Hue varies with orientation (west and east). Intensity changes with tilt angle ($5\deg, 30\deg$). Bispectral normals capture object-scale detail at different magnifications (5x and 10x). Light directions left to right: (0.10, -0.99, 0), (0.49 -0.87 0), (0.49 -0.87 0), (-0.49, -0.87, 0)} 
\end{figure}

\paragraph{\textbf{Multiple Diffuse Maps:}} We compute weights $w_i = 1-\alpha$ where $\alpha$ is an angular difference between  the input light direction $\mathbf{l}$ and $\mathbf{l}_i$ for each associated diffuse reflectance map, and generate a per-pixel color by interpolating  between diffuse reflectance maps. These are not precise approximations for iridescence (Figure~\ref{fig:micro}), but demonstrate effects not available in traditional NPR. Compare the \hbox{3-D}-like quality of the scale in Figure~\ref{fig:bispectral_butterfly} (b-d) with the traditional approach Figure~\ref{fig:bispectral_butterfly} (a) that uses constant chroma and visible-only structure. Figure~\ref{fig:bispectralshading} shows hue variation at different magnifications. 

\section{User Study}
\label{sec:userstudy}
We conducted a user study to validate that our multispectral analysis methods: (1) cannot be achieved with current NPR techniques, (2) improves precision in real-world life science research tasks and (3) have effective parameter controls. We access how material properties
in authentic multispectral data, including layered shape, enable new forms of analysis beyond the aesthetic-based capabilities of current NPR. We formulate quantitative metrics to measure user accuracy performing the same research task with traditional NPR and our algorithms in a \emph{between subjects user study}. We also report qualitative feedback on user experience using parameter controls.

\subsection{Participants and Methodology}
\label{sec:method}

We recruited $39$ participants. Among these, $21$ were self-categorized as experts in one of four disciplines: biology ($9$), anthropology ($4$), forensic science ($3$) and paleontology ($5$).  Eighteen ($18$) self-categorized as novices with background knowledge in these fields, but less practical experience. Participants in the pilot study did \emph{\textbf{not}} participate in the user study. Participants did not receive compensation.

We designed two research tasks based on pilot study findings -  \emph{Feature Selection} and \emph{Guard Cell Counting}. No training was required. Tasks were completed remotely over a web-based interface to our client-server implementation.  Simple instructions were presented as four general steps: examine the specimen (which was rendered in either a traditional or multispectral style), locate specific features, adjust pan-zoom and slider controls for feature enhancement, select the features in the rendered image or choose \emph{not found}. 

\textbf{Quantitative metrics}  were used to evaluate selection accuracy. Selections were categorized as correct if the selection circle intersected feature pixels ($100 \%$ accuracy) or occurred  within an allowable distance $\mathbi{d}$ from the feature (with $\%$ accuracy decreasing exponentially with distance). Incorrect selections occurred outside $\mathbi{d}$ or when the \emph{not found} button was selected. We use a $25$ pixel selection radius with $\mathbi{d} = 50$ pixels for a $1480 \times 1168$ resolution image.   Users also provided \textbf{qualitative   feedback} by rating the effectiveness of parameter controls on a scale of $0-5$. Participants were not given time constraints but took on average $10$ - $15$ minutes per task. Most participants ($71.79 \%$) participated in only one task and $21.21 \%$ participated in both. We now describe our research tasks.

\paragraph{Feature Selection} is used to identify biological structures in specimens. Participants were instructed to find shape features in a rendered image. Features were visualized separately as enhancement map icons. Three datasets were used. Each dataset was assigned four features  - half visible and half near-infrared. The $32$ participants were randomly assigned spectral band shading or a traditional NPR algorithm. Traditional NPR included Toon, Gooch and Exaggerated shading, and multi-pass joint bilateral filtering combined with Lambertian shading. Assignments were distributed evenly among experts and novices such that half ($8$ experts, $8$ novices)  received spectral band shading and half (8 experts, 8 novices) received traditional NPR shading ($2$ experts and $2$ novices for each traditional algorithm). User accuracy per algorithm was evaluated.

\paragraph{Guard Cell Counting} is used to identify organisms with more than two paired (homologous) sets of chromosomes. The number of chromosome sets in somatic cells (ploidy level) is directly tied to stoma size (area covered) and stomatal density (number of stomata per unit sq area of a leaf). These values differ from species to species(~\cite{MISHRA1997689}). As chromosome sets increase, stomata size increases and density decreases. Participants were instructed to locate and select guard cells by placing and resizing a marker to completely cover the cell in micro-image ($20x$ objective) visualizations. An example guard cell was displayed separately as a reference.  There were $18$ participants, randomly assigned to one of three visualization types: visible photo, traditional MLIC, our bispectral MLIC shading, Lambertian shading with bilateral filtering  on our $\mathbf{n}_{combined}$ and traditional Lambertian shading with bilateral filtering on $\mathbf{n}_{vis}$. Images were evenly distributed among experts and novices ($3$ novices and $3$ experts per image type). Counts were used to determine stoma size and stomatal density, and ultimately the number of sets of chromosome pairs and results were compared with published results (~\cite{Nesom1983GalaxG}) from manual methods (Figure~\ref{fig:bispectral_stoma_diffuseshading}a). We also consulted experts from our pilot study for confirmation.

\subsection{Evaluation Metrics}

\begin{figure}[h]
\centering
\includegraphics[width=1.0\hsize]{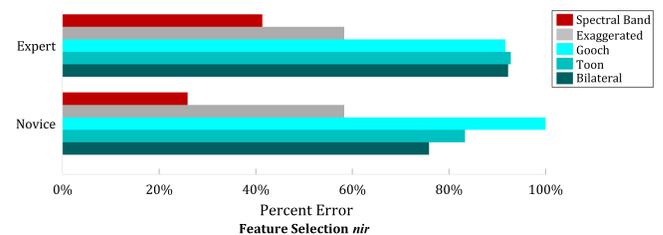}
\setlength{\belowcaptionskip}{-10pt} 
\caption{\label{fig:user1-graph1}%
	We achieve analysis not possible with traditional NPR. Percentage error when detecting near-infrared features.}
\end{figure}

\begin{figure}[h]
\centering
\includegraphics[width=1.0\hsize]{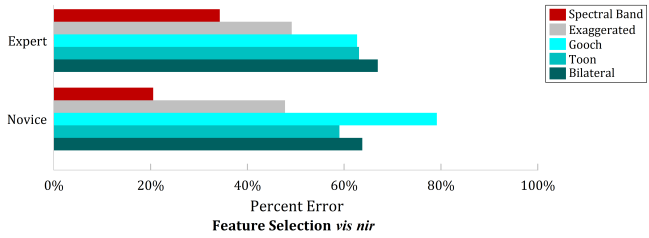}
\setlength{\belowcaptionskip}{-10pt} 
\caption{\label{fig:fselect-erroroverall}%
	Our solution produces more precise results than traditional NPR. Percentage error when detecting visible and near-infrared and features.}
\end{figure}

\emph{Feature selection} results show our NPR\textbf{\emph{plus}} \emph{multispectral analysis} system \textbf{permits scientific analysis not possible with current NPR}. We computed the average error ($\% accuracy$, Section~\ref{sec:method}) over all near-infrared features for each algorithm. Both experts and novices detected near-infrared sub-surface features at a lower error rate  than traditional methods. Figure~\ref{fig:user1-graph1} \emph{top} shows expert error (spectral band $41.34 \%$, exaggerated $58.33\%$, Gooch $91.67\%$, Toon $92.81 \%$, Bilateral $92.25 \%$). Novice error is shown on the bottom (spectral band $25.87 \%$, exaggerated $58.33\%$, Gooch $100 \%$, Toon $83.33 \%$, Bilateral $75.87\%$). A similar trend occurred when considering both visible and near-infrared features. Figure~\ref{fig:fselect-erroroverall} \emph{top} shows expert error (spectral band $34.28\%$, exaggerated $49.16 \%$, Gooch $62.68 \%$, Toon $63.07 \%$, Bilateral $66.96 \%$). Novice error is shown on bottom (spectral band $20.53 \%$, exaggerated $47.78 \%$, Gooch $79.17 \%$, Toon $59.01 \%$, Bilateral $63.75\%$).
\begin{figure}[h]
\centering
\includegraphics[width=0.8\hsize]{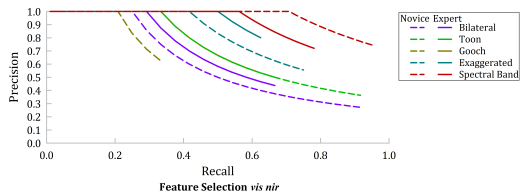}
\setlength{\belowcaptionskip}{-10pt} 
\caption{\label{fig:fselect-precisionrecall}%
	Our solution increases accuracy in Feature selection.}
\end{figure}

\begin{figure}[h]
\centering
\includegraphics[width=0.45\hsize]{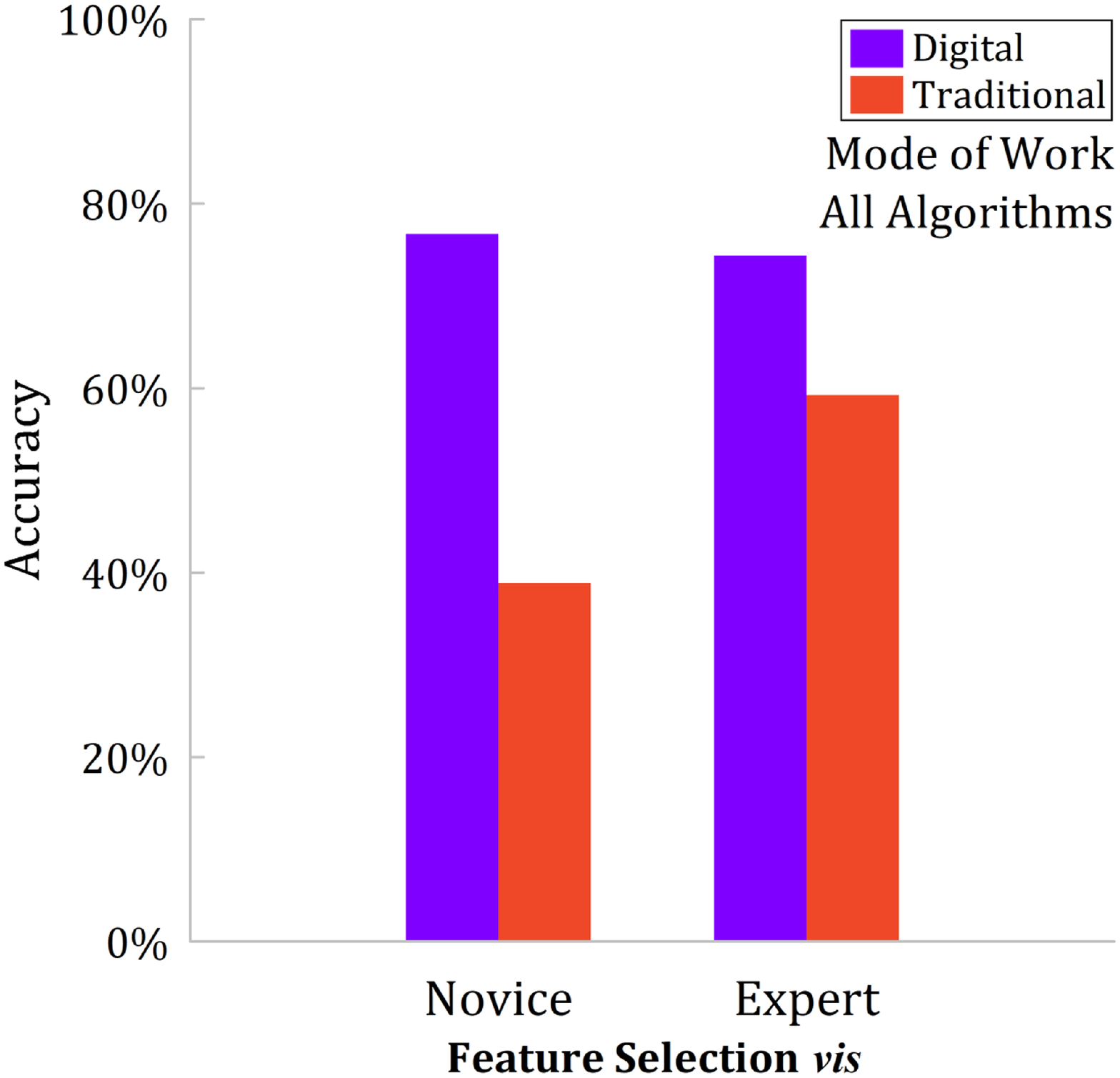}
\includegraphics[width=0.45\hsize]{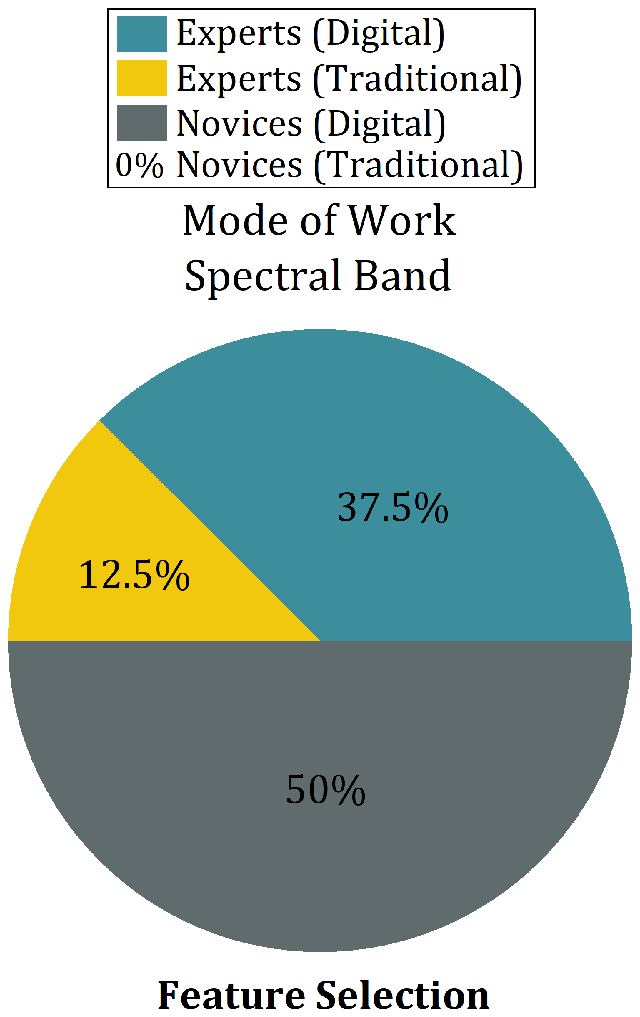}
\setlength{\belowcaptionskip}{-10pt} 
\caption{\label{fig:modesofwork}%
(left) Participants who primarily work with digital media outperformed those who primarily work with traditional physical media. (right) Participant distribution: spectral band shading feature selection.}
\end{figure}

Our algorithms \textbf{improve precision in real research tasks}. Figure~\ref{fig:fselect-precisionrecall} shows a feature ranking exercise. The $y$ axis  plots precision (accuracy) of correct selections ordered from high to low. The $x$ axis plots the fraction of correct selections over the total correct features listed. Incorrect selections were excluded.  Participants using spectral band shading were more accurate for a longer recall. Participants who worked primarily with digital media - images and software -  were more accurate  than those who worked with traditional mediums - printed photographs and x-rays on light tables. Figure~\ref{fig:modesofwork} \emph{left} compares expert (media $74.42 \%$, traditional $59.30 \%$) and novice (media $76.76 \%$, traditional $38.89 \%$) accuracy based on modes of work.  All novices using spectral band shading had digital media experience, while our expert pool was split $12.5 \%$ traditional and $37.5 \%$ digital (Figure~\ref{fig:modesofwork} \emph{right}). This explains why novices outperformed experts when using spectral band shading. 

\begin{figure}[h]
\centering
\includegraphics[width=1.0\hsize]{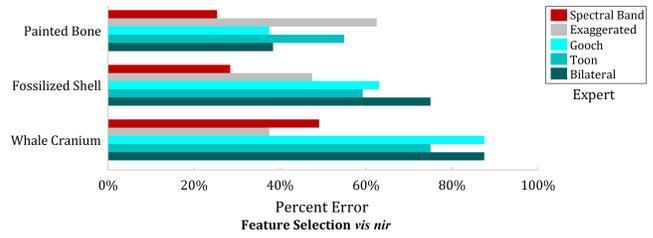}
\setlength{\belowcaptionskip}{-10pt} 
\caption{\label{fig:user1-graph5-datasets}%
	Dataset characteristics influence the precision of NPR algorithms.}
\end{figure}

\begin{figure}[h]
\centering
\includegraphics[width=0.45\hsize]{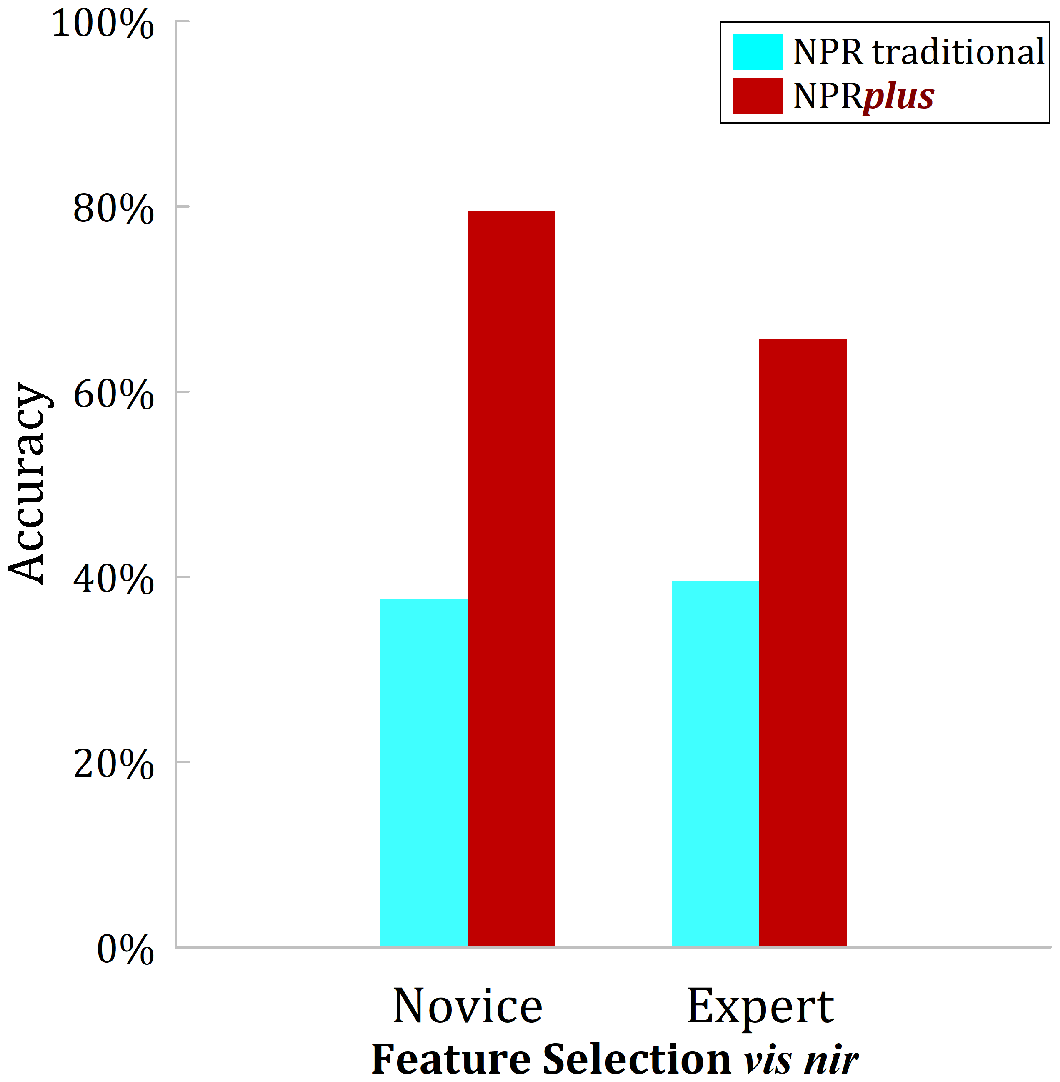}
\includegraphics[width=0.45\hsize]{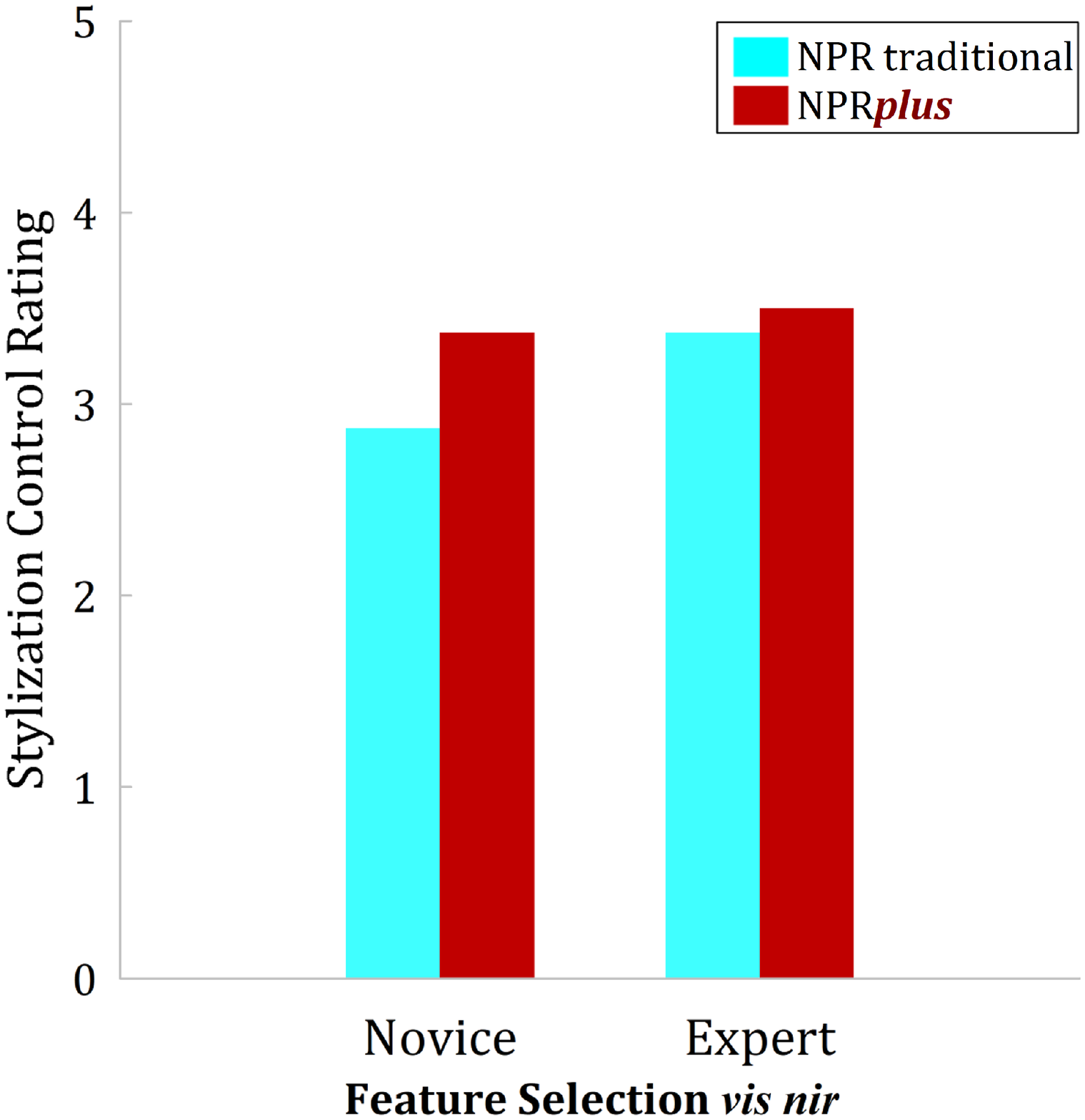}
\setlength{\belowcaptionskip}{-10pt} 
\caption{\label{fig:user3-graph2-percentageerror}%
	Our NPR\textbf{plus} approach is more effective than traditional NPR. (left) Accuracy for feature selection. (right) Participant ratings for contrast and shape enhancement controls on a scale of 0-5. Feature selection task.}
\end{figure}

Figure~\ref{fig:user1-graph5-datasets} shows that dataset characteristics influence the selection error. Exaggerated shading, the most effective classic method had a $12\%$ higher error rate for the painted bone than it did for the monotone shell as participants could not find features underneath the red paint. Gooch shading had high error rates throughout. Overall, our methods were more effective for feature selection when compared to traditional methods combined across all features (Figure~\ref{fig:user3-graph2-percentageerror}). Experts were $22\%$ more effective, and novices were $40\%$ more effective.

\emph{Guard Cell Counting} with enhanced bispectral detail was more precise than traditional NPR or manual methods. We correctly predicted the number of sets of chromosome pairs in our specimen (~\cite{Nesom1983GalaxG}) as diploid (two). Stoma counts were recorded as the number of markers. Marker heights and widths were converted to microns before computing stoma area and stomatal density. Table 1 records the number of correctly identified stoma, and error between computed stomatal density and published results. Novice performance varied greatly. Some selected only a couple of guard cells with high precision producing low SD error (MLIC: $22.31$ and Bilateral\textbf{\emph{plus}}: $66.05$), while others selected many, with moderate correct selections but good precision, resulting in low overall recalls (highest: $0.19$ for MLIC).  Multispectral information was effective as novices had the lowest RMSE and MAPE error for bilateral filtering methods ($220.95$, $0.44$ respectively for bilateral\textbf{\emph{plus}} and $233.43$, $0.38$ respectively for MLIC\textbf{\emph{plus}}). Experts performed consistently, with a higher recall ($0.37$) for bispectral shading than  visible photographs ($0.34$). RMSE and MAPE for the photograph ($414.86$ $499.25$ $0.87$ respectively)  decreased  for the rendered visualizations with the lowest error for bispectral shading ($130.24$ $156.82$ $0.28$ respectively). Interestingly,  MAPE consistently decreased from $0.87$ for a photograph to $0.28$ for the best case with bilateral shading. We anticipate use of fluorescent properties of the stoma for automatic stoma detection for predicting chromosome pairings.

\begin{table}[h]
\setlength{\tabcolsep}{3pt}
\begin{tabular}{p{0.35\linewidth} | l | l l l}
\hline
\multicolumn{5}{c}{\textbf{Novices}} \\
\hline
 \multirow{2}{*}{Algorithm} & \multirow{2}{*}{Recall} & \multicolumn{3}{c}{Error metric}  \\\cline{3-5}
 & & SD & RMSE  &  MAPE \\
  \hline
 Visible photograph & 0.18 & 590.40 & 734.32 & 1.21  \\
 MLIC & \textbf{0.19} & \textbf{22.31} & 351.60 & 0.73  \\
 Bilateral & 0.05 & 112.90 & 282.09 & 0.54 \\
 Bilateral\textcolor{maroon}{\textbf{\textit{plus}}} & 0.03 & 66.95 & \textbf{220.85} & 0.44 \\
 MLIC\textcolor{maroon}{\textbf{\textit{plus}}} & 0.10 & 220.71 & 233.43 & \textbf{0.38} \\
\end{tabular}
\begin{tabular}{p{0.35\linewidth} | l | l l l}
\hline
\multicolumn{5}{c}{\textbf{Experts}} \\
\hline
 \multirow{2}{*}{Algorithm} &  \multirow{2}{*}{Recall} & \multicolumn{3}{c}{Error metric}  \\\cline{3-5}
 & & SD & RMSE  &  MAPE \\
  \hline
 Visible photograph & 0.34 & 414.86 & 499.25 & 0.87  \\
 MLIC & 0.26 & 200.73 & 384.74 & 0.78  \\
 Bilateral & 0.16 & 299.43 & 341.16 & 0.64 \\
 Bilateral\textcolor{maroon}{\textbf{\textit{plus}}} & 0.19 & 188.13 & 199.37 & 0.35 \\
 MLIC\textcolor{maroon}{\textbf{\textit{plus}}} & \textbf{0.37} & \textbf{130.24}  & \textbf{156.82} & \textbf{0.28} \\
 \hline
\end{tabular}
\label{fig:table-gruardcellcounting}
\setlength{\abovecaptionskip}{20pt} 
\caption{Our NPR\textbf{plus} solution reduces error in guard cell size measurement task.}
\end{table}

\subsection{Study Limitations}
Experts did not have a strong preference for our parameter controls compared to traditional ones, while novices clearly gave higher ratings to our controls (Figure~\ref{fig:user3-graph2-percentageerror} \emph{right}). This divergence is likely due to the meticulous way experts verified selections, coupled with the broader flexibility in our parameter set requiring a higher learning curve. In guard cell counting,  the narrow field of focus and image quality could be improved with HDR imaging, and merging multiple images at different depths of focus.

\begin{figure*}[t]
\centering
\vspace{0.5pt}
\setlength{\tabcolsep}{0.4pt}
 \begin{tabular}{cccccccccc}  
  \includegraphics[height=0.61in]{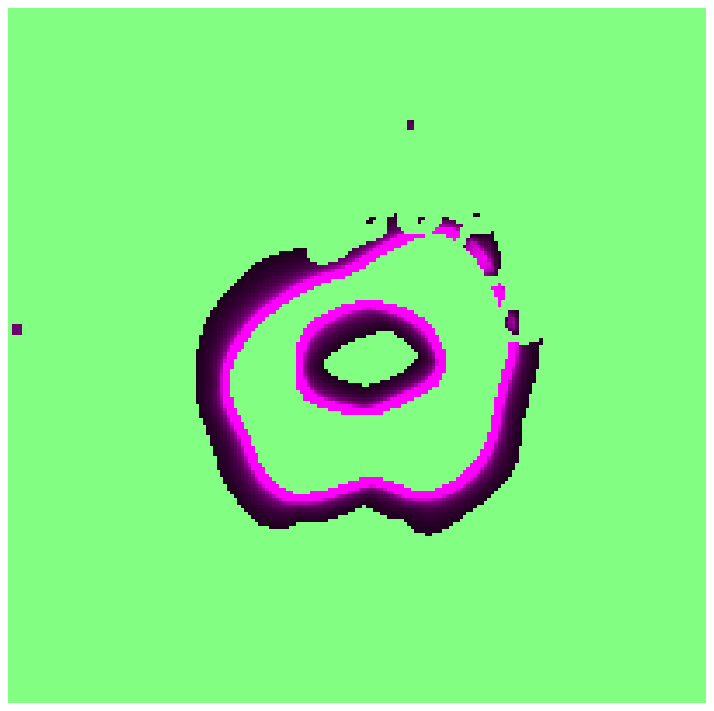}
 &\includegraphics[height=0.61in]{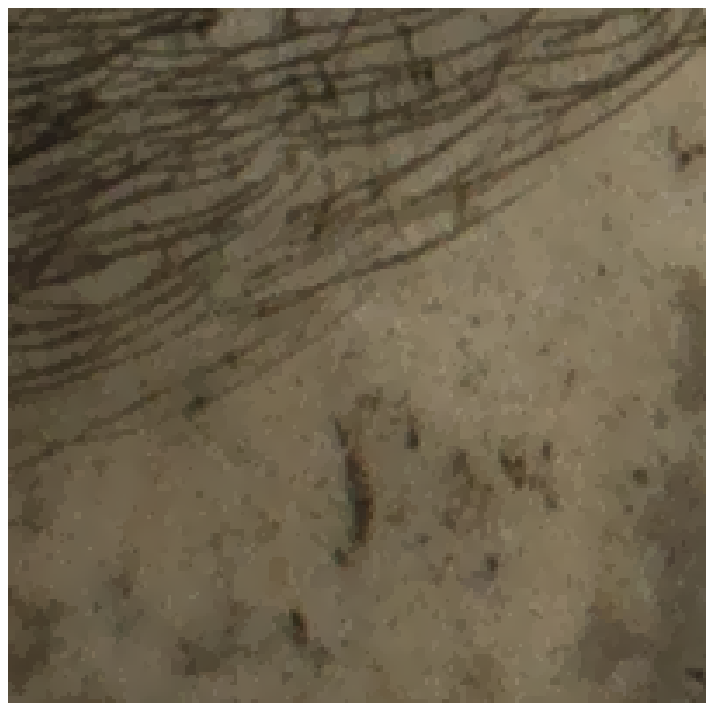}
 &\includegraphics[height=0.61in]{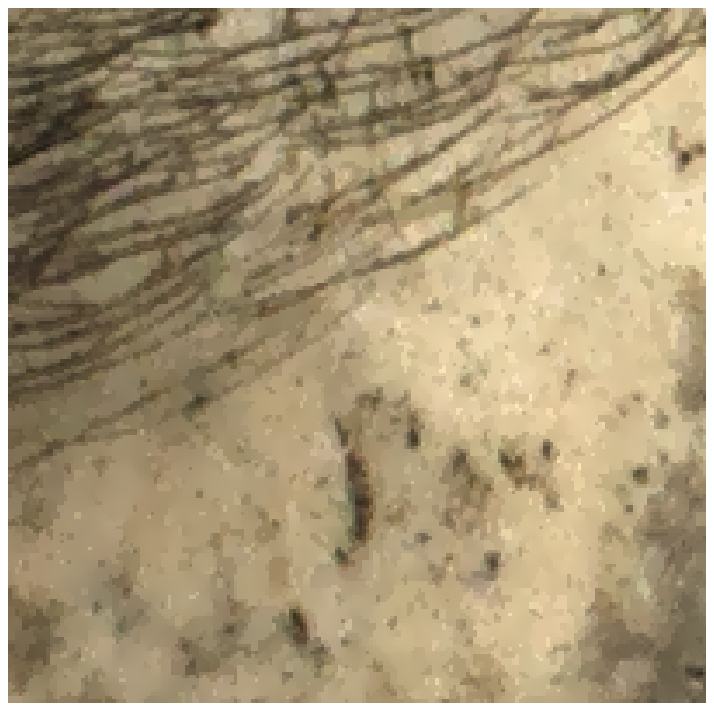}          
 &\includegraphics[height=0.61in]{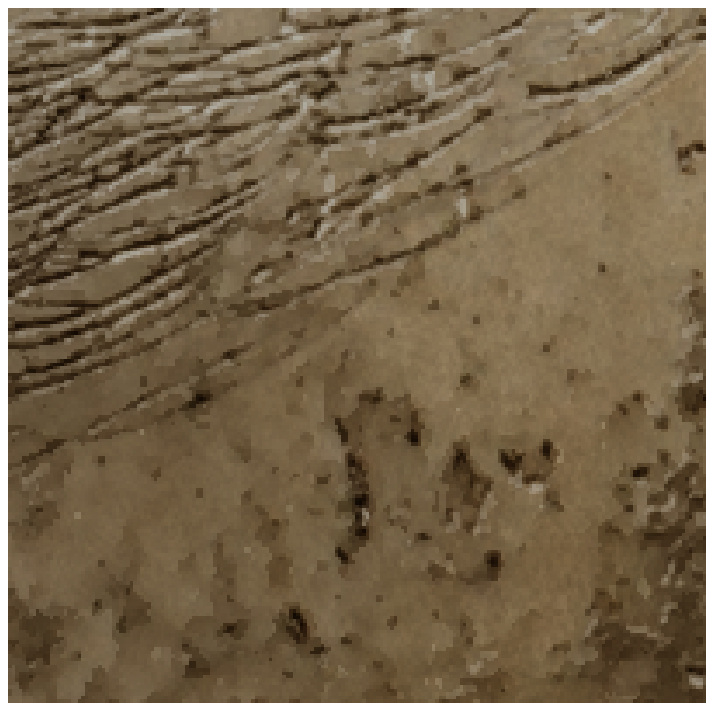}
 &\includegraphics[height=0.61in]{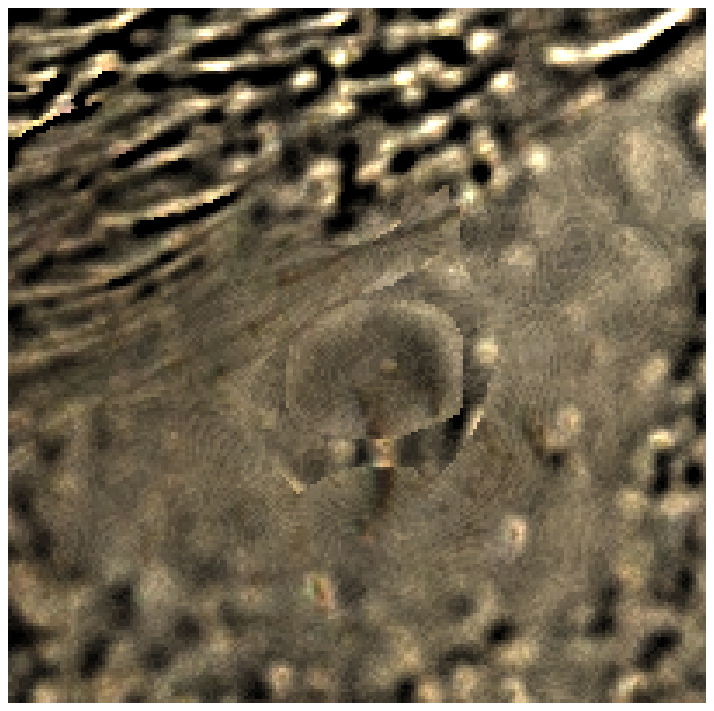}
 &\includegraphics[height=0.61in]{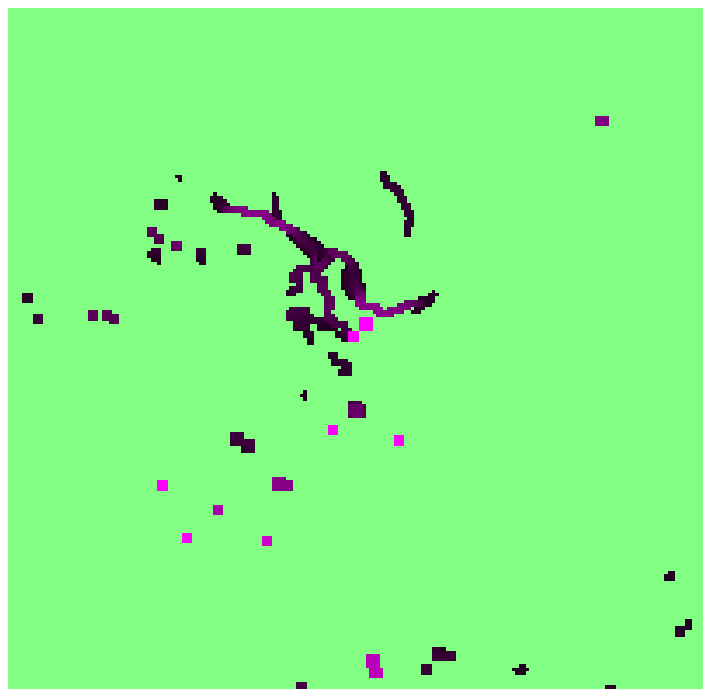}
 &\includegraphics[height=0.61in]{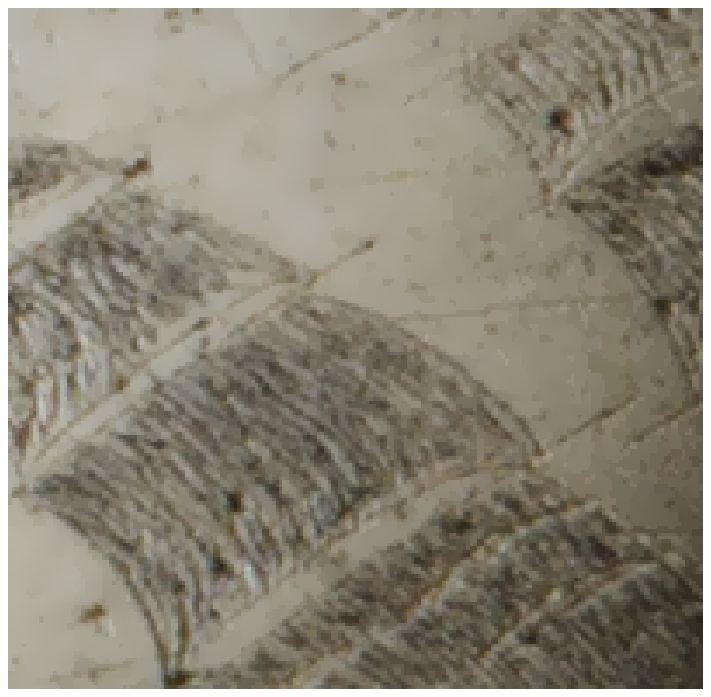}
 &\includegraphics[height=0.61in]{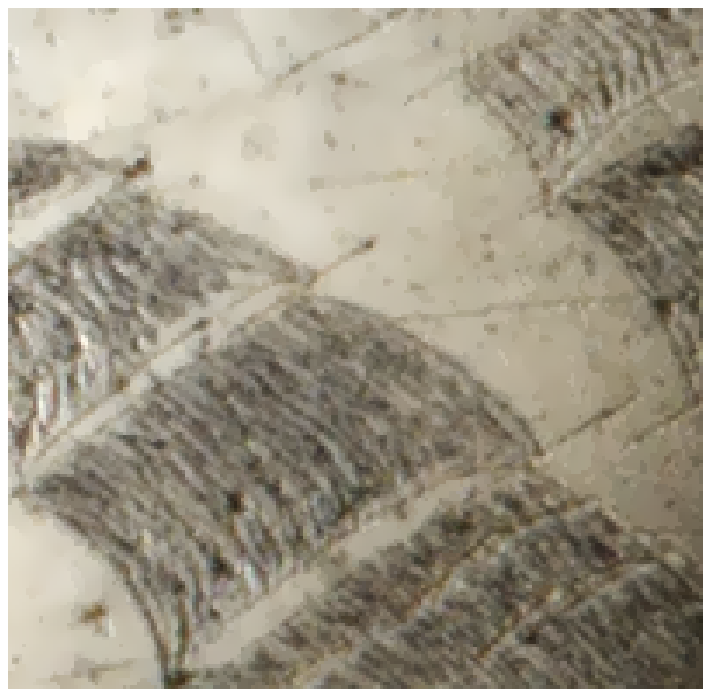}
 &\includegraphics[height=0.61in]{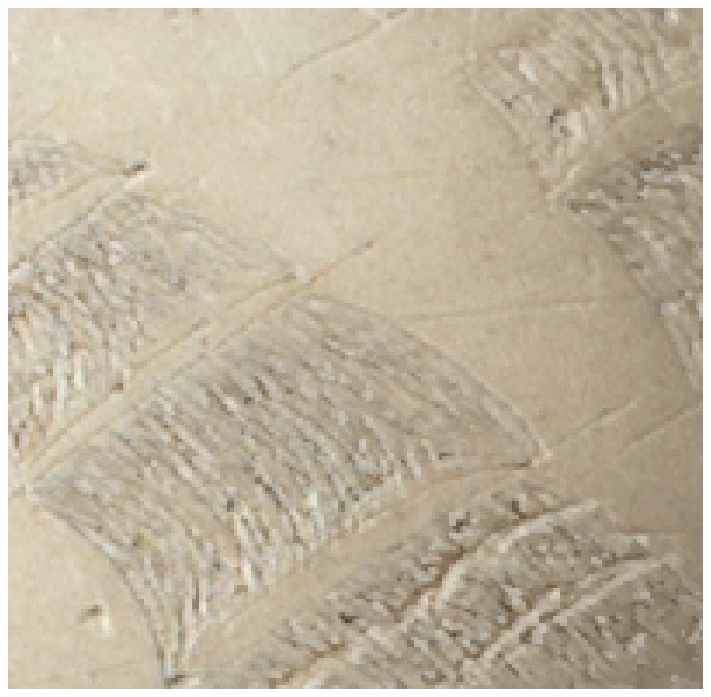}
 &\includegraphics[height=0.61in]{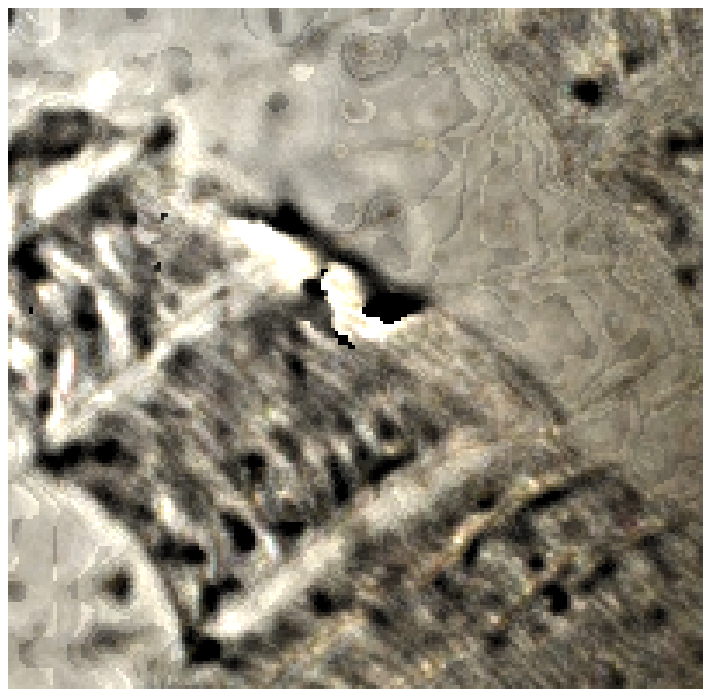}
 \end{tabular}%
\caption{Our new near-infrared enhancement maps (green) for (left to right) Lambertian Shading, Blinn-Phong Shading, Exaggerated Shading and our new Spectral Band Shading for two different details. We capture invisible near-infrared shape features (magenta) that visible color methods do not. Whale cranium with scrimshaw, Department of Mammalogy, AMNH.}
\setlength\abovecaptionskip{-0.7\baselineskip}
\setlength\belowcaptionskip{-8pt}
\label{fig:subsurface}
\end{figure*}

\subsection{Applications}
\label{sec:applications}

\begin{figure}[h]
\centering
\def\imh{0.875in}
\setlength{\tabcolsep}{1.75pt}
\begin{tabular}{cc}

  \includegraphics[height=\imh]{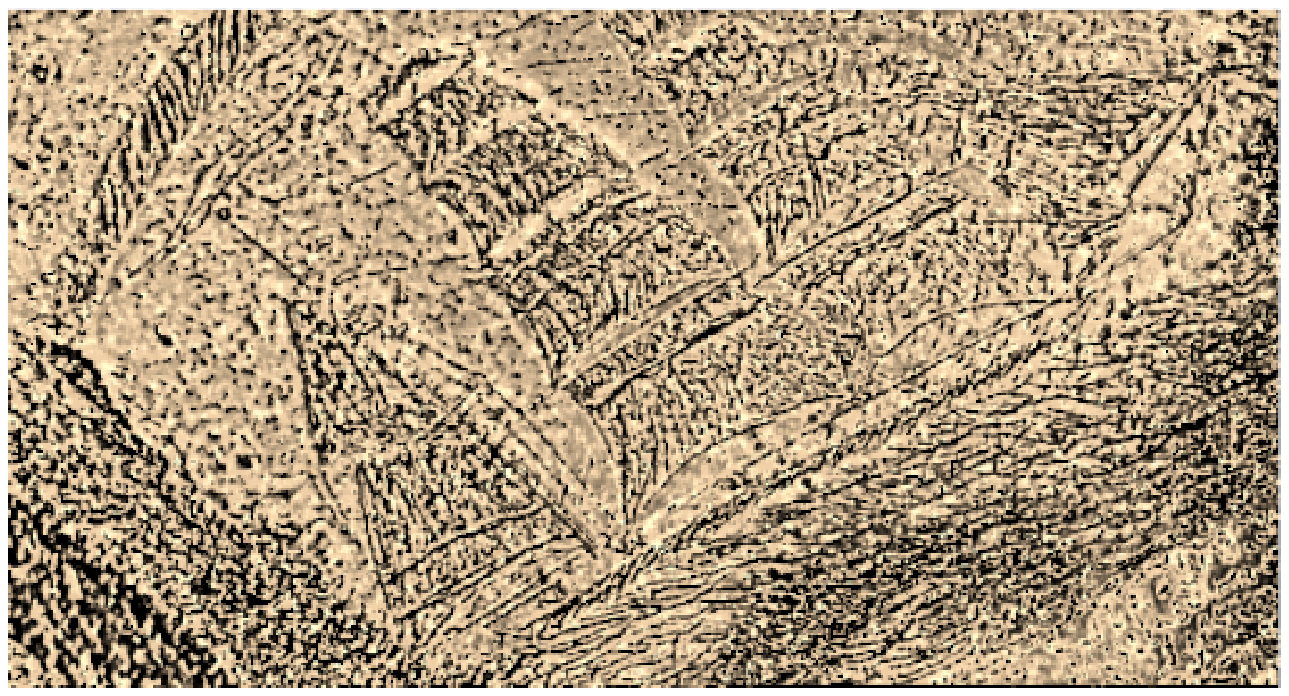}
& \includegraphics[height=\imh]{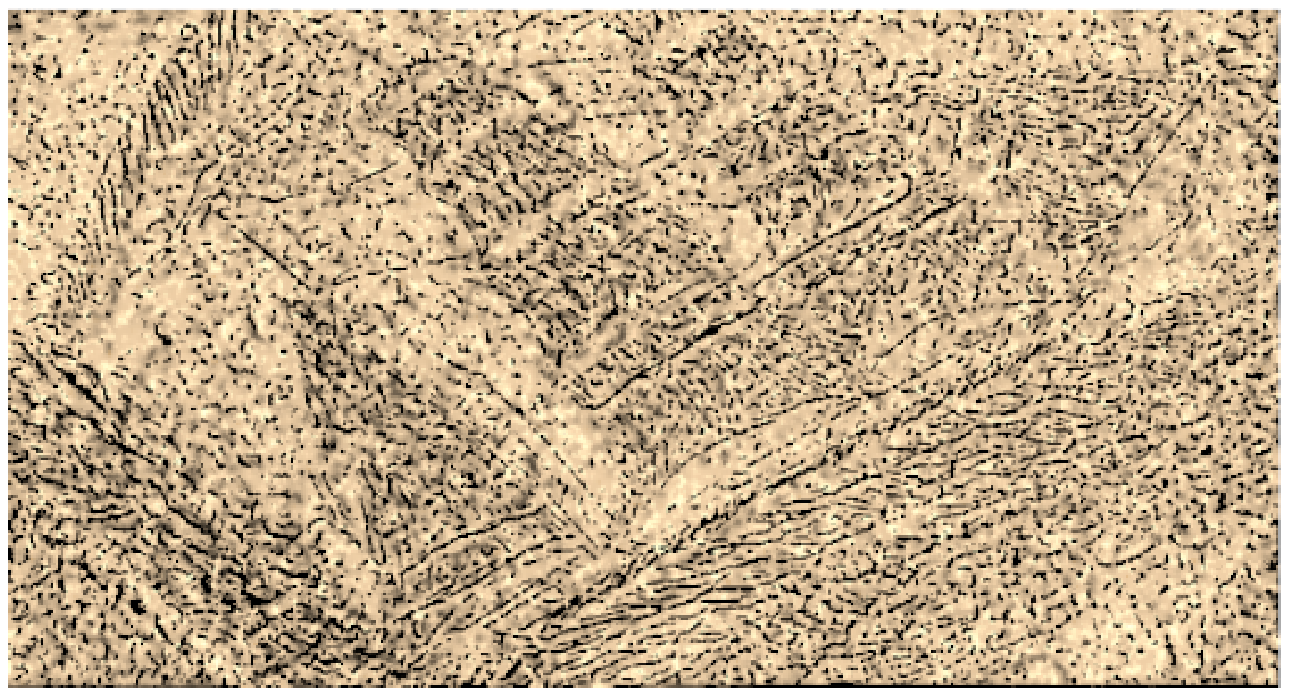}
\\
  \includegraphics[height=\imh]{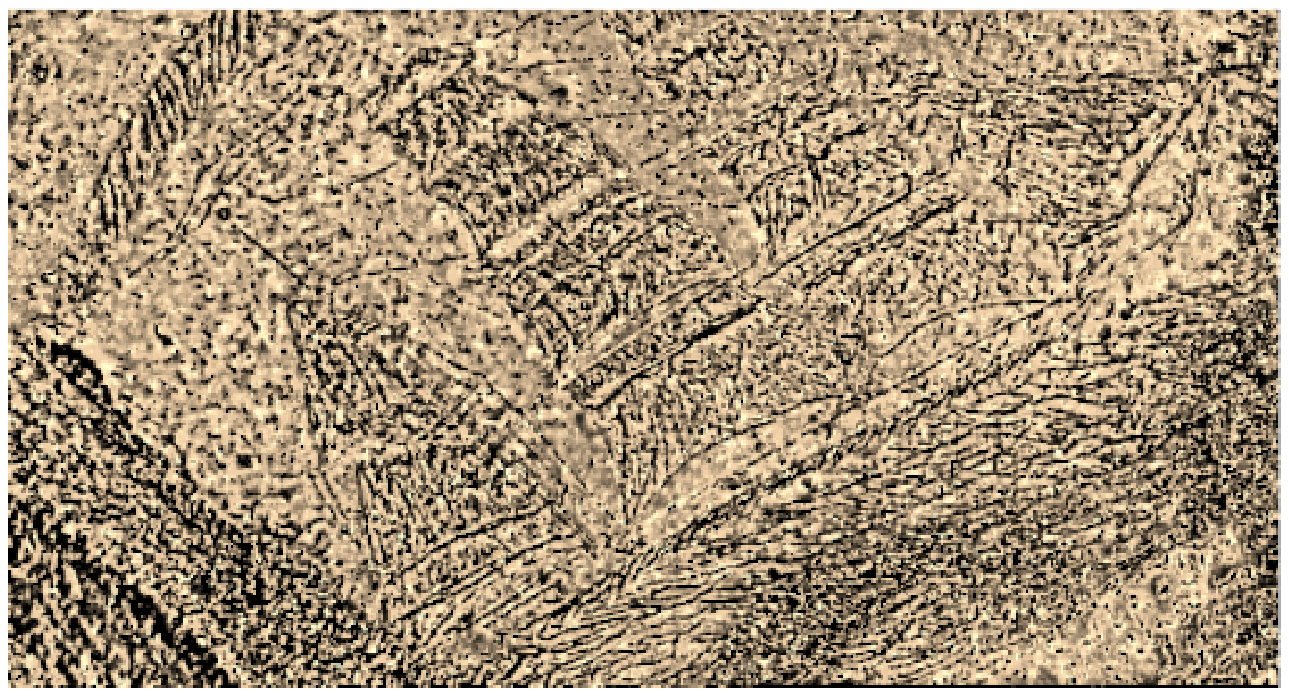}
& \includegraphics[height=\imh]{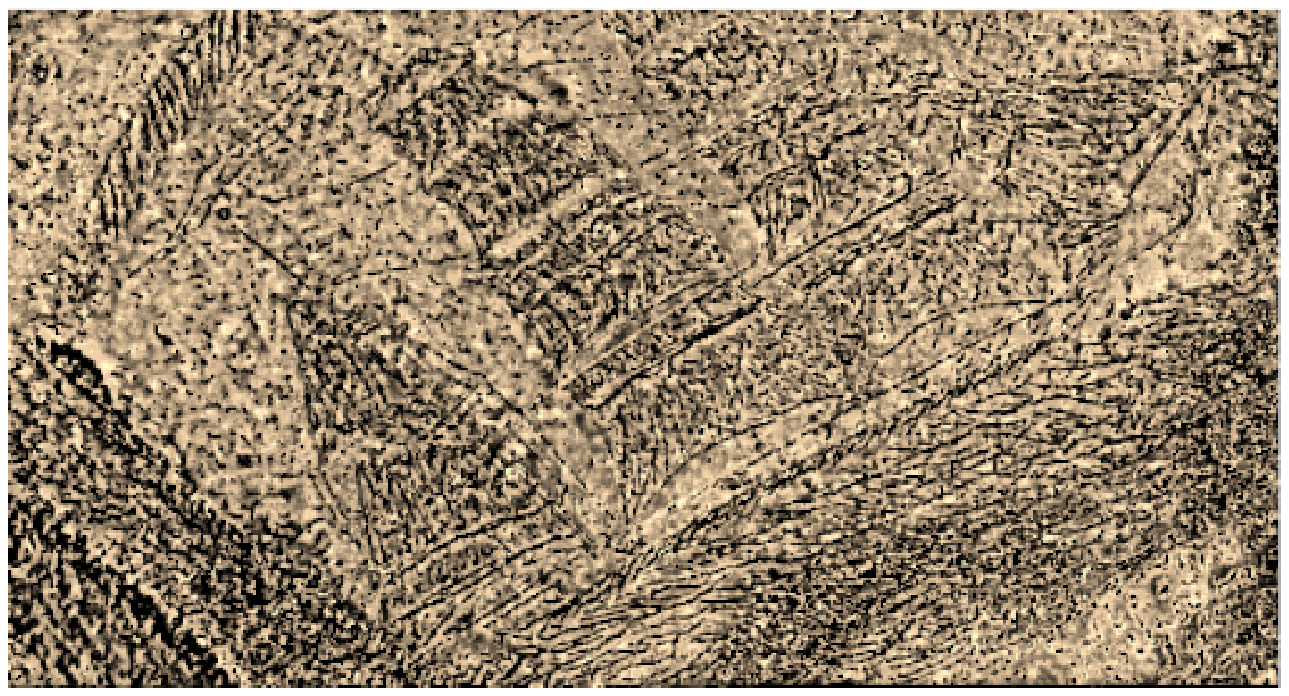}

\end{tabular}
\setlength{\belowcaptionskip}{-10pt} 
\setlength\abovecaptionskip{1pt}
\caption{(top) Exaggerated shading with (left) multiscale visible curvature shading and (right) multiscale curvature shading and multispectral lines. (bottom) Spectral band shading with (left) multiscale, multispectral curvature shading and visible lines and (right) multiscale multispectral curvature shading and multispectral lines. Whale cranium with scrimshaw, Department of Mammalogy, AMNH.}
\label{fig:sailboat}
\end{figure}

\begin{figure}[h]
\centering
\def\imh{0.83in}
\setlength{\tabcolsep}{0.3pt}
\begin{tabular}{cccc}

$400-700nm$ & $720nm$ & $830nm$ & $1000nm$\\

\includegraphics[height=\imh]{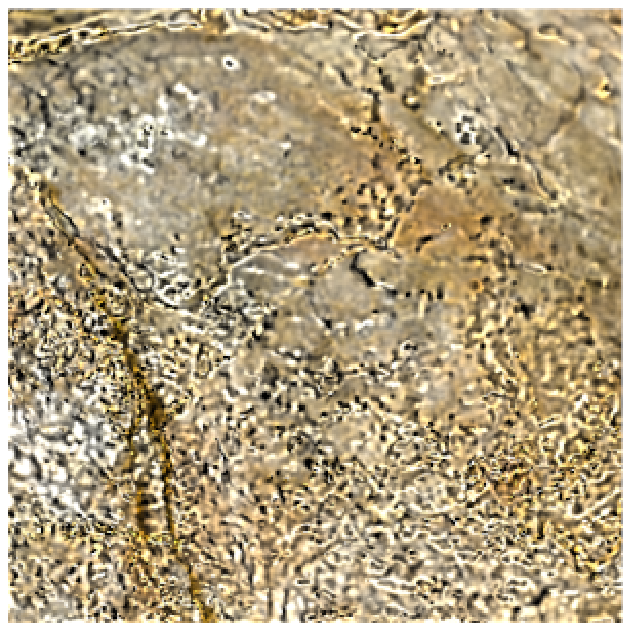}
&\includegraphics[height=\imh]{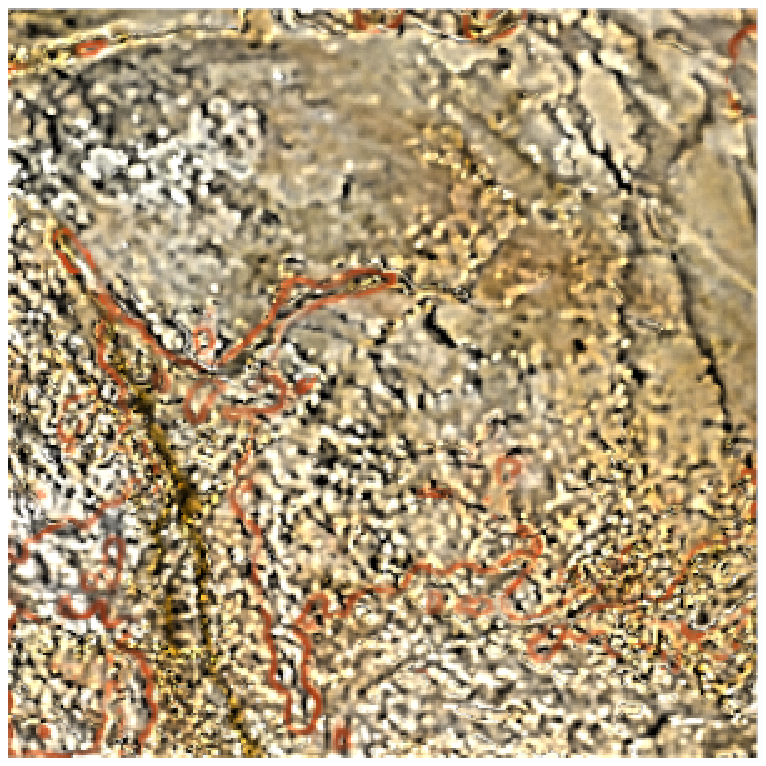}
&\includegraphics[height=\imh]{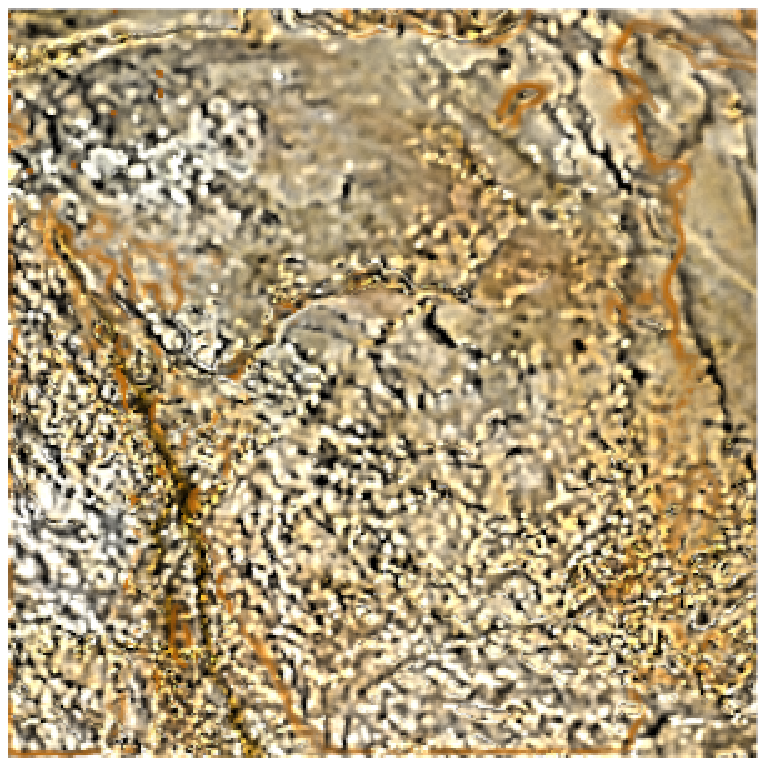}
&\includegraphics[height=\imh]{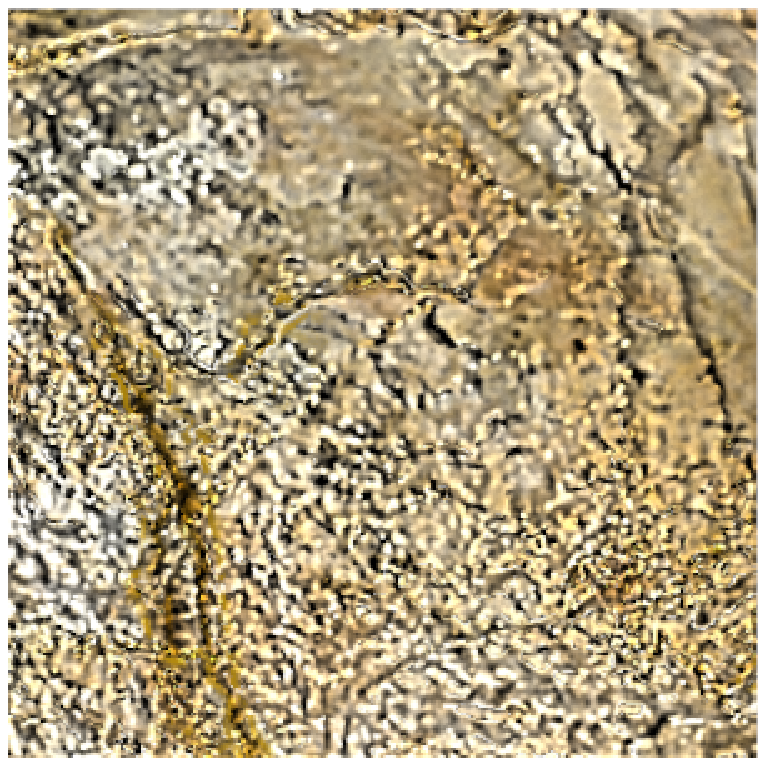}
\\

\end{tabular}
\setlength{\belowcaptionskip}{-8pt}
\setlength\abovecaptionskip{1pt}
\caption{Spectral Band Control. DPC No. 11835, Archaeolemur, Field No. 92-M-257, Fossil Primates Division, Duke University Lemur Center.}
\label{fig:sbsbandcontrol}
\end{figure}

\begin{figure}[h]
\centering
\includegraphics[width=1.0\hsize]{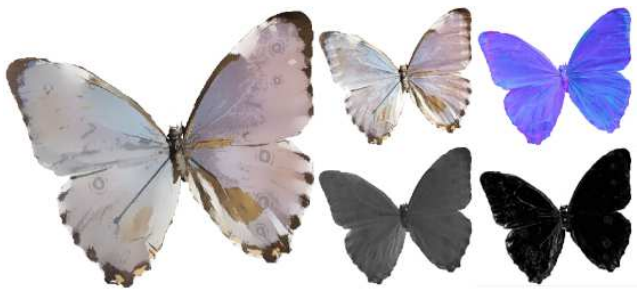}
\setlength{\belowcaptionskip}{-10pt}
\caption{\label{fig:irblending}%
Infrared data reveals features on the undersurface of the wing. Morpho sulkowsly butterfly, McGuire Center for Lepidoptera and Biodiversity, FLMNH. }
\end{figure}

Study results show our pipeline is generalizable to disciplines like biology, forensics and paleontology where visualizing micro structures, sub-surface detail and latent imprints (residual color, bitemarks, fingerprints) are important. Moreover, our system makes complex analysis accessible in simple formats for a broader life science research community.  Multispectral datasets provide new information for NPR (Figure~\ref{fig:subsurface}). Figure~\ref{fig:sailboat} illustrates multiscale, multispectral curvature shading and line extraction for analyzing engraved bone carvings (\emph{scrimshaw}). Spectral band shading with curvature shading emphasizes shapes (grooves and hill features) and suppressed markings. Figure~\ref{fig:sbsbandcontrol} demonstrates spectral band control for studying pigments that change appearance at different wavelengths. Figure~\ref{fig:irblending} shows pipeline flexibility. Near-infrared blending enhances circular patterns on the underside of the butterfly wing. The artistic effect was created by filtering visible color and near-infrared spectra and applying Lambertian shading. Per pixel, per channel blending of quantized near-infrared \hbox{2-D} spectra creates a toon-like effect.



\noindent\textbf{Related Work} We leverage acquisition systems from prior work that use narrow-band pass filters~\cite{Habel2012,Toque09,HordleyFM04} to record material properties and incorporate multispectral shape~\cite{Brusco2006,minim2012}. Our acquired data can be classified with multichannel image data types that store both color and surface shape~\cite{Malzbender01} in the form of normals~\cite{Rushmeier99} computed from multi-illumination photography. These data types permit interactive relighting for detail enhancement and have been used to analyze cultural heritage artifacts~\cite{tolerfranklinvastcourse2010_full}. Our focus is processing and rendering that maximizes the captured data by combining stylization information across spectra, a methodology not available even for volumetric rendering in NPR~\cite{Rusinkiewicz06,Schein04,Burns05}. 

Our shading methods relate to multiscale  NPR decomposition that combines shape details across \emph{multilight} images~\cite{Fattal2007}. Exaggerated shading uses normal stacks to maximize contrast across all frequencies and orientations~\cite{Rusinkiewicz06}. Multiscale curvature shading uses derivatives of surface normals to darken indentations and lighten ridges~\cite{Kindlmann03,Miller94,Zhukov98,rgbn07}. Linear shape cues delineate depth discontinuities~\cite{Dooley90}, crest lines~\cite{Ohtake04}, and suggestive contours~\cite{DeCarlo03}. Our \emph{near-infrared enhancement maps} enable multiscale feature enhancement across material layers.

Photo-enhancement methods~\cite{ZhangSM08} sharpen image details using wavelet decomposition methods that isolate and combine visible color with near-infrared information. Variants permit haze removal, skin smoothing, illuminant detection and material classification~\cite{Susstrunk2010}. Tristimulus estimation from full spectra data reduces color errors in realistic images~\cite{Darling2011,HordleyFM04}. Mesh refinement methods use near-infrared data to modify visible \hbox{3-D} shape~\cite{Choe2016RefiningGF,Choe2014}. Our contrast operators emphasize variations in the sub-surface detail ignored in these solutions.


\section{Conclusion}

We presented principles and tools for processing multispectral data for NPR. Our user study demonstrated the utility of our pipeline for novel analysis not possible with current NPR. Our project will facilitate data-driven NPR. Our shape validation experiment will inform new shape recovery methods for composite materials. Future work will  extend our algorithms to authentic color reconstruction, relighting, and reconstruction techniques that enable \hbox{3-D} simulation of diverse biological materials.


\section*{Acknowledgements}
This material is based upon work supported by the National Science Foundation under Grant No. 1510410. This work was conducted at the UF Graphics Imaging and Light Measurement Lab (GILMLab).



\bibliographystyle{ACM-Reference-Format}
\bibliography{ms.bib}

                  \appendix 
\section{RSNCC Multimodal Registration}
\label{append:RSNCC}

Given two multispectral (or multimodal) images, $I_1$ and $I_2$, with pixel coordinates $p = (x, y)^{T}$, let $w_p = (u_p, v_p)^T$ denote the displacement mapping $p$ in $I_1$ to $p + w_p$ in $I_2$ and let   $I_{1,p}$ and $I_{2,p}$ denote the color vectors of $I_1$ and $I_2$ for pixel $p$ respectively. The RSNCC function in Equation~\ref{equ:matchcost} computes matching cost $\mathnormal{E}$: 

\begin{equation}
\mathnormal{E} = \left(p,\omega_p\right) =  \rho \left(1- \lvert\Phi_I\left(p,\omega_p \right) \rvert \right) + \tau \rho\left(1-\lvert\Phi_{\nabla I} \left(p, \omega_p \right) \rvert\right)
\label{equ:matchcost}
\end{equation}

\noindent where $\rho\left(x\right)$ is a robust function, $\Phi_{I}\left(p,\omega_p\right)$ is the normalized cross correlation between a patch centered at $p$ in $I_1$  and patch $p + \omega_p$ in $I_2$ in intensity (or color) space. $\Phi \nabla_{I}\left(p, \omega_p \right)$ is a similar term in gradient space. $\tau$ is a weight that combines the color and gradient terms.

The first matching phase is a global optimization,  Equation~\ref{equ:global}, that estimates a homography matrix $\mathnormal{H}$ for image-wise translation, rotation and scaling.

\begin{equation}
\mathnormal{E}\left(\mathnormal{H}\right) = \sum\limits_{p} \mathnormal{E}\left(p, \omega_p\right)
\label{equ:global}
\end{equation}

This is followed by the local optimization in Equation~\ref{equ:local}, a pixel-wise residual displacement estimation incorporating regularization terms

\begin{equation}
\mathnormal{E}\left(\mathbf{w}\right) = \sum\limits_{p} \mathnormal{E^{RSNCC}}\left(p, \omega_p\right) + \lambda_{1} \sum\limits_{p} \psi \left(\parallel \nabla\omega_p \parallel^{2}\right) + \lambda_2 \sum\limits_{q \in \mathnormal{N}\left(p\right)} \parallel \omega_p - \omega_q \parallel
\label{equ:local}
\end{equation}

\noindent where $\mathbf{w} = \left(\mathbf{u}^{T}, \mathbf{v}^{T}\right)^{T}$ is the vector form of $\omega_p$, $\mathbf{u}$ and $\mathbf{v}$ are vectors of $u_p$ and $v_p$ and $\lambda_1$ and $\lambda_2$ are two parameters. 

Dense matching occurs in a coarse-to-fine manner by iteratively updating $\mathnormal{E}\left(\mathbf{w}\right)$ at each level, and then propagating the results to the next level in the sequence for variable initialization. Optical flow methods and iterative reweighted least squares are used to minimize and solve~\ref{equ:local}.


\end{document}